\documentclass[%
superscriptaddress,
preprint,
 amsmath,amssymb,
 aps,
]{revtex4-1}

\usepackage{graphicx}
\usepackage{overpic}
\usepackage{morefloats}
\usepackage{dcolumn}
\usepackage{bm}

\usepackage[usenames,dvipsnames]{color}
\usepackage[normalem]{ulem}

\renewcommand{\vec}[1]{\mathbf{#1}}

\begin{document}


\title{Phase-field crystal description of active crystallites: elastic and inelastic collisions}

\author{Lukas Ophaus}
\thanks{These two authors contributed equally}
\affiliation{Institut f\"ur Theoretische Physik, Westf\"alische
Wilhelms-Universit\"at M\"unster, Wilhelm-Klemm-Strasse 9, 48149
M\"unster, Germany}
\affiliation{Center of Nonlinear Science (CeNoS), Westf\"alische
Wilhelms-Universit\"at M\"unster, Corrensstrasse 2, 48149 M\"unster,  Germany}

\author{Johannes Kirchner}
\thanks{These two authors contributed equally}
\affiliation{Institut f\"ur Theoretische Physik, Westf\"alische
Wilhelms-Universit\"at M\"unster, Wilhelm-Klemm-Strasse 9, 48149
M\"unster, Germany}

\author{Svetlana V. Gurevich}
\thanks{ORCID: 0000-0002-5101-4686}
\affiliation{Institut f\"ur Theoretische Physik, Westf\"alische
Wilhelms-Universit\"at M\"unster, Wilhelm-Klemm-Strasse 9, 48149
M\"unster, Germany}
 \affiliation{Center of Nonlinear Science (CeNoS), Westf\"alische
Wilhelms-Universit\"at M\"unster, Corrensstrasse 2, 48149 M\"unster,  Germany}

\author{Uwe Thiele}
 \thanks{ORCID: 0000-0001-7989-9271}
\email{u.thiele@uni-muenster.de}
\homepage{http://www.uwethiele.de}
 \affiliation{Institut f\"ur Theoretische Physik, Westf\"alische
Wilhelms-Universit\"at M\"unster, Wilhelm-Klemm-Strasse 9, 48149
M\"unster, Germany}
 \affiliation{Center of Nonlinear Science (CeNoS), Westf\"alische
Wilhelms-Universit\"at M\"unster, Corrensstrasse 2, 48149 M\"unster,  Germany}

\date{\today}

\begin{abstract}
The active Phase-Field-Crystal (aPFC) model combines elements of the Toner-Tu theory for self-propelled particles and the classical Phase-Field-Crystal (PFC) model that describes the transition between liquid to crystalline phases. In the liquid-crystal coexistence region of the PFC model, crystalline clusters exist in the form of localized states that coexist with the homogeneous background. At sufficiently strong self-propulsion strength (or activity) they start to travel. We employ numerical path continuation and direct time simulations to first investigate the existence regions of different types of localized states for a one-dimensional system. The results are summarized in morphological phase diagrams in the parameter plane spanned by activity and mean concentration. Then we focus on the interaction of traveling localized states studying their collision behavior. As a result we distinguish 'elastic' and 'inelastic' collisions. In the former, localized states recover their properties after a collision while in the latter they may annihilate or form resting or traveling bound states. In passing, we describe oscillating and modulated traveling localized states that have as the steadily traveling localized states no counterpart in the classical PFC model.
\end{abstract}

\maketitle


\section{\label{sec:intro}Introduction}
The formation of patterns has always been an intriguing phenomenon for scientists and laypersons alike. Regular spatial, temporal and spatio-temporal patterns universally occur in nature ranging from physical, chemical and biological systems to geological and even social systems \cite{Ball1999,Lam1998,Pismen2006,Misbah2017}. Classifying macroscopic physical systems, one can distinguish, on the one hand, passive systems that are normally closed and develop towards thermodynamic equilibrium. The resulting structures can show spatial patterns like, e.g., regular crystal lattices, and one normally relates them to self-assembly as typical structure lengths directly result from the properties of individual constituents. 
On the other hand, there are active or out-of equilibrium systems that are usually open systems and develop under permanent energy conversion. They normally show self-organized and dissipative structures whose typical length scales depend on transport coefficients and rate constants \cite{CrHo1993rmp}. Such structures do not persist when the driving fluxes are switched off. One example are systems composed of active particles~\cite{wadaPRL99,rappelPRL83,vicsekPRE74,sumpter} that transform, e.g., chemical into mechanical energy resulting in their self-propelled directed motion~\cite{Marchetti.RevModPhys.85,BDLR2016rmp}. Ensembles of such particles can show fascinating collective phenomena. Long- and short-range interactions between the particles can result in polar ordering and a synchronization of the particle motion~\cite{uchidaPRL106,golestanianSoftMatter7}. The coordinated collective motion of many such active particles is often called swarming~\cite{Marchetti.RevModPhys.85}. Striking examples are schools of fish and flocks of birds, sometimes referred to as ``living particles''. Beside the naturally occurring examples of self-propelled organisms, also artificial active particles are developed. For instance, there exist microswimmers that employ light~\cite{PalacciScience,jiangPRL105}, ultrasound~\cite{WangACSNano2012,VoWi2020preprint} or chemical energy \cite{howsePRL99} to fuel their ``engines''.

Similar to the case of equilibrium systems, for the active particles one can distinguish different phases of the ensemble behavior. Parameters like the particle density, the strength of the active driving of individual particles and the strength and specific type of interactions between neighboring particles determine whether one encounters motility-induced clustering or random motion of individual particles or moving swarms with a high degree of spatial order. Such ordered swarms may be seen as a crystalline state while the disordered states may be seen as gas-like or liquid-like~\cite{BDLR2016rmp,MaVC2018arpc}. Phase separations between such gas- and liquid-like states can be induced by solely changing the particle motility~\cite{Ginot2015prx,SSWK2015prl,CaTa2015arcmp}. At very high densities, also highly ordered resting~\cite{Thar2002,Thar2005} and traveling~\cite{PalacciScience,theurkauff2012prl,LibchaberPRL2015,ginot2018aggregation,ToTR2005ap,MSAC2013prl,BDLR2016rmp} crystalline states can be formed.

Systems of active particles are often investigated with particle-based models employing large-scale direct numerical simulations. Besides, there exists a wide range of continuum models for active media~\cite{ToTR2005ap,Marchetti.RevModPhys.85,Menz2015prspl,RKBH2018pre}. For instance, the Tuner-Tu model is a prominent continuum theory for ``flocking''~\cite{ToTu1995prl,TonerTu} that expands the Navier-Stokes description of classical fluids by various active terms that break the Galilean invariance by distinguishing the frame of reference of resting particles. Elements of the Tuner-Tu model are combined with the Phase-Field-Crystal (PFC) model to form a simple model for active interacting particles~\cite{MenzelLoewen}. The PFC model is widely employed as a generic continuum model to study the dynamics of atomistic and colloidal crystallization on microscopic length and diffusive time scales \cite{ElderGrantPRL88,EmmerichPFC,TGTP2009prl,ERKS2012prl}. In the context of continuum models for pattern formation \cite{CrHo1993rmp}, the PFC model represents an equivalent with mass-conserving dynamics of the Swift-Hohenberg (SH) equation, a generic equation for the formation of steady patterns close to a short-wave instability. Therefore a PFC model is sometimes also called the conserved Swift-Hohenberg (cSH) equation \cite{TARG2013pre,Knob2016ijam}.

The resulting active Phase-Field-Crystal model (aPFC)~\cite{MenzelLoewen} describes transitions between a liquid phase and resting and traveling crystalline phases of ensembles of active particles \cite{MenzelLoewen,MenzelOhtaLoewenPhysRevE.89,ChGT2016el,OphausPRE18}. On the technical level, the aPFC model couples the passive cSH equation that describes the dynamics of a density-like conserved order parameter field with the nonconserved dynamics of a polarization field that represents the coarse-grained orientation and direction of motion of the self-propelled particles. Employing the aPFC model, a variety of different periodic \cite{MenzelLoewen,ChGT2016el} and localized \cite{OphausPRE18} active crystals is described in the literature. The additional influence of hydrodynamic interactions is also studied ~\cite{MenzelOhtaLoewenPhysRevE.89} by incorporating an additional velocity field into the model.

The main subject of the present work is the existence and interaction of active crystallites, i.e., localized crystalline patches, also known as localized states of the aPFC equation. In general, localized states are experimentally observed and modeled in various areas of biology, chemistry and physics \cite{MathBio,BioPatterns,coulletPRL84,ChemWaves,BurkeKnoblochLSgenSHe,PuBA2010ap,AA-LNP-08}. Examples include localized patches of vegetation patterns in a bare background \cite{TM_PRL_99,MERON2004}, local arrangements of free-surface spikes of magnetic fluids occurring in the bistable region closely below the onset of the Rosenzweig instability \cite{richterPRL94} and localized spot patterns in nonlinear optical systems \cite{SchaepersPRL2000} and oscillating localized states (oscillons) in vibrated layers of colloidal suspensions \cite{LiouPRL1999}.

In the context of solidification described by PFC models, localized states are observed in and near the thermodynamic coexistence region of liquid and crystal state.  Crystalline patches of various size and symmetry can coexist with a liquid environment depending on control parameters as mean density and temperature \cite{RATK2012pre,TARG2013pre,EGUW2018springer}. For instance, increasing the mean density, the crystallites increase in size as further density peaks (depending on context also called ``bumps'', or ``spots'') are added at the crustal-liquid interface. Ultimately, the whole domain fills up and for a finite domain, branches of localized states in a bifurcation diagram terminate on the branch of space filling periodic states. Within their existence region, different branches of localized states form a ladder-like structure of ``snaking'' branches \cite{BurkeKnoblochSnakingChaos2007,SandstedeSnakes}. There is an important difference between systems with and without a conservation law like the PFC model and the SH model, respectively. Namely, the snaking curves of localized states are slanted \cite{TARG2013pre, BoCR2008pre,Dawe2008sjads,LoBK2011jfm,PACF2017prf} and aligned \cite{K_IMA16,BurkeKnoblochSnakingChaos2007,ALBK2010sjads,LSAC2008sjads}, respectively. 

A first investigation of localized states in an active PFC model is given in Ref.~\cite{OphausPRE18} that determines the linear stability analysis of homogeneous states, analyzes the onset of motion of periodic and localized states in one dimensional domains and provides corresponding bifurcation diagrams. The present work deepens this analysis by providing a more extensive analysis of the parameter regions where the various localized states exist accompanied by selected bifurcation diagrams. The gained knowledge is then employed in an exploration of the collision behavior of localized states. This ranges from one-to-one collisions of traveling localized states to interactions of ensembles of localized states. This allows us to introduce concepts as the critical free path necessary for fully elastic collisions.

Our work is structured as follows. In Section~\ref{sec:model} we briefly present the active PFC model and discuss its analytical and numerical treatment. After revising the linear stability of the liquid phase and the overall phase behavior in section~\ref{sec:LSA}, spatially localized states are discussed in Section~\ref{sec:LS}. Their bifurcation structure is analyzed focusing on two main control parameters, namely, mean density and strength of self-propulsion. The analysis of the collision behavior is presented in Section~\ref{sec:scattering}. The final Section~\ref{sec:conc} provides a conclusion and outlook.

\section{The Model}\label{sec:model}
\subsection{Governing equations}\label{sec:level2}
%
The active PFC model is introduced by Menzel and L\"owen in Ref.~\cite{MenzelLoewen}. Its order parameters are the scalar field $\psi(\mathbf{r},t)$ that corresponds to a scaled shifted density and the vector field $\mathbf{\ensuremath{P}}(\mathbf{r},t)$ referred to as ``polarization''. It denotes the local coarse grained orientational order of the active particles that is identical to the net direction of self-propulsion. In particular, $\psi(\mathbf{r},t)$ is the modulation about a fixed mean density $\bar{\psi}$, hence $\int_\Omega\psi\, \mathrm{d^n r}=0$, where $\mathbf{r}\in \Omega \subset \mathbb{R}^\mathrm{n}$ and $\Omega$ is the considered domain in $n$-dimensional space. Here it is presented in nondimensional form.

The aPFC model~\cite{MenzelLoewen} combines a conserved dynamics for the density
\begin{equation}
  \partial_{t}\psi =  \nabla^{2}\frac{\delta\mathcal{F}}{\delta\psi}-v_{0}\nabla\cdot\mathbf{P},\label{eq:dtpsigrad}
\end{equation} 
and a non-conserved dynamics of the polarization field
\begin{equation}
\partial_{t}\mathbf{P} = \nabla^{2}\frac{\delta\mathcal{F}}{\delta\mathbf{P}}-D_{\mathrm{r}}\frac{\delta\mathcal{F}}{\delta\mathbf{P}}-v_{0}\nabla\psi. \label{eq:dtPgrad}
\end{equation} 
Here, the strength of self-propulsion $v_0$ is at the same time the strength of the only coupling of the fields.  The used coupling has the simplest allowed form that does not break the conservation of $\psi$~\cite{ChGT2016el,MenzelLoewen,OphausPRE18}, i.e., its evolution still follows a continuity equation $\partial_t\psi=-\nabla\cdot\vec{j}$ with a flux $\vec{j}$.

The polarization $\mathbf{P}$ undergoes translational and rotational diffusion with the rotational diffusion constant $D_{\mathrm{r}}$. A source term is proportional to the density gradient. Note, that advection with the material flux $\vec{j}$ is not considered, i.e., it is assumed that diffusive processes dominate the polarization dynamics.

In the limiting case of a passive system ($v_0=0$), the dynamics of $\psi(\mathbf{r},t)$ and $\mathbf{\ensuremath{P}}(\mathbf{r},t)$ is variational, i.e., the two equations reduce to a respective gradient dynamics on the underlying free energy functional $\mathcal{F}[\psi,\vec{P}]$. If the functional does not contain any energetic coupling (as is the case here), the dynamics of $\psi$ and $\mathbf{\ensuremath{P}}$ completely decouple.
The free energy functional
\begin{equation}
\mathcal{F}=\mathcal{F}_{\mathrm{pfc}}+\mathcal{F}_{\mathbf{P}}\label{eq:functional}
\end{equation}
is composed of the standard Phase-Field-Crystal functional~\cite{ElderGrantPRL88, ElderGrantPRE70, EmmerichPFC}  (identical to the Swift-Hohenberg functional \cite{CrHo1993rmp}) with cubic nonlinearity 
\begin{equation}
 \mathcal{F}_{\mathrm{pfc}}[\psi] = \int \mathrm{d^nr}\left\{ \frac{1}{2}\psi\left[\epsilon+\left(1+\nabla^{2}\right)^{2}\right]\psi+\frac{1}{4}(\psi+\bar{\psi})^{4}\right\},
 \label{eq:functionalpsi}
\end{equation}
and the orientational part
\begin{equation}
\mathcal{F}_{\mathbf{P}}[\vec{P}]=\int \mathrm{d^nr} \left(\tfrac{C_1}{2}\mathbf{P}^{2}+\tfrac{C_2}{4}\mathbf{P}^{4}\right).
\label{eq:functionalP}
\end{equation}
The functional (\ref{eq:functional}) describes transitions between a uniform and a periodically patterned state~\cite{EmmerichPFC}. The negative interfacial energy density ($\sim-|\nabla\psi|^2$, obtained by partial integration from the $\psi\Delta\psi$ term) favors the creation of interfaces and is only limited by the stiffness term ($\sim(\Delta\psi)^2$).  The quartic bulk energy contains the  temperature-like parameter $\epsilon$. For negative values (low temperatures, i.e., for an undercooled liquid), $\psi$ forms a periodic (crystalline) state, while high temperatures result in a uniform (liquid) phase. 

The orientational part [Eq.~(\ref{eq:functionalP})] with $C_1<0$ and $C_2>0$ represent a double-well potential that results in a spontaneous polarization of absolute value $\sqrt{-C_1/C_2}$ and arbitrary orientation. Here, we concentrate on the already rich behavior obtained for positive $C_1>0$ and vanishing $C_2=0$ as our analysis shall directly connect to former studies~\cite{MenzelLoewen,MenzelOhtaLoewenPhysRevE.89,ChGT2016el,OphausPRE18}. At positive $C_1>0$ diffusion tends to reduce the polarization.

Introducing the functional Eqs.~(\ref{eq:functional}) into Eqs.~(\ref{eq:dtpsigrad}) and~(\ref{eq:dtPgrad}) and executing the variations yields the coupled evolution equations
\begin{align}
\partial_{t}\psi &= \nabla^{2}\left\{\left[\epsilon+\left(1+\nabla^{2}\right)^{2}\right]\psi+\left(\bar{\psi}+\psi\right)^{3}\right\}-v_{0}\nabla\mathbf{\cdot P}, \label{eq:dtpsi} \\
\partial_{t}\mathbf{P} &= C_1\nabla^{2}\mathbf{P} - D_{\mathrm{r}}C_1\mathbf{P}-v_{0}\nabla\psi,
\label{eq:dtP}
\end{align}
where the equation for $\mathbf{P}$ is fully linear.
%
\subsection{\label{sec:level2}Steady and stationary states}
%
In the following, we study Eqs.~(\ref{eq:dtpsi}) and (\ref{eq:dtP}) in one spatial dimension. Then, the polarization becomes a scalar field $P(x,t)$ that describes the coarse grained strength of local ordering and self-propulsion as well as its sense of direction. Beside fully time-dependent behavior studied by direct numerical simulation, we analyze stable and unstable steady and stationary states of the system described by Eqs.~(\ref{eq:dtpsi}) and ~(\ref{eq:dtP}). Here, we denote states at rest as ``steady'' while uniformly traveling states are denoted as ``stationary''.  The latter are steady in a frame moving with some constant velocity $c$. After a coordinate transformation into the comoving frame, i.e., replacing $\partial_{t}\psi=c\partial_x\psi$ and $\partial_{t}P=c\partial_x P$ we once integrate Eq.~(\ref{eq:dtpsi}) and obtain the system of ordinary differential equations (ODEs)
\begin{align}
0 =& \partial_{x}\left\{\left[\epsilon+\left(1+\partial_{xx}\right)^{2}\right]\psi+\left(\bar{\psi}+\psi\right)^{3}\right\}-v_{0}P -c \psi-j, \label{eq:steadystatePSI} \\
0 =& C_1 \partial_{xx}P-D_{\mathrm{r}}C_1 P - v_{0}\partial_{x}\psi-c\partial_{x}P. \label{eq:steadystateP}
\end{align}
Here, the integration constant $j$ represents the constant flux in the comoving frame. For resting states velocity $c$ and flux $j$ are zero while for traveling states they have to be determined together with the solution profiles.

The trivial liquid (uniform) state $(\psi_0=0, P_0=0)$ always solves Eqs.~(\ref{eq:steadystatePSI}) and (\ref{eq:steadystateP}) and its linear stability is discussed in the next section. Changes in stability of the liquid state are related to bifurcations where resting and traveling crystalline (periodic) states emerge that themselves can in secondary bifurcations give rise to branches of crystallites (localized states). Bifurcation diagrams show how the various branches of such spatially-modulated states $(\psi=\psi(x), P=P(x))$ are related. To determine them we transform Eqs.~(\ref{eq:steadystatePSI}) and (\ref{eq:steadystateP}) into a system of seven first order ODEs, employ periodic boundary conditions, an integral condition for mass conservation and an integral condition that projects out the translational symmetry mode and employ the numerical pseudo-arclength continuation techniques \cite{KrauskopfOsingaGalan-Vioque2007,DWCD2014ccp,EGUW2019springer} bundled in the toolbox \textsc{AUTO07p} \cite{DoKK1991ijbc,DoedelOldeman2009}. For detailed explanations and hands-on tutorials see \cite{cenosTutorial,EGUW2019springer}. In the context of SH and PFC (cSH) models such methods are extensively used, e.g., in Refs.~\cite{BurkeKnoblochLSgenSHe,ALBK2010sjads,TARG2013pre,TFEK2019njp}. However, they are less frequently applied to the aPFC model \cite{OphausPRE18} and other models of active matter \cite{StGT2020c,TSJT2020pre}.

For the direct numerical simulation we use a pseudo-spectral method. After choosing initial conditions, Eqs. (\ref{eq:dtpsi}) and (\ref{eq:dtP}) are integrated forward in time via a semi-implicit Euler method, while spatial derivatives and nonlinearities are calculated in Fourier space and in real space, respectively.

\section{Overall phase behavior}
\label{sec:LSA}
%
\begin{figure}
 \centering
 \includegraphics[width=0.7\textwidth]{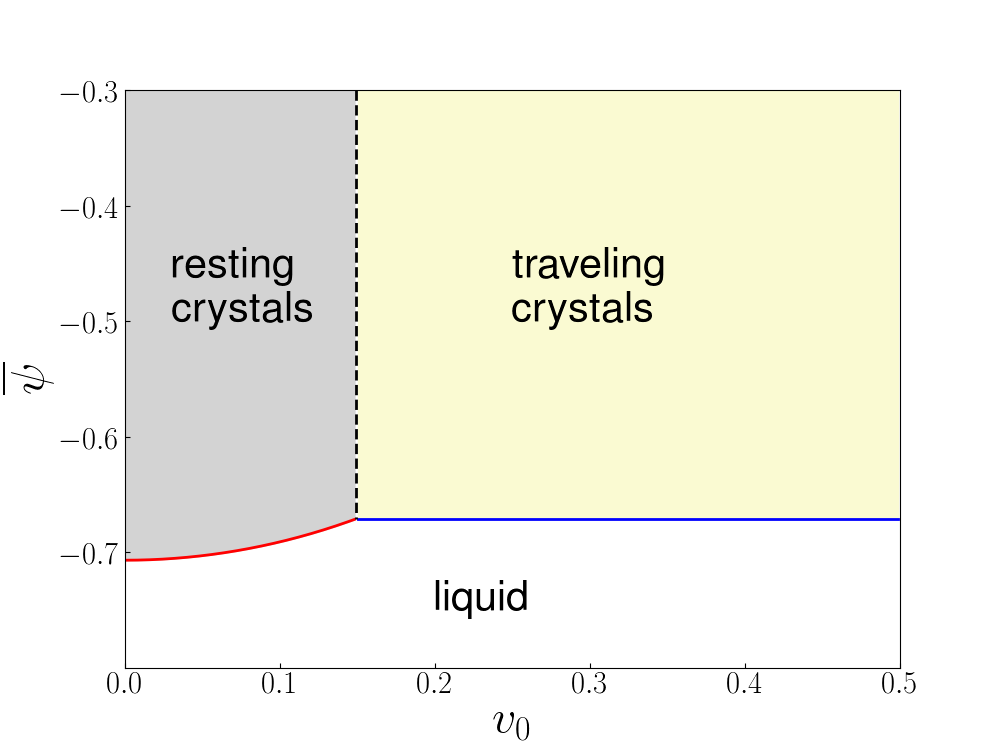}
 \caption{Linear stability and morphological phase diagram indicating where in the parameter plane spanned by self-propulsion strength $v_0$ and mean concentration $\bar{\psi}$ the three basic domain filling states occur. Namely, these are liquid (uniform) and resting and traveling crystalline (periodic) states. The red line separates the liquid state (no shading) and the resting crystalline state (dark gray shaded), whereas the blue line separates the liquid state and the traveling crystalline state (yellow shaded). The vertical dashed line at $v_0\approx 0.158$ indicates the transition between resting and traveling crystal. The remaining parameters are $\epsilon=-1.5$, $D_{\mathrm{r}}=0.5$, $C_1=0.1$, $C_2=0$.}
 \label{fig:apfc_phasediagram}
\end{figure}

As mentioned above, the trivial homogeneous liquid state $(\psi_0,P_0)=(0,0)$ solves Eqs.~(\ref{eq:steadystatePSI}) and (\ref{eq:steadystateP}) for any mean density $\bar\psi$ and self-propulsion strength $v_0$. However, it may be linearly stable or unstable as can be deduced by a linear stability analysis, i.e., by a linearization of the 
Eqs.~(\ref{eq:steadystatePSI}) and (\ref{eq:steadystateP}) in small fluctuations about the trivial state. This is done in Ref.~\cite{OphausPRE18} where the resulting dispersion relations are discussed in detail (also cf.~\cite{MenzelLoewen,ChGT2016el}). Here, we only present in Fig.~\ref{fig:apfc_phasediagram} the resulting stability and morphological phase diagram. Note that the present results are obtained with the approximate critical wavenumber, $q_\mathrm{crit}=1$. On the scale of Fig.~\ref{fig:apfc_phasediagram} the result can not be distinguished from the exact one.

One distinguishes two cases: (i) the eigenvalue at instability onset is real-valued, the primary bifurcation is a pitchfork bifurcation and the emerging nonlinear state corresponds to a resting crystal (red line and dark gray shaded region in Fig.~\ref{fig:apfc_phasediagram}); and (ii) the eigenvalue at onset is complex-valued, the primary bifurcation is a Hopf bifurcation and the emerging state is a traveling crystal (blue line and yellow shaded region in Fig.~\ref{fig:apfc_phasediagram}).

Above onset, the border between resting and traveling crystals can not be determined by a linear analysis of the trivial state. However, it can be determined employing a velocity expansion close to the drift-pitchfork bifurcation responsible for the transition \cite{OphausPRE18}. The resulting condition for the onset of motion of all types of steady states is 
\begin{equation}
 \Vert \psi\Vert_{2} - \Vert P\Vert_{2} = 0.
 \label{eq:Fredholm}
\end{equation}
Here $\Vert \psi\Vert_{2}$ and $\Vert P\Vert_{2}$ are the $L^2$-norms of the density and polarization field of the resting state, respectively. It is remarkable that the critical activity ($v_0^c\approx 0.15$ in Fig.~\ref{fig:apfc_phasediagram}) is virtually independent of mean density. This is similar to results of Ref.~\cite{ChGT2016el} on the dependence of this border on an effective undercooling.

The corresponding border (vertical dashed line in Fig.~\ref{fig:apfc_phasediagram}) is obtained by tracking the criterion~(\ref{eq:Fredholm}) while numerically continuing the fully nonlinear states. Practically, Eq.~(\ref{eq:Fredholm}) is added as additional integral condition allowing to directly track the onset of motion in the $(v_0, \bar\psi)$-plane. The continuation
result perfectly agrees with the onset of motion obtained by direct time simulations \cite{OphausPRE18}. Note, that due to periodic boundary conditions and fixed domain size the traveling crystal keeps its spatial periodicity. We emphasize that the criterion~(\ref{eq:Fredholm}) is valid for any steady state and is used in the next section to detect drift-pitchfork bifurcations of various localized states. The corresponding critical values of $v_0$ are very close but not identical to the one for periodic states.
%
\section{Localized States}
\label{sec:LS}
%
Similar to the passive PFC model \cite{TARG2013pre}, the active PFC model exhibits a wide range of different localized states (LSs) \cite{OphausPRE18}. These are finite patches of a periodic structure embedded in a homogeneous background, i.e., finite size crystals in coexistence with the liquid phase. For a discussion how these LSs are related to the liquid-crystal phase transition in the thermodynamic limit, i.e., how the Maxwell construction emerges when the domain size is increased towards infinity see Ref.~\cite{TFEK2019njp}.
Here, we refer to the LSs as crystallites. In contrast to the passive PFC model, where all states are at rest, for the active PFC model, resting and uniformly traveling crystallites are described, with a transition at finite critical values of the activity $v_0$ \cite{OphausPRE18}. The onset of motion occurs at finite critical values of the activity $v_0$ through a drift-pitchfork or a drift-transcritical bifurcation. It can be predicted using criterion~(\ref{eq:Fredholm}). Selected first bifurcation diagrams for LSs are given in Ref.~\cite{OphausPRE18}. Here, we scrutinize the emergence of resting and traveling LSs in detail and determine their regions of existence. This information is subsequently used in section~\ref{sec:scattering} to study the collision behavior of traveling LSs.

As in Ref.~\cite{OphausPRE18} we focus on activity strength $v_0$ and mean density $\bar{\psi}$ as main control parameters. As the onset of motion is largely independent of $\bar{\psi}$ (cf.~Fig.~\ref{fig:apfc_phasediagram}), we base our analysis on bifurcation diagrams with control parameter $v_0$ obtained at different fixed $\bar{\psi}$. We begin with the passive case.

\subsection{Slanted snaking in the passive case}
\label{sec:ls-passive}

\begin{figure}
 \centering
  \includegraphics[width=0.7\textwidth]{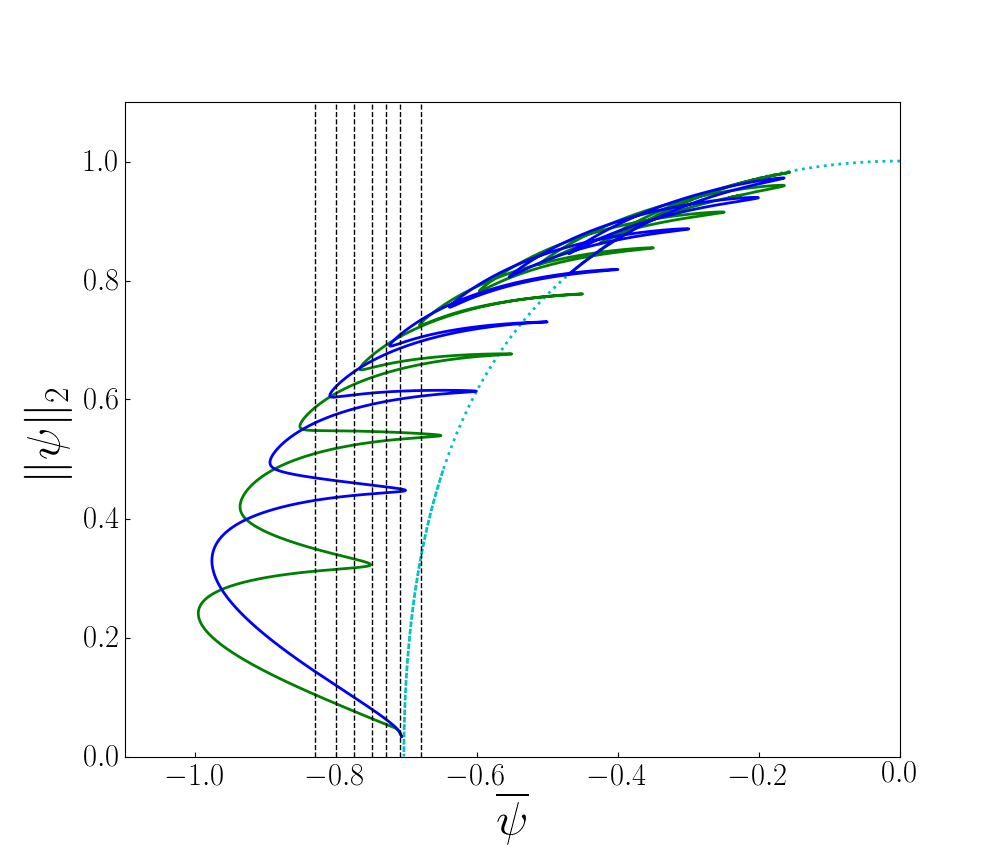}
  \caption{Shown is an example of slanted homoclinic snaking of steady localized states for the passive PFC model, i.e., Eqs.~(\ref{eq:steadystatePSI}) and (\ref{eq:steadystateP}) with $v_0=0$. The bifurcation diagram shows the norm in dependence of mean density $\bar\psi$. At $\bar{\psi}\approx-0.71$ a branch of crystalline (periodic) states with 15 peaks (dotted cyan line) bifurcates off the liquid state (line of zero norm) in a supercritical pitchfork bifurcation. Slightly afterwards two branches of crystallites (LS) branch off subcritically from the branch of periodic states. They consist of symmetric localized steady states with an odd (green line) and an even (blue line) number of peaks, respectively. Undergoing a sequence of non-aligned saddle-node bifurcations they snake towards larger density and norm until ending again on the branch representing the spatially extended crystal. The vertical dashed lines mark the particular fixed values of $\bar{\psi}=$-0.83, -0.80, -0.775, -0.75 ,-0.73, -0.71 and -0.68 for which bifurcation diagrams in $v_0$ are presented below. Parameters are $\epsilon=-1.5$, $D_{\mathrm{r}}=0.5$, $C_1=0.1$, $C_2=0$ and a domain size of $L=100$.}
 \label{fig:pfc-bif}
\end{figure}

For the passive PFC model, the existence of steady LSs and their intricate bifurcation structure of slanted homoclinic snaking is studied for 1d~\cite{TARG2013pre,EGUW2019springer,TFEK2019njp} and 2d systems \cite{EGUW2019springer,TFEK2019njp}. They are obtained by solving Eqs.~(\ref{eq:steadystatePSI}) and (\ref{eq:steadystateP}) with $v_0=0$. Here, Fig.~\ref{fig:pfc-bif} presents as a reference case a corresponding bifurcation diagram employing the mean density $\bar\psi$ as control parameter. The effective temperature is $\epsilon=-1.5$ and the domain size is chosen to $L=100$ as in \cite{TARG2013pre}. As solution measure we employ the $L^2$-norm of the density profile as it is well suited measure to distinguish the solutions. For typical corresponding plots of chemical potential, Helmholtz free energy and grand potential see Refs.~\cite{TARG2013pre,TFEK2019njp}.

The liquid (homogeneous) state $\psi(x)=0$ has zero norm and is represented by the horizontal line. It is stable at small densities (strongly negative $\bar\psi$) and is destabilized at a critical mean density of $\bar{\psi}_c\approx-0.71$ where a branch of crystalline (periodic) states bifurcates in a supercritical pitchfork bifurcation in agreement with the linear stability result for the liquid state (cf.~$v_0=0$ in Fig.~\ref{fig:apfc_phasediagram}). For the chosen domain size these profiles have 15 density peaks.
Above this $\bar{\psi}$ value, fluctuations perturbing the liquid state exponentially grow and a spatially extended crystals or LS form. The supercritical branch itself is only stable in a very small parameter range close to the primary bifurcation. Then, in a secondary bifurcation two subcritical branches of LSs simultaneously emerge. For diverging domain size the secondary bifurcation coincides with the primary one. The two branches consist of LSs that gain stability at a first saddle-node bifurcation and then loose and gain stability 'periodically' as the branches snake towards larger density acquiring larger norms in the process. The LSs on both branches show parity symmetry (reflection symmetry w.r.t.\ the central peak or trough) and show an even and an odd number of density peaks, respectively. In each 'period' of the snaking they symmetrically gain two peaks until the LSs fill the domain and the branches end again on the branch of periodic states. Beyond this bifurcation the latter branch is stable. Further, branches of unstable asymmetric LS exist that connect the two branches of symmetric LS. Here, we characterize individual LS by the number $n$ of peaks they contain, i.e., we speak of $n$-peak solutions. For more details and information how the entire bifurcation structure develops with changing temperature see Ref.~\cite{TARG2013pre}.

Here, the slanted snaking structure of Fig.~\ref{fig:pfc-bif} serves as starting point for an extensive investigation of active crystallites. Next, we present bifurcation diagrams in dependence of $v_0$ at several fixed values of $\bar{\psi}$. They are indicated in Fig.~\ref{fig:pfc-bif} by vertical dashed lines. This shall allow us to understand how the numerous branches of traveling LSs come into existence.

\subsection{Crystallites in the active case}
\label{sec:cont_v0}

\begin{figure}
 \centering
  \includegraphics[width=0.7\textwidth]{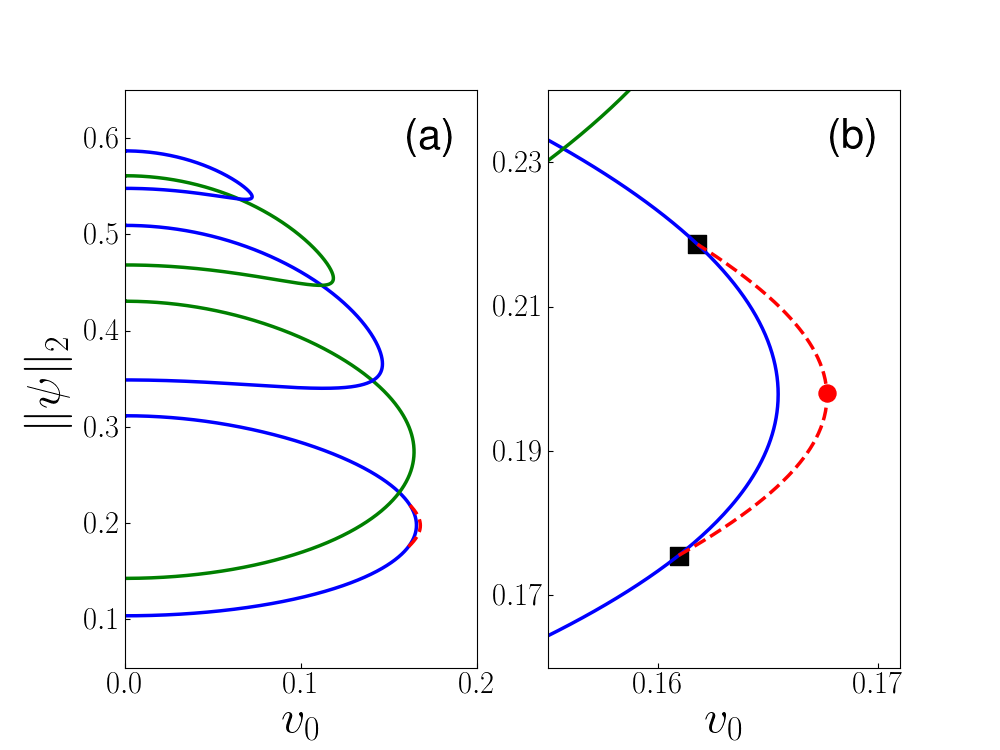}
  \caption{Panel (a) shows the bifurcation diagram as a function of $v_0$ at fixed $\bar{\psi}=-0.83$. Line styles of resting states and remaining parameters are as in Fig.~\ref{fig:pfc-bif}. Branches of traveling localized states are gives as red dashed lines. Panel (b) gives a magnification of the rightmost saddle-node bifurcation of steady LSs  as there the traveling LSs emerge. The filled black square symbols indicate the loci of drift-pitchfork bifurcations while the red circle marks an example solution whose profiles is given in Fig.~\ref{fig:profiles}~(a). Remaining parameters are $\epsilon=-1.5$, $D_{\mathrm{r}}=0.5$, $C_1=0.1$, $C_2=0$, and $L=100$.}
 \label{fig:ubar_0_83}
\end{figure}
\begin{figure}
   \includegraphics[width=1.\textwidth]{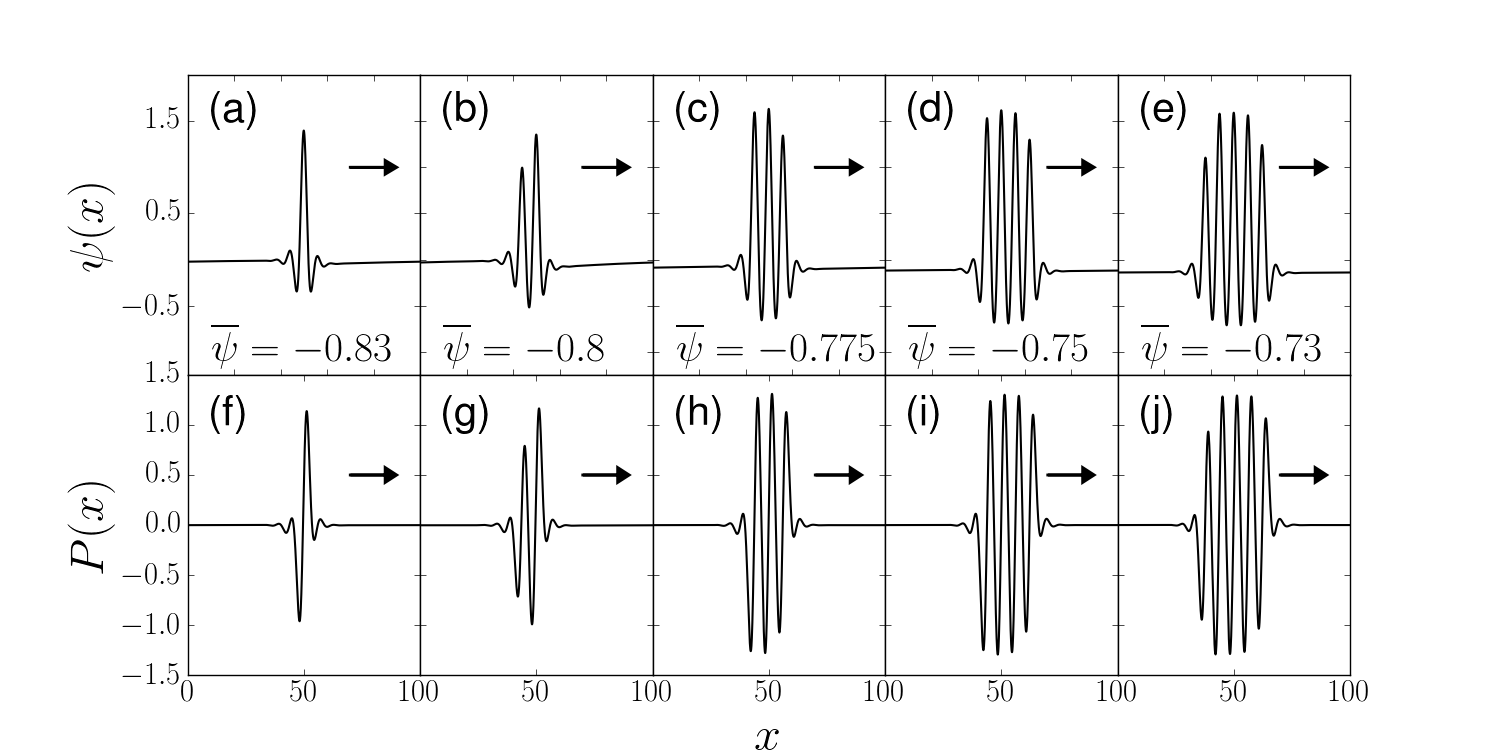}
  \caption{Density (top) and polarization (bottom) profiles $\psi(x)$ and $P(x)$ of various traveling LS for increasing values of $\bar{\psi}$ at fixed $v_0=0.1677$. The peaks travel with a constant velocity. The direction of motion is indicated by the black arrows.  Panels (a) and (f) show a profile from the traveling one-peak branch in Fig.~\ref{fig:ubar_0_83}, (b) and (g) show a traveling two-peak state from Fig.~\ref{fig:ubar_0_80}. (c)/(h), (d)/(i) and (e)/(j) depict larger traveling LSs. Respective values of $\bar{\psi}$ are given in the panels.
  Remaining parameters as in Fig.~\ref{fig:ubar_0_83}. 
  }
 \label{fig:profiles}
\end{figure}

From here on, we only consider the active PFC model. Next, we determine steady and stationary states by solving Eqs.~(\ref{eq:steadystatePSI}) and (\ref{eq:steadystateP}) where for any $v_0\neq 0$ the velocity $v$ of the comoving frame is an unknown that has to be determined as a nonlinear eigenvalue of the problem. The rich sets of LSs obtained in section~\ref{sec:ls-passive} for the passive case are now employed as reference at the corresponding values of \(\bar{\psi}\). First, the relatively small value of $\bar{\psi}=-0.83$ is considered where Fig.~\ref{fig:pfc-bif} shows ten LSs (cf.~intersections with $\bar{\psi}=-0.83$). Each of them is taken as starting solution for a continuation run as a function of $v_0$, the criterion~(\ref{eq:Fredholm}) is monitored to detect drift bifurcations. After branch switching at these bifurcations also branches of traveling states are continued. The resulting bifurcation diagram is presented in Fig.~\ref{fig:ubar_0_83} with selected solution profiles given in Fig.~\ref{fig:profiles}.

We see that the resulting bifurcation curves of steady LS form five separate branches corresponding to different peak numbers between one and five. As before, blue and green branches correspond to odd and even symmetry. With increasing activity the two sub-branches of each branch approach each other till they annihilate at respective saddle-node bifurcations.  This happens first (at smallest $v_0$) for the LS with the largest number of peaks. If a time simulation at large values of $v_0$ is initialized with such a steady LS, the density peaks decay and the crystallite melts into the homogeneous background liquid. The smaller crystallites are slightly more robust, i.e., their annihilating saddle-node bifurcation is at larger $v_0$.

Only on the branch of one-peak LS one encounters drift-pitchfork bifurcations, namely, one supercritical drift-pitchfork bifurcation on each sub-branch, both at nearly identical critical values $v_\mathrm{c}$ slightly above $0.16$. The two drift bifurcations are connected by a branch of traveling one-peak LS that is also limited by a saddle-node bifurcation [see magnification in Fig.~\ref{fig:ubar_0_83}(b)]. All other resting LSs melt before motion can sets in, i.e., all other saddle-node bifurcations of steady state branches are located at a $v_0<v_\mathrm{c}$.
This is confirmed by evaluating the criterion~(\ref{eq:Fredholm}): Only the branch of the one-peak LS reaches sufficiently far for a zero crossing in the difference of the norms to occur. The two drift-pitchfork bifurcations marked by black squares in Fig.~\ref{fig:ubar_0_83}(b) exactly correspond to the zero crossings of the criterion.

Closely inspecting the typical profiles $\psi(x)$ and $P(x)$ of traveling LSs in Fig.~\ref{fig:profiles} we notice that the density profile has lost its reflection symmetry and is now slightly asymmetric. In addition, the relative position of $\psi(x)$ and $P(x)$ has changed: For resting LS, peak maxima in $\psi$ exactly coincide with zeros of $P$ and the resulting inversion symmetry implies that the total polarization (integral of $\psi$ times $P$ over the width of the structure) vanishes, i.e. there is no net polarization. This is not the case for traveling LS where the relative position of $\psi(x)$ and $P(x)$ is shifted and, in consequence, there is a finite net polarization.


\begin{figure}
  \includegraphics[width=0.7\textwidth]{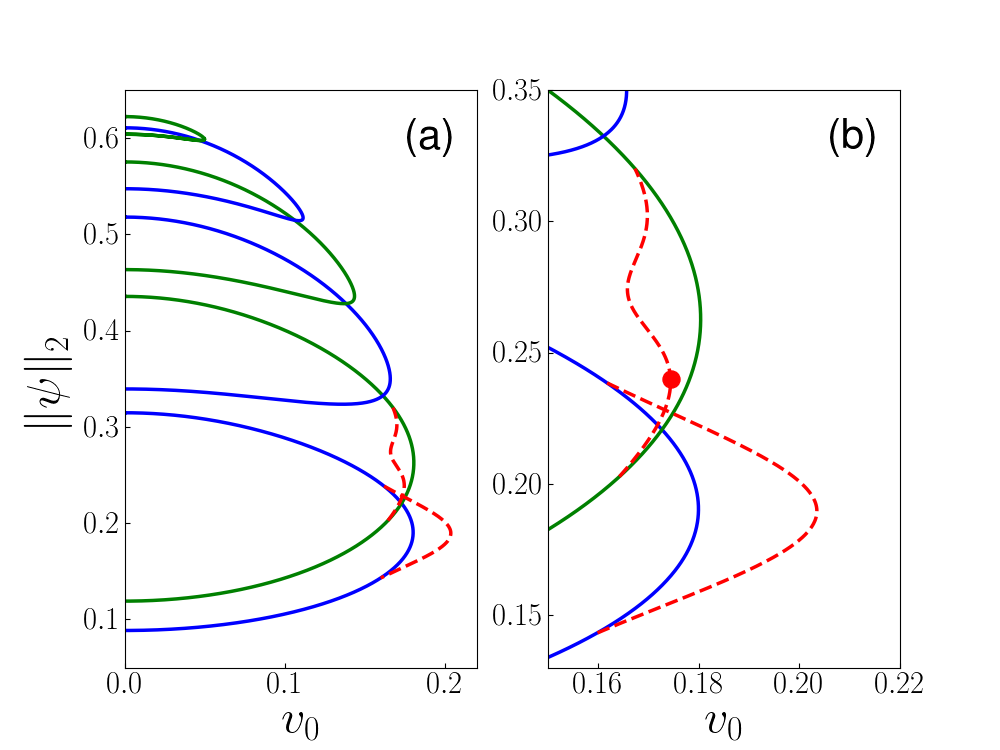}
  \caption{
Panel (a) shows the bifurcation diagram as a function of $v_0$ at fixed $\bar{\psi}=-0.80$ while (b) gives a zoom onto the branches of traveling LSs. Line styles, symbols and remaining parameters are as in Fig.~\ref{fig:ubar_0_83}.}
 \label{fig:ubar_0_80}
\end{figure}

Now we increase the mean concentration to $\bar{\psi}=-0.80$, i.e., there are now 12 steady LSs at $v_0=0$ (cf. Fig.~\ref{fig:pfc-bif}) as an additional branch of resting six-peak LS has appeared above the other branches. The resulting bifurcation diagram over $v_0$ is given in Fig.~\ref{fig:ubar_0_80}. The branches of steady $n$-peak LS are qualitatively similar to the previous case of \(\bar{\psi}=-0.83\) (Fig.~\ref{fig:ubar_0_83}), only the annihilating saddle-node bifurcations have moved to slightly larger $v_0$ and the range of represented norms is larger. More importantly the saddle-node bifurcation where the traveling one-peak LS ends has moved markedly to the right, i.e., the $v_0$-range of their existence is much larger then before. Most remarkably, a second branch of traveling LS has come into existence that consist of two-peak states. Its structure is more involved than the one of the one-peak states as it undergoes three saddle-node bifurcations. However, these states only exist in a small $v_0$-range and the whole branch lies within the region limited by the branch of steady two-peak LS. The profiles of an exemplary traveling two-peak LS are shown in Fig.~\ref{fig:profiles}(b,g).

\begin{figure}
 \centering
 \includegraphics[width=0.7\textwidth]{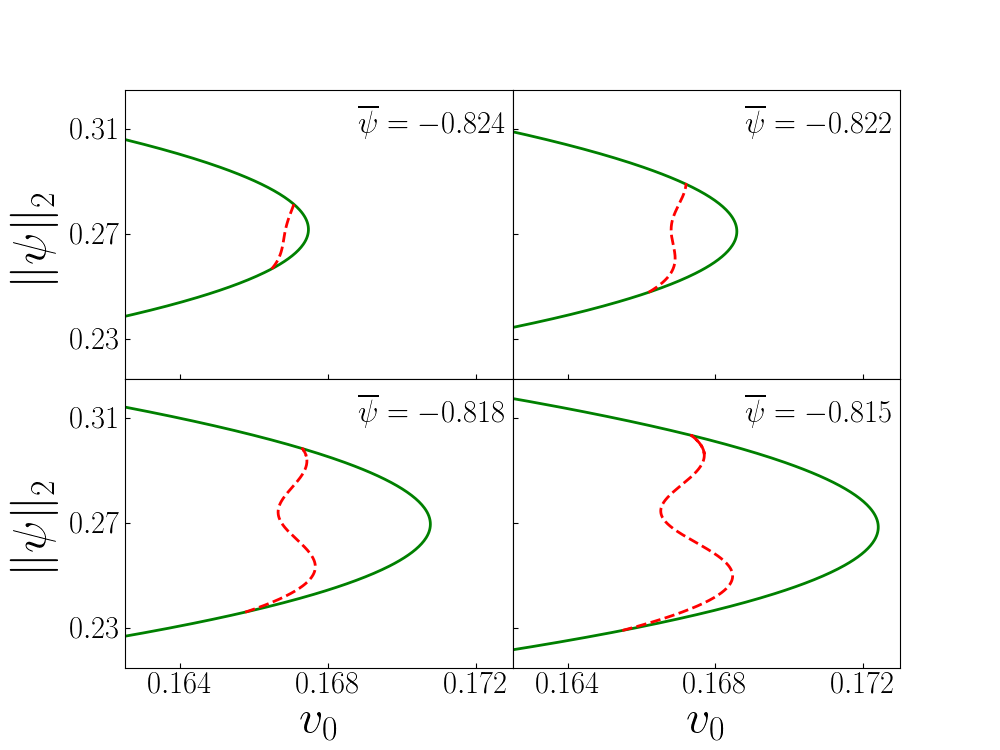}
 \caption{To demonstrate how traveling two-peak LS emerge, a sequence of bifurcation diagrams of two-peak LS is given for mean densities between the ones of Fig.~\ref{fig:ubar_0_83} and Fig.~\ref{fig:ubar_0_80}, namely, for  $\bar{\psi}=-0.824,-0.822,-0.818$ and $-0.815$ (from top left to bottom right). Line styles and remaining parameters are as in Fig.~\ref{fig:ubar_0_83}.
}
 \label{fig:doubleonset}
\end{figure}

As the branch of traveling two-peak LS in Fig.~\ref{fig:ubar_0_80} has a nontrivial form, we further investigate its metamorphosis with increasing \(\bar{\psi}\). Fig.~\ref{fig:doubleonset} shows four intermediate stages between Figs.~\ref{fig:ubar_0_83} and~\ref{fig:ubar_0_80}. From this, it is clear 
that the traveling two-peak branch emerges (as does the branch of traveling one-peak states) via the simultaneous creation of two drift-pitchfork bifurcations at the saddle-node bifurcation of the resting LS (cf.\ top left panel of Fig.~\ref{fig:doubleonset} at \(\bar{\psi}=-0.824\)). As \(\bar{\psi}\) increases, saddle-node bifurcations appear on the branch through a hysteresis bifurcation and a transition from a supercritical to a subcritical drift-pitchfork bifurcation.

\begin{figure}
  \includegraphics[width=0.7\textwidth]{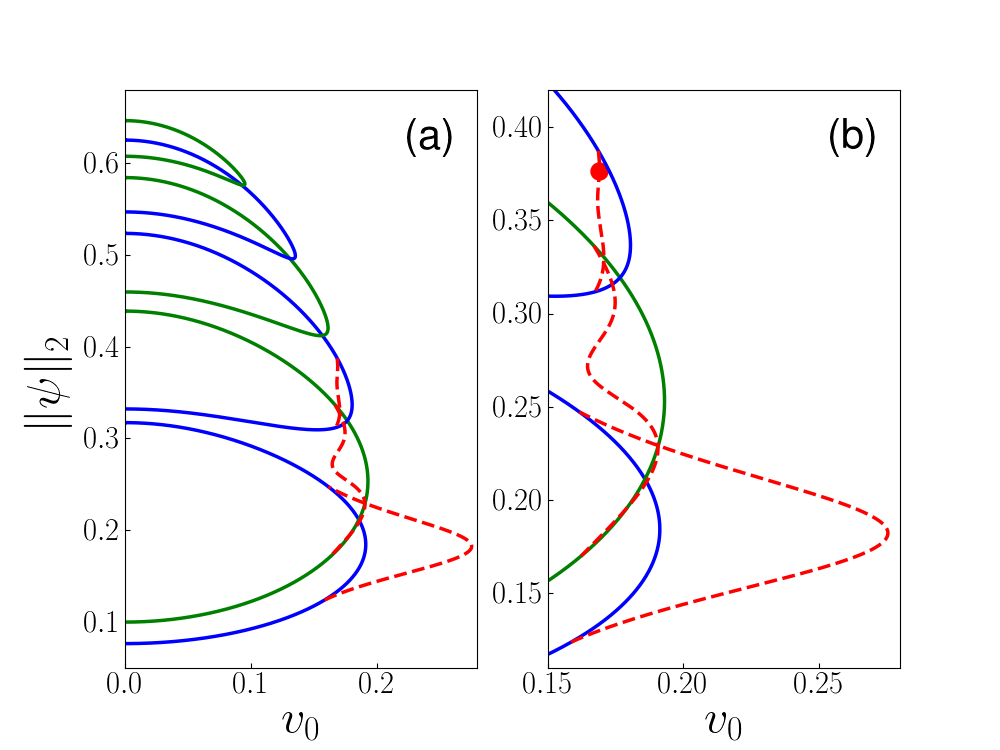}
  \caption{Panel (a) shows the bifurcation diagram as a function of $v_0$ at fixed $\bar{\psi}=-0.775$ while (b) gives a zoom onto the branches of traveling one-peak, two-peak and three-peak LSs. Line styles, symbols and remaining parameters are as in Fig.~\ref{fig:ubar_0_83}.}
 \label{fig:ubar_0_775}
\end{figure}

Increasing the mean density to $\bar{\psi}=-0.775$ we find the bifurcation diagram in Fig.~\ref{fig:ubar_0_775}. In contrast to the previous case, the branch of resting three-peak LSs now reaches activities high enough to result in the emergence of a branch of traveling three-peak LS. Its creation is identical to that of the branch of traveling two-peak LS, cf. Fig.~\ref{fig:doubleonset}. However, the traveling three-peak LS only exist in a very narrow range of $v_0$. An exemplary solution profile of a traveling three-peak LS is given in Fig.~\ref{fig:profiles}.


\begin{figure}
   \includegraphics[width=0.7\textwidth]{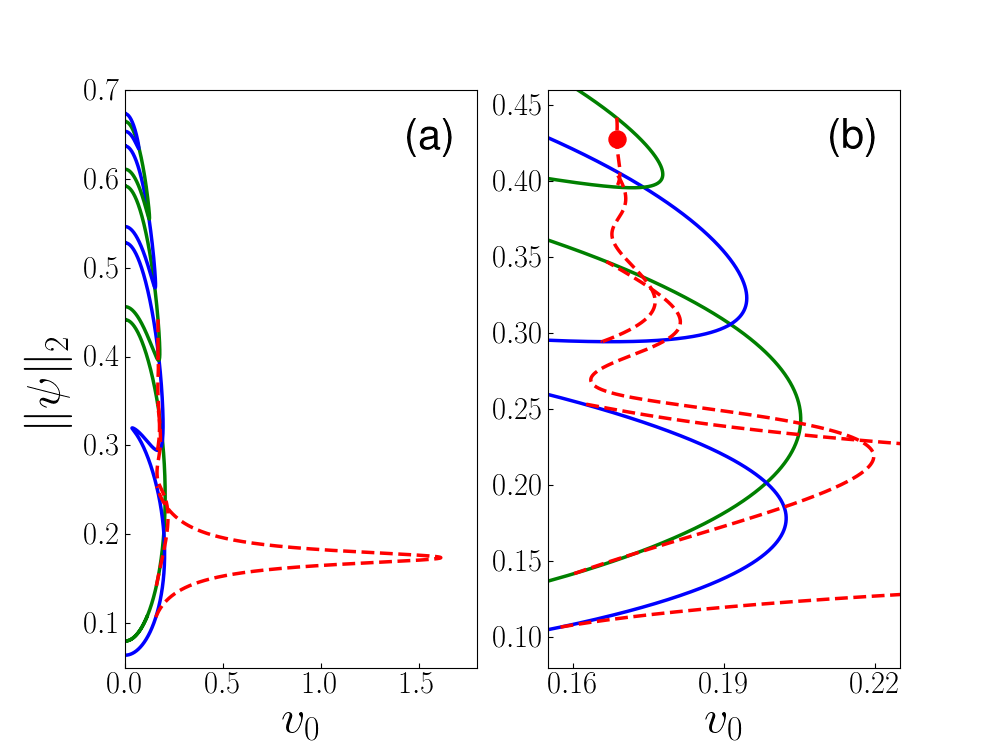}
  \caption{Panel (a) shows the bifurcation diagram as a function of $v_0$ at fixed $\bar{\psi}=-0.75$ while (b) gives a zoom onto the branches of traveling one- to four-peak LS. Line styles, symbols and remaining parameters are as in Fig.~\ref{fig:ubar_0_83}. Corresponding velocities of traveling LSs are given in Fig.~\ref{fig:ubar_0_75_vel}.
    }
 \label{fig:ubar_0_75}
\end{figure}

\begin{figure}
  \includegraphics[width=0.7\textwidth]{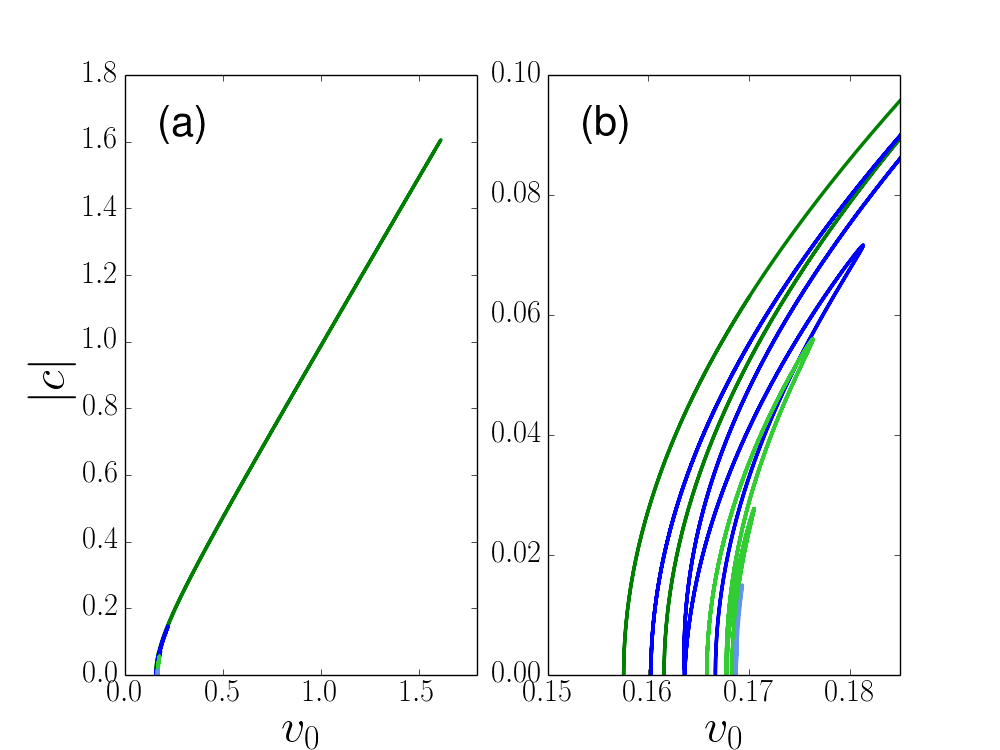}
 \caption{Panel (a) gives the absolute value of the drift velocity $|c|$ as a function of $v_0$ for the traveling LS at $\bar{\psi}=-0.75$, i.e., from Fig.~\ref{fig:ubar_0_75}. (b) gives a zoom onto the range where the drift bifurcations occur. Traveling LS with an odd number of peaks are given as lines in greenish colors (one-peak LS: green, three-peak LS: lime) and LS of even peaks in blueish (two-peak LS: blue, four-peak: short light blue line most right).}
 \label{fig:ubar_0_75_vel}
\end{figure}

A further increase to \(\bar{\psi}=-0.75\) results in the appearance of a branch of steady seven-peak LS and a branch of traveling four-peak LS by mechanisms as discussed before (see Fig.~\ref{fig:ubar_0_75}).  Selected solution profiles are again shown in Fig.~\ref{fig:profiles}. Most remarkably, the ranges of existence of the traveling one-peak and two-peak LS have grown to a large extent while all the other traveling LSs are confined to a small $v_0$-range between \(v_0 \approx 0.16\) and \(v_0 \approx 0.19\). In particular, the branch of traveling four-peak LS is practically vertical. The saddle-node bifurcation where the traveling one-peak LS terminates has moved out to $v_0\approx1.6$. The drift velocities $|c|$ of all traveling LSs in Fig.~\ref{fig:ubar_0_75} are compared in Fig.~\ref{fig:ubar_0_75_vel}. We see that for all traveling LSs, in the vicinity of the onset of motion at respective parameter values $v_c$, the velocity $|c|$ increases like $|v_0-v_c|^{1/2}$ as expected for a drift-pitchfork bifurcation, whereas at large $|v_0-v_c|$ the increase becomes linear. The values $v_c$ increase from about $0.157$ to $0.169$ as the number of peaks increases from one to four. Note that the $|c|$ of the traveling two- and three-peak LS become zero at their end points (as expected) but also at one respective point in the middle of the branch (at $v_0$ approximately $0.164$ and $0.168$, respectively, see Fig.~\ref{fig:ubar_0_75_vel}(b)). This indicates that the appearance of the two saddle-node bifurcations on the traveling LS branches also relates to the appearance of a drift-transcritical bifurcation, where the branch of traveling LS crosses a branch of asymmetric steady LS (not shown here, cf.~Ref.~\cite{OphausPRE18}).


\begin{figure}
  \includegraphics[width=0.7\textwidth]{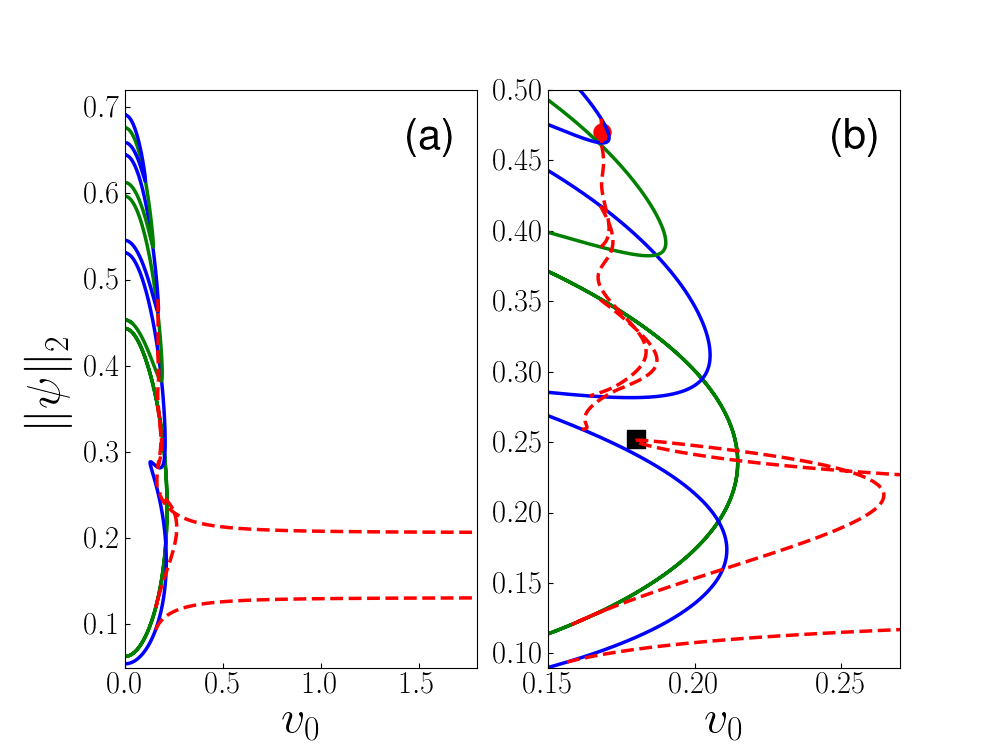}
 \caption{Panel (a) shows the bifurcation diagram as a function of $v_0$ at fixed $\bar{\psi}=-0.73$ while (b) gives a zoom onto the branches of traveling one- to five-peak LS. Traveling one-peak LS exist now for arbitrarily large activity. Here, the black square indicates the newly formed saddle-node bifurcation resulting from the pinch-off bifurcation. Remaining line styles, symbols and parameters are as in Fig.~\ref{fig:ubar_0_83}.}
 \label{fig:ubar_0_73}
\end{figure}

\begin{figure}
 \centering
 \includegraphics[width=0.7\textwidth]{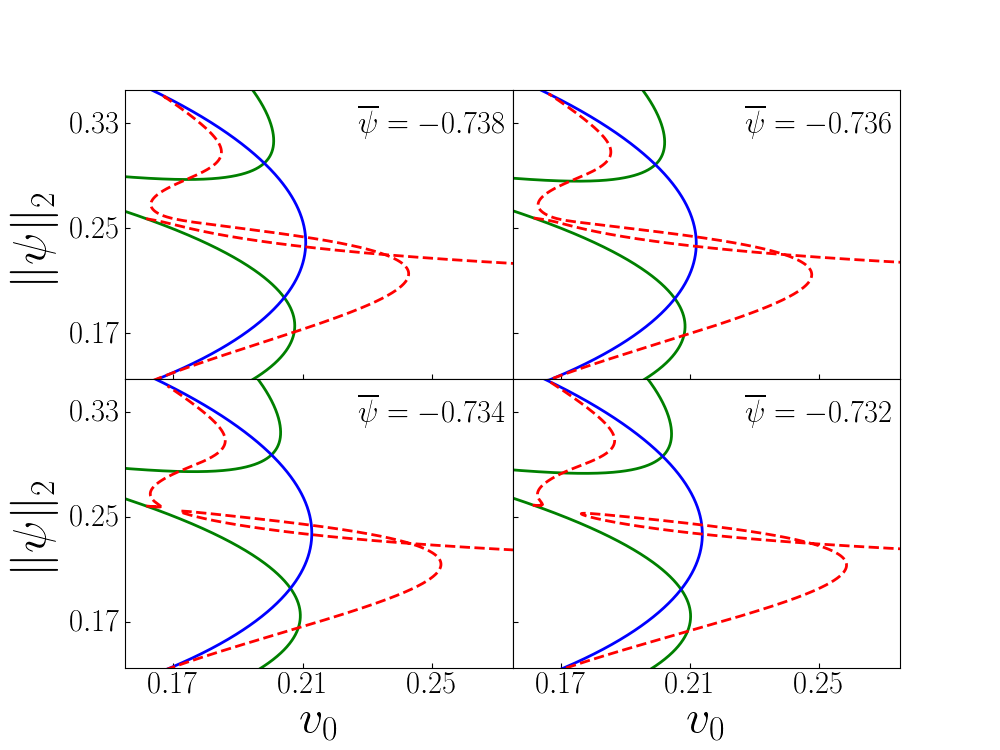}
 \caption{A series of bifurcation diagrams focusing on the pinch-off bifurcation involving the branches of traveling one- and two-peak LSs. Diagrams for $\bar{\psi}=-0.738,-0.736,-0.734$ and $-0.732$ (from top left to bottom right) are shown. Line styles and remaining parameters are as in Fig.~\ref{fig:ubar_0_83}.}
 \label{fig:separation}
\end{figure}

At $\bar{\psi}=-0.73$, see Fig.~\ref{fig:ubar_0_73}, branches of resting one- to five-peak LS exists and each of them is source of a branch of traveling LS, i.e., they also exist with one- to five-peaks. Two further remarkable changes have occurred. First, the bifurcation structure of the branches of resting one-peak and three-peak has changed. They are now connected by an additional saddle-node bifurcation at $v_0\approx0.18$ (Fig.~\ref{fig:ubar_0_73}(b)). Details of the pinch-off transition occurring between $\bar{\psi}=-0.736$ and $\bar{\psi}= 0.734$ at $v_0\approx0.17$ can be discerned in Fig.~\ref{fig:separation}, where a series of bifurcation diagrams focuses on the relevant changes occurring between Figs.~\ref{fig:ubar_0_75} and~\ref{fig:ubar_0_73}. We see that for increasing $\bar{\psi}$ the two nearly aligned parts of the two branches approach each other until, ultimately, they touch and reconnect. There is now an ``open loop'' that is disconnected from the branches of resting LS. In the pinch-off, two saddle-node bifurcations are created. Their separation increases with increasing \(\bar{\psi}\) and the one on the left hand side soon merges into the remaining drift-pitchfork bifurcation.

Second and most importantly, the position of the saddle-node bifurcation where the branch of traveling one-peak LS terminates at high $v_0$ has diverged to $v_0\to\infty$, i.e., these traveling states now exist at arbitrarily high activities (and can reach corresponding velocities).


\begin{figure}
  \includegraphics[width=0.7\textwidth]{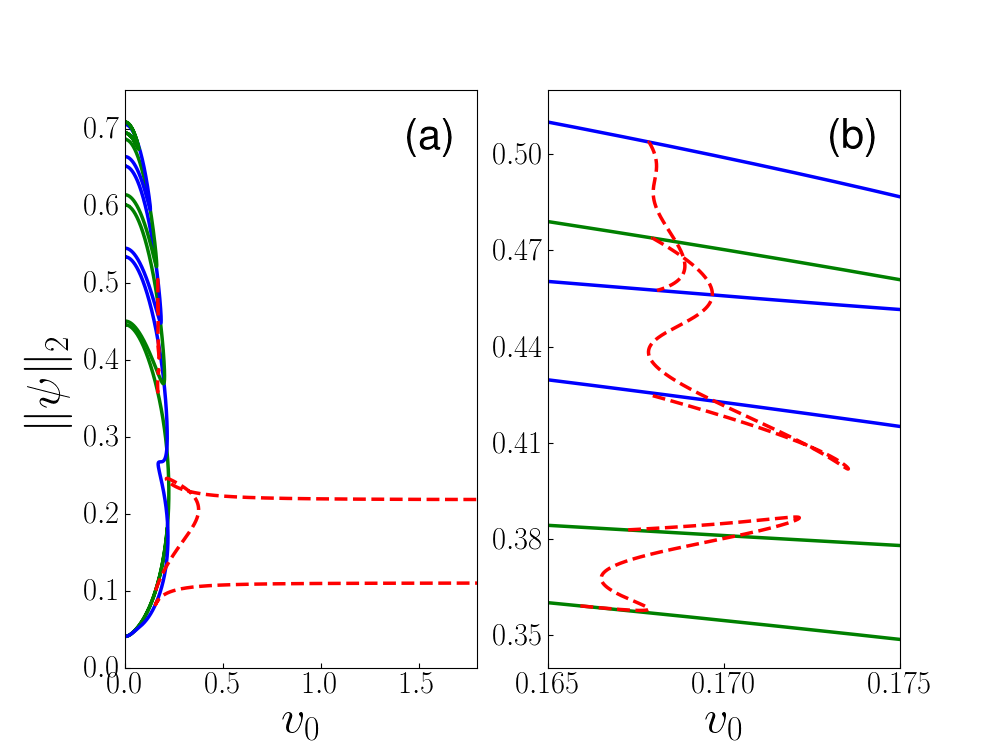}
  \caption{Panel (a) shows the bifurcation diagram as a function of $v_0$ at fixed $\bar{\psi}=-0.71$ while (b) gives a zoom onto the branches of traveling three- to five-peak LSs. Traveling one-peak LSs still exist for arbitrarily high activity. Remaining line styles, symbols and parameters are as in Fig.~\ref{fig:ubar_0_83}.}
 \label{fig:ubar_0_71}
\end{figure}

\begin{figure}
 \centering
 \includegraphics[width=0.7\textwidth]{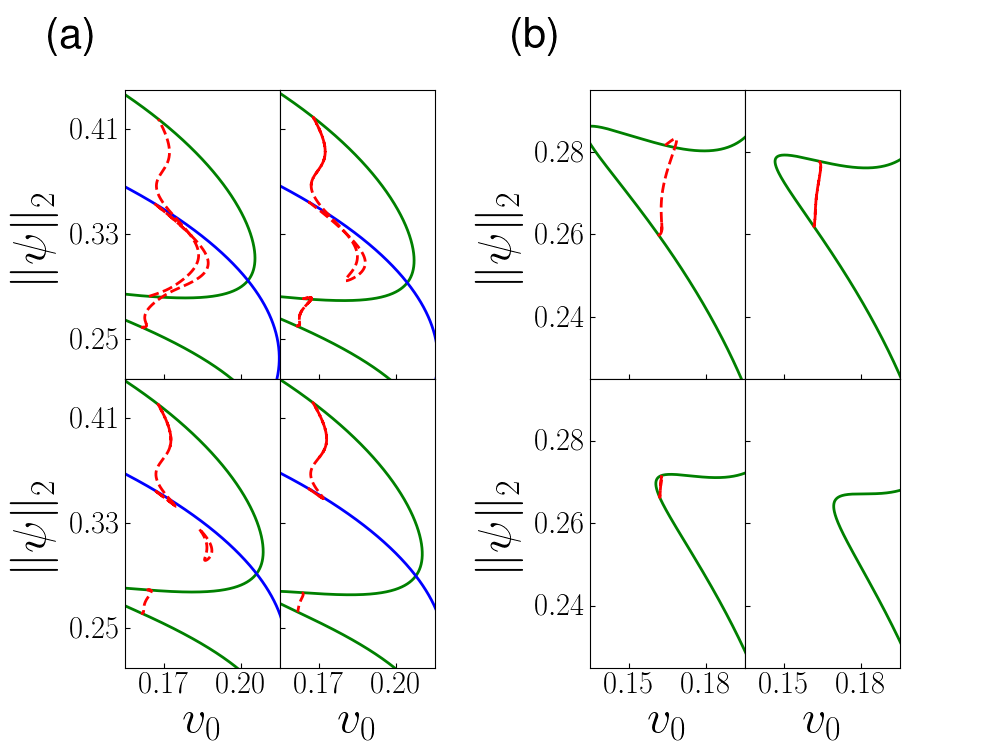}
  \caption{Details are given for the involved bifurcation processes occurring between $\bar{\psi}=-0.73$ (Fig.~\ref{fig:ubar_0_73}) and $\bar{\psi}=-0.71$ (Fig.~\ref{fig:ubar_0_71}). Panel (a) focuses on the pinch-off bifurcation of traveling two- and three-peak LSs and shows magnifications for $\bar{\psi}=-0.730,-0.727,-0.724$ and $-0.721$ (from top left to bottom right) where even isolas of traveling LSs are created. Panel (b) shows how the traveling two-peak LS disappears with increasing $\bar{\psi}=-0.728,-0.722,-0.715$ and $-0.710$ (from top left to bottom right). Line styles and remaining parameters are as in Fig.~\ref{fig:ubar_0_83}.
}
\label{fig:branchswitching}
\label{fig:decay}
\end{figure}

The final increase in mean density that we consider here, brings us to $\bar{\psi}=-0.71$, see Fig.~\ref{fig:ubar_0_71}. Compared to \(\bar{\psi}=-0.73\)  (Fig.~\ref{fig:ubar_0_73}), an additional branch of resting eight-peak LS has appeared. Also, the bifurcation structure of the branches of traveling LSs is now partly dissolving due to further pinch-off bifurcations. To obtain an impression see the sequence of magnifications at different $\bar{\psi}$ given in Fig.~\ref{fig:branchswitching}(a). Note that in the process an isola of traveling LS is created by two pinch-off bifurcations. Upon further increase of $\bar{\psi}$ this isola shrinks and disappears in another codimension-2 bifurcation. Figure~\ref{fig:branchswitching}~(b) illustrates with a sequence of magnifications at different $\bar{\psi}$ how another branch of traveling LS first straightens, then shrinks in length by moving towards a saddle-node bifurcation of resting LS that finally entirely absorbs it. In this way, many of the multi-peak traveling LS disappear. However, branches of traveling one- and two-peak LS are robust and determine the behavior at large $v_0$ (cf. Fig.~\ref{fig:ubar_0_71}). 

The study of individual bifurcation diagrams has shown how the various branches of resting and traveling localized states emerge, expand, reconnect, shrink and vanish when changing the mean density. These glimpses gained at particular values of the mean density shall now be amplified by considering the behavior in the parameter plane spanned by mean density $\bar{\psi}$ and activity $v_0$.

\subsection{Existence ranges of resting LS}

\begin{figure}
 \centering
 \includegraphics[width=0.7\textwidth]{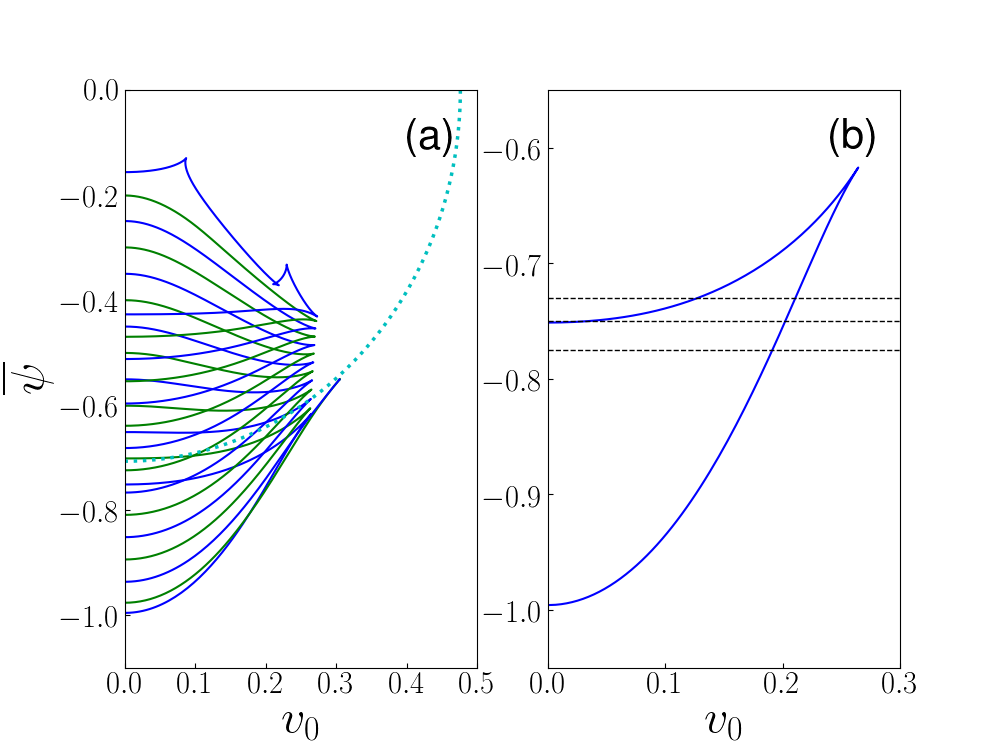}
 \caption{Panel (a) gives the loci of all saddle-node bifurcations for resting one- to fifteen-peak LSs. LSs with odd (even) peak number are given in blue (green), while the primary bifurcation where the resting periodic state emerges is given as dotted cyan line (corresponding to red line in Fig.~\ref{fig:apfc_phasediagram}). Panel (b) focuses on the loci of the folds of the one-peak LS and highlights by horizontal lines the particular values $\bar{\psi}=-0.775, -0.75$, and $-0.73$ for which Figs.~\ref{fig:ubar_0_775}, \ref{fig:ubar_0_75}, and~\ref{fig:ubar_0_73}, respectively, present bifurcation diagrams with $v_0$ as control parameters. The remaining parameters are as in Fig.~\ref{fig:ubar_0_83}. }
 \label{fig:ex_resting}
\end{figure}


To obtain an overview where the various resting LSs occur in parameter space, we track in Fig.~\ref{fig:ex_resting} the loci of all saddle-node bifurcations of branches of LSs in the plane spanned by activity $v_0$ and mean concentration $\bar{\psi}$. At fixed $\bar{\psi}$ the solid lines give the position of folds of corresponding branches in bifurcation diagrams with $v_0$ as control parameter, i.e., in Figs.~\ref{fig:ubar_0_83} to~\ref{fig:branchswitching}. Considering a vertical cut, i.e., fixing a particular constant activity $v_0$, intersections with the blue and green lines give the loci of folds in bifurcation diagrams with $\bar{\psi}$ as control parameter, i.e., in  Fig.~\ref{fig:pfc-bif}. The dotted cyan line tracks the primary bifurcation where the branch of periodic states emerges from the liquid state.

Roughly speaking, each type of resting symmetric LS exists within the region limited by the corresponding solid line. Most of their folds only exist up to $v_0\approx0.25$, one survives till
$v_0\approx0.3$ where it annihilates at the primary bifurcation where the resting periodic state emerges changes from subcritical to supercritical. All branches of LS cease to exist at slightly larger $v_0$.  The two spikes and the small swallow tail structure in the blue line at largest $\bar{\psi}$ indicate that it depends on the value of $v_0$ to which periodic state the LSs snake up to. For the passive PFC model (\(v_0=0\)) at the parameters chosen here, the snaking branches terminate on the periodic 15-peak state (cf.~Fig.~\ref{fig:pfc-bif}) whereas for activities larger than  \(v_0\approx  0.12\), a periodic 16-peak state is approached. We note in passing that Fig.~\ref{fig:ex_resting}~(a) has a very similar appearance as Fig.~6 of Ref.~\cite{TARG2013pre} that shows the loci of bifurcations for the passive PFC model in a plane spanned by $\bar{\psi}$ and the effective temperature $\epsilon$. Here the role of the latter is taken by the activity $v_0$. The implication could be that there exist nonequilibrium analogues of a tricritical point and binodal lines for the present active PFC case. This shall be pursued elsewhere.

Figure~\ref{fig:ex_resting}~(b) enlarges a part of Fig.~\ref{fig:ex_resting}~(a) only focusing on the one-peak LS. Starting at low \(\bar{\psi}\) and increasing \(\bar{\psi}\) the fold moves to the right, hence the branch of steady one-peak LS exists in a larger range of \(v_0\).At \(\bar{\psi}\approx -0.745\) an additional fold appears at $v_0=0$ and moves towards larger $v_0$ with further increasing \(\bar{\psi}\). This fold connects branches of one- and three-peak LS as can be seen in Fig.~\ref{fig:ubar_0_75}. At $v_0=-0.26$ the two folds annihilate in a hysteresis bifurcation. The dashed horizontal lines indicate values corresponding to bifurcation diagrams in Fig.~\ref{fig:ubar_0_775} (\(\bar{\psi}=-0.775\)), Fig.~\ref{fig:ubar_0_75} (\(\bar{\psi}=-0.75\)) and Fig.~\ref{fig:ubar_0_73} (\(\bar{\psi}=-0.73\)).

\subsection{Existence range of traveling LS}
\label{sec:loc_singledouble}
%
\begin{figure}
 \centering
  \includegraphics[width=0.7\textwidth]{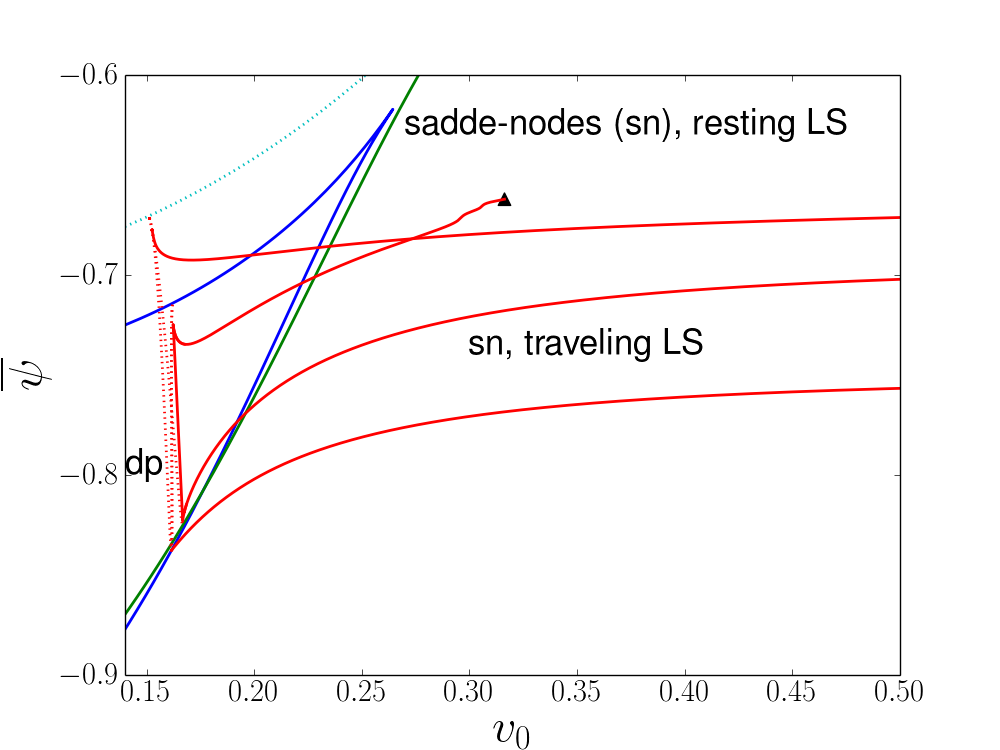}
  \caption{Shown are the loci of saddle-node bifurcations of resting LSs (``sn'',  solid blue and green lines for one-peak and two-peak LS, respectively) and traveling LSs (``sn'', solid red lines) as well as of the drift-pitchfork bifurcations where LSs start to move (``dp'', dotted red lines). The loci of the saddle-node bifurcation of the resting periodic state are given as dotted cyan line. The black triangle marks the point where the investigation of the corresponding branch is terminated as profiles have become seemingly chaotic. A magnification of the region where the drift-pitchfork bifurcations occur is given in Fig.~\ref{fig:ex_singledouble_zoom1}. Remaining parameters are as in Fig.~\ref{fig:ubar_0_83}.}
 \label{fig:ex_singledouble}
\end{figure}
 
\begin{figure}
 \includegraphics[width=0.95\textwidth]{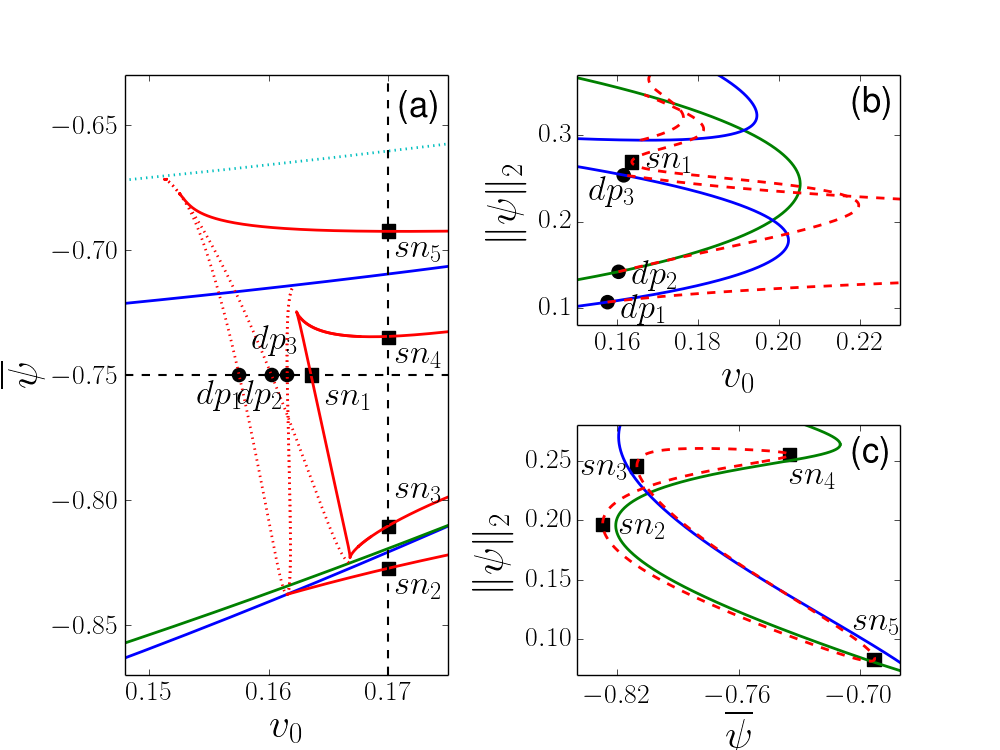}
 \caption{Panel (a) provides a magnification of the region in Fig.~\ref{fig:ex_singledouble} where the drift-pitchfork bifurcations occur. Horizontal and vertical dashed lines highlight the values $\bar{\psi}=-0.75$ and $v_0=0.17$ for which panels (b) and (c) gives the respective one-parameter bifurcation diagrams. Positions of selected drift-pitchfork and saddle-node bifurcations are labeled ``$dp_i$'' and ``$sn_i$'', respectively.}
\label{fig:ex_singledouble_zoom1}
\end{figure}

\begin{figure}
 \includegraphics[width=0.7\textwidth]{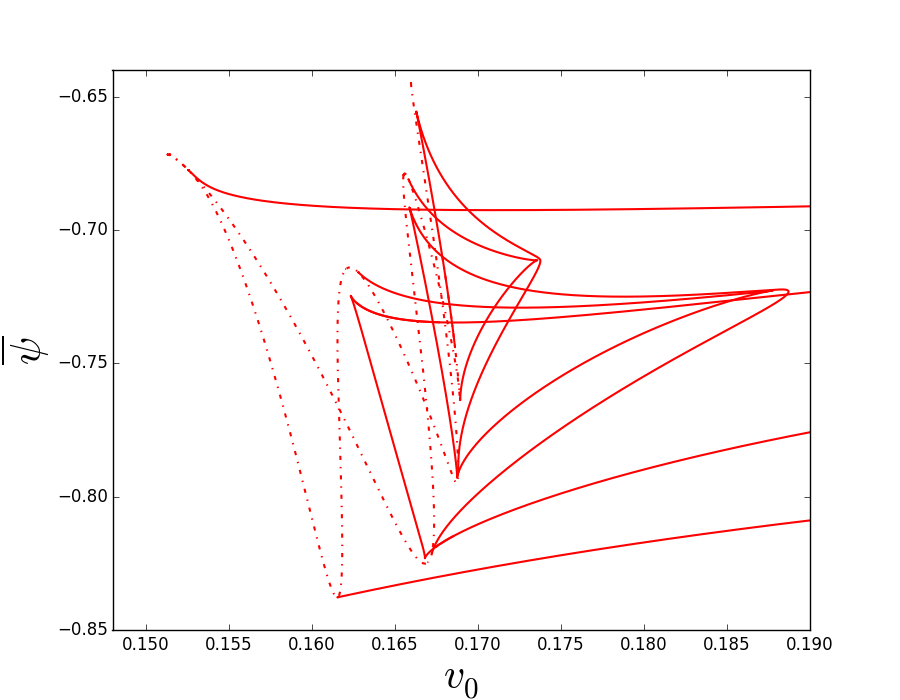}
 \caption{Loci of the saddle-node and drift-pitchfork bifurcations of traveling localized one-, two-, three- and four-peak solutions (not distinguished) from two-parameter continuation. Saddle-node bifurcations are given by straight lines and drift-pitchfork bifurcations by dotted lines. Remaining parameters are as in Fig.~\ref{fig:ubar_0_83}.}
\label{fig:ex_traveling}
\end{figure}

Next, we employ two-parameter continuation to track bifurcations of traveling LSs in the $(v_0,\bar{\psi})$-plane to obtain an overview of their existence ranges. In this way we analyze on the one hand the loci of saddle-node bifurcations that normally indicate the end of the existence range. On the other hand, we also follow drift-pitchfork bifurcations that mark the onset of the existence of traveling LSs. Figure~\ref{fig:ex_singledouble} gives results for both, one- and two-peak LSs. Several properties that we have observed during the continuation in either \(v_0\) or \(\bar{\psi}\) are condensed in this diagram. The solid lines give the loci of saddle-node-bifurcations. Specifically, for one- and two-peak resting LSs in blue and green, respectively, while the loci for all traveling LSs are in red. Dashed lines mark the loci of the drift-pitchfork bifurcations, i.e., the onset of motion. Roughly speaking, traveling LSs exist in the region between the drift-pitchfork and the saddle-node bifurcations of traveling LSs.

The tracks of the various bifurcations in the $(v_0,\bar{\psi})$-plane can, of course, be related to the previously discussed bifurcation diagrams. Figure~\ref{fig:ex_singledouble_zoom1}~(a) shows a magnification  of Fig.~\ref{fig:ex_singledouble} in the region close to the onset of motion highlighted by dashed horizontal and vertical lines the values \(\bar{\psi}=-0.75\) and \(v_0=0.17\), respectively. 
Crossings with the horizontal line denote the positions of bifurcations in the bifurcation diagram with control parameter $v_0$ in Fig.~\ref{fig:ex_singledouble_zoom1}~(b). In an analogous way the vertical line relates to Fig.~\ref{fig:ex_singledouble_zoom1}~(c). Close inspection of the three panels indeed shows their full consistence.

Therefore, we have gained exact knowledge about the parameter region in which a given state exists. Considering, e.g., the branch of the one-peak traveling LS, Fig.~\ref{fig:ex_singledouble_zoom1}~(b) shows that the drift-pitchfork bifurcation named \(dp_1\) in Fig.~\ref{fig:ex_singledouble_zoom1}~(a) marks the onset of this branch when changing $v_0$ at fixed $\bar{\psi}$ and Fig.~\ref{fig:ex_singledouble_zoom1}~(b) and (c) show that the saddle-node bifurcations named \(sd_1\) and \(sd_3\) in Fig.~\ref{fig:ex_singledouble_zoom1}~(a) mark onset and disappearance of this branch when changing \(v_0\) at fixed $\bar{\psi}$ and $\bar{\psi}$ at fixed \(v_0\), respectively.

More effects observed in the previous sections can be linked to results of the bifurcation tracking. For instance, the divergence of the locus of the saddle-node-bifurcation of the branch of one-peak LS (cf.~Fig.~\ref{fig:ubar_0_73}) corresponds in Fig.~\ref{fig:ex_singledouble} to the lowest solid red line's approach of a horizontal asymptote. The two-parameter continuation also confirms, that drift-pitchfork bifurcations are created pairwise together with a branch of traveling LSs between them at saddle-node-bifurcations of resting LSs. In Fig.~\ref{fig:ex_singledouble_zoom1}~(a) this is particularly well visible at about $(v_0,\bar{\psi})\approx(0.1615, -0.8378)$ where red dotted lines (drift-pitchfork bifurcations including $dp_1$ and $dp_3$) and a red solid line (saddle-node-bifurcation of traveling LSs with $sn_2$) simultaneously emerge from the blue solid line (saddle-node-bifurcation of resting LSs) - all for one-peak states. Note that such a structure can also occur at the saddle-node-bifurcation of resting periodic states.

To complete the overview and show connections between the different bifurcations and their origin, Fig.~\ref{fig:ex_traveling} not only includes the saddle-node bifurcations and drift-pitchfork bifurcations of one- and two-peak LSs (as Figs.~\ref{fig:ex_singledouble} and \ref{fig:ex_singledouble_zoom1}) but adds the corresponding loci of traveling three-peak and four-peak LSs. Their range of existence is limited to a smaller region as compared to the one- and two-peak LS. 

The discontinued branch in Fig.~\ref{fig:ex_singledouble} with the end marked by a black triangle, corresponds in this region to states that seem chaotic. This is related to the fact that the homogeneous background of the LS is linearly unstable above $\psi\approx-0.6710$. We have not further investigated this region.

\subsection{Morphological phase diagram}

\begin{figure}
\centering
\begin{overpic}[width=9cm]{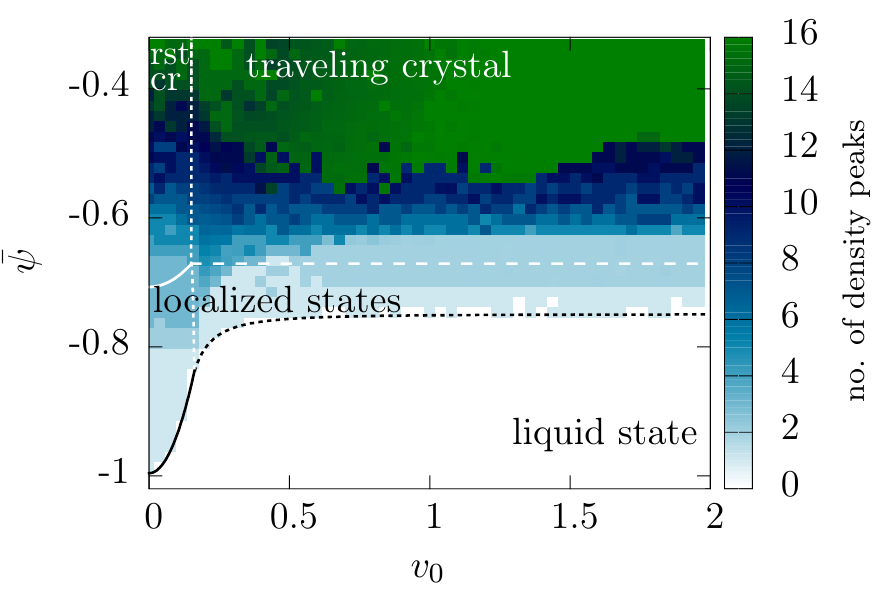}
\put(-1,65){(a)}
\end{overpic}
\begin{overpic}[width=9cm]{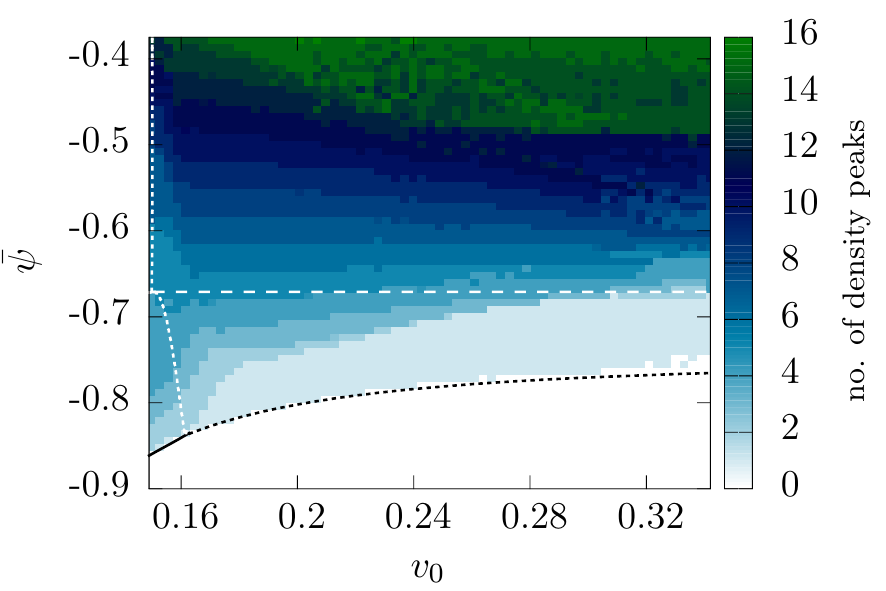}
\put(-1,65){(b)}
\end{overpic}

\caption{\label{fig:phasediagram_1dLSA1d} (a) Morphological phase diagram in the plane spanned by activity $v_0$ and mean density $\bar{\psi}$. The region of stable liquid state is white, while crystalline structures of various size exist in the colored areas. The color bar indicates the number of density peaks obtained in time simulations. Resting and traveling LSs are marked by shades of blue while domain filling periodic patterns show as green. The various lines in the diagram, the initial conditions of simulations, peak counting and parameter sampling are described in the main text. (b) Zoom into the $v_0$-range close to the onset of motion. Steps in shading illustrate the gradual growth of LS for increasing $\bar{\psi}$. Remaining parameters as in Fig.~\ref{fig:ubar_0_83}.}
\end{figure}

In Fig.~\ref{fig:phasediagram_1dLSA1d}~(a) we provide a morphological phase diagram in the plane spanned by activity $v_0$ and mean density $\bar{\psi}$. It summarizes the ranges of existence of resting and traveling crystalline and LSs by combining information gained through linear and nonlinear analyses and direct numerical simulation. The accompanying Fig.~\ref{fig:phasediagram_1dLSA1d}~(b) gives a magnification of the $v_0$-range close to the onset of motion. The blue and green colored areas indicate where spatially modulated states exist. Thereby the shading encodes the number of density peaks that evolve in the periodic domain of $L=100\approx 16\,L_c$ where $L_c=2\pi$. White regions correspond to the homogeneous liquid state ($\psi=0$). Spatially extended states that fill the entire domain with 16 peaks are shown in green and occur at large $\bar{\psi}$.

The initial condition for all time simulations is one density peak in the shape of a Gaussian with oscillatory tails $(\propto\exp(-x^2/5)\cos(x))$. It is randomly placed on a background of white noise with a small amplitude noise as polarization. After a sufficiently long transient, the number of density peaks is counted for a time intervall. Shown is then the median of counted peaks (always a natural number), as also oscillatory states arise in some small parameter ranges (see Sec.~\ref{sec:1d_oscillatingstates}). The parameter increments between simulations in Fig.~\ref{fig:phasediagram_1dLSA1d}~(a) are $\Delta v_0=0.04$ and $\Delta \bar{\psi}=0.016$ while for panel (b) $\Delta v_0=0.003$ and $\Delta \bar{\psi}=0.008$ is used.

The curved solid white line and the horizontal dashed white line give the linear stability border of the liquid phase as determined in section~\ref{sec:LSA} and shown in Fig.~\ref{fig:apfc_phasediagram}. In its vicinity, LSs of different sizes as indicated by the various shades of blue coexist with the liquid state. The area of existence of LSs is bound from below by the loci of the folds of the resting and the traveling one-peak LS that reach the lowest values of $\bar{\psi}$. Their loci as tracked by two-parameter continuation is depicted as solid and dotted black lines, respectively, cf.~Sec.~\ref{sec:loc_singledouble}. It is remarkable how well the results from time simulations match the existence bounds predicted by parameter continuation. Also note how well the steps in shading in panel (b) indicate the gradual increase in size of the LS with increasing $\bar{\psi}$.
%
Other two-parameter continuations track the onset of motion where resting states undergo a drift-pitchfork bifurcation. The vertical dotted white line marks the onset of motion for crystals. The lower part ($\bar{\psi}<-0.69$) of the nearly vertical line (see magnification in Fig.~\ref{fig:phasediagram_1dLSA1d}~(b)) represents the onset for the one-peak LS. 

\subsection{Oscillatory states}
\label{sec:1d_oscillatingstates}

\begin{figure}
 \centering
 \vspace*{-0.5cm}
  \includegraphics{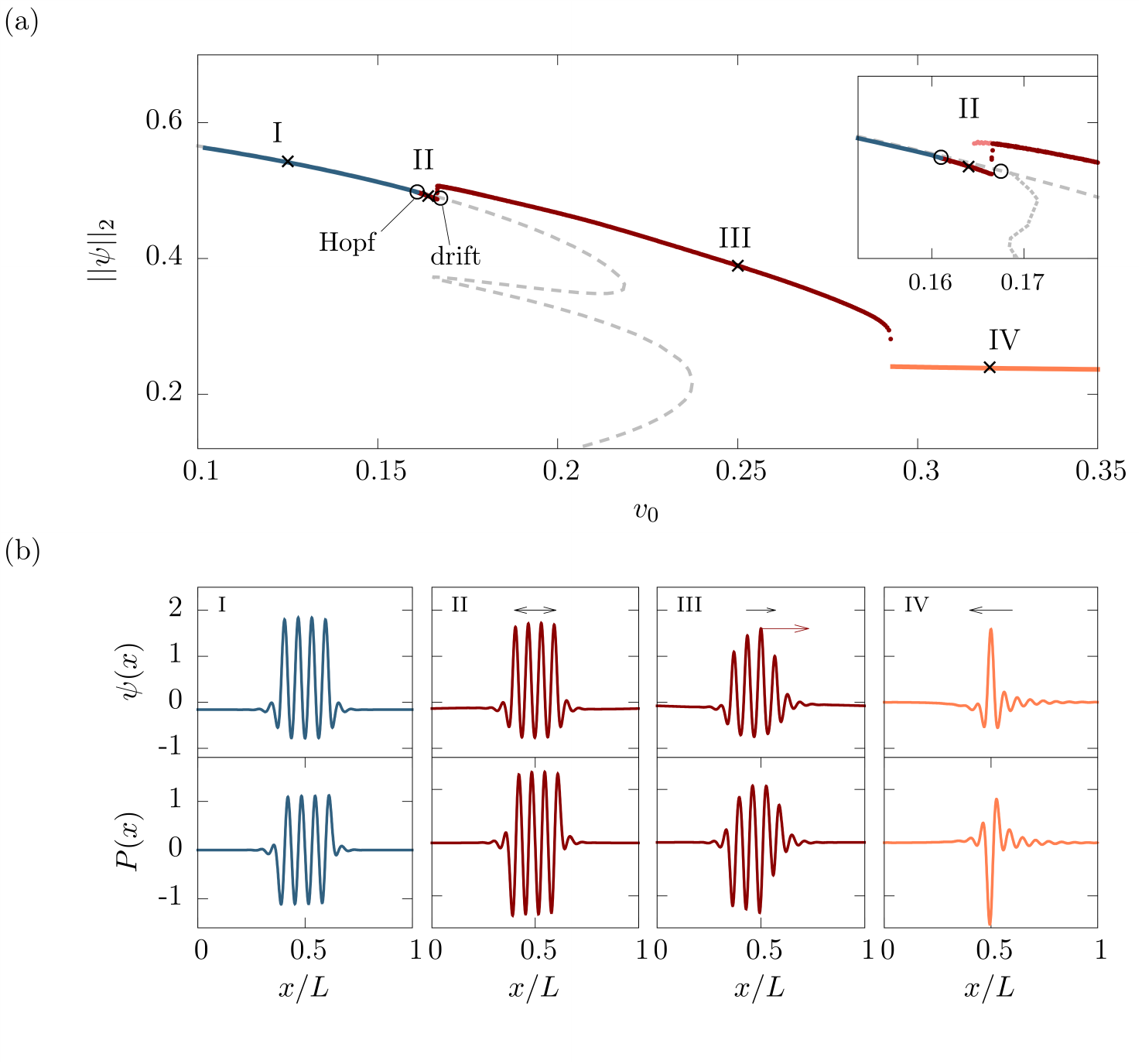}
  \vspace*{-1cm}
  \caption{(a) Bifurcation diagram of four-peak LS obtained by continuation (gray dashed line) and time simulations (solid lines): The (time-averaged) norm $||\psi||_2$ is shown as a function of activity $v_0$. The branch of resting states undergoes a Hopf bifurcation at $v_\mathrm{Hopf}\approx0.16$ and a drift instability at $v_\mathrm{c}>v_\mathrm{Hopf}$ (black circles). A branch of oscillatory states appears and undergoes a fold bifurcation at $v_0\approx0.28$ beyond which traveling one-peak LS (orange) are found. The inset enlarges the bifurcation structure. Following the branch of drifting oscillatory states in negative $v_0$-direction (light red branch in inset) reveals a hysteresis and the subcritical nature of its drift bifurcation. The branch of steadily traveling four-peak LS (gray dotted, unstable, only shown in inset) emerges at the drift bifurcation at $v_\mathrm{c}$. Panel~(b) shows profiles of selected states at loci marked in panel (a). Directions of motion are indicated by arrows. In particular: (I)~resting four-peak LS; (II)~oscillating four-peak LS; (III)~modulated traveling four-peak LS; (IV)~traveling one-peak LS. See Fig.~\ref{fig:oscillation_spacetime} for space-time plots of (II) and (III).  $\bar{\psi}=-0.68$ and remaining parameters as in Fig.~\ref{fig:ubar_0_83}.
}
\label{fig:hopf_bif}
\end{figure}

\begin{figure}
 \hspace*{1.3cm}
  \begin{overpic}[width=6.9cm]{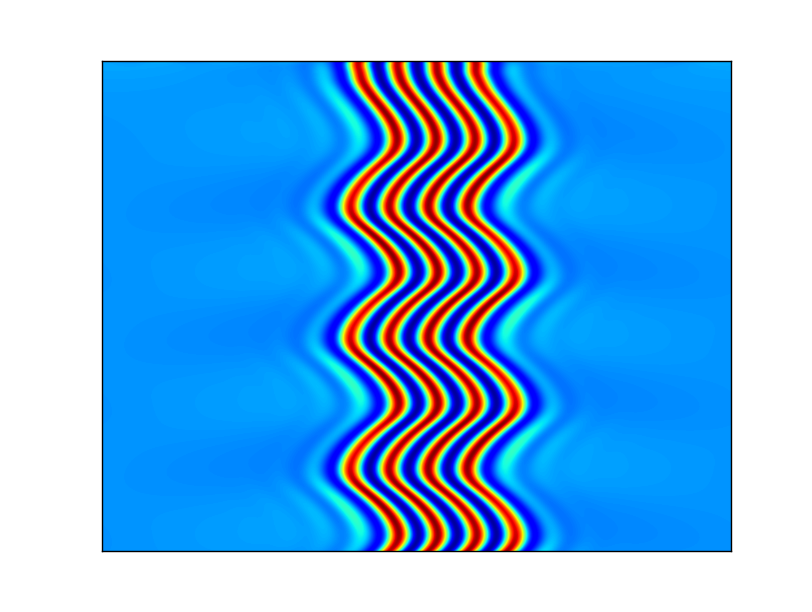}
  \put(-12,70){(a)}
  \put(-9,35){\rotatebox{90}{\makebox(0,0){$t$}}}
  \put(7.5,66){0}
   \put(1.5,51){500}
   \put(-1.5,36){1000}
   \put(-1.5,21){1500}
   \put(-1.5,7){2000}
   \put(10,0){0}
   \put(48,0){0.5}
   \put(89,0){1}
   \put(46,-8){$x/L$}
  \end{overpic}
  \hspace*{.3cm}
  \begin{overpic}[width=6.9cm]{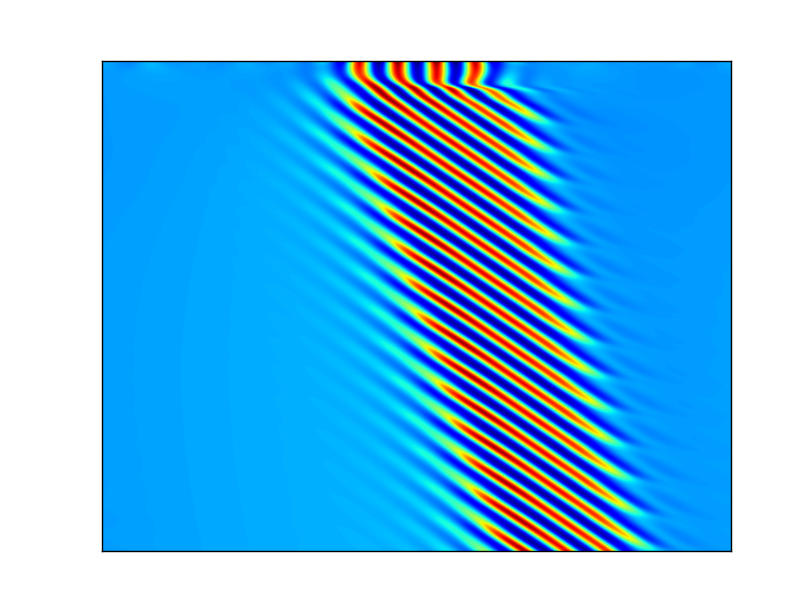}
  \put(-11,70){(b)}
  \put(-7,35){\rotatebox{90}{\makebox(0,0){$t$}}}
  \put(8,66){0}
   \put(1,51){150}
   \put(1,36){300}
   \put(1,21){450}
   \put(1,7){600}
   \put(10,0){0}
   \put(48,0){0.5}
   \put(89,0){1}
   \put(46,-8){$x/L$}
  \end{overpic}
  \vspace*{0.7cm}
  \caption{Panels (a) and (b) show space-time representation of the time evolution of four-peak LS for states~II ($v_\mathrm{Hopf}<v_0=0.164<v_\mathrm{c}$) and III ($v_0=0.25>v_\mathrm{c}$ ) in Fig.~\ref{fig:hopf_bif}, respectively.
  }
  \label{fig:oscillation_spacetime}
\end{figure}

The parameter scan performed to create the morphological phase diagram in Fig.~\ref{fig:phasediagram_1dLSA1d} has allowed us to identify additional classes of moving states beside the steadily traveling ones. These are on the one hand oscillating LSs that move back and forth without a net drift and modulated traveling LSs. The latter have an envelope that drifts with a steady velocity $c_\mathrm{group}$ while individual peaks travel faster at $c_\mathrm{phase}>c_\mathrm{group}$, i.e., they are created at the back of the LS and vanish at its tip. Such states exist with different peak numbers close to the linear stability border of the uniform phase. We detect Hopf bifurcations on branches of resting LSs at activities $v_0=v_\mathrm{Hopf}$ lower than $v_\mathrm{c}$ where the drift sets in. At $v_\mathrm{Hopf}$ branches of oscillating states emerge that can themselves undergo drift bifurcations.

Figure~\ref{fig:hopf_bif}~(a) depicts a corresponding bifurcation diagram for four-peak LS using $v_0$ as control parameter while panel (b) shows selected density and polarisation profiles. The diagram is obtained combining numerical continuation of steady and steadily traveling LS and time simulations of the oscillating and modulated traveling LS. At low activity, the LS is at rest (blue line, e.g., state~I). At about $v_\mathrm{Hopf}\approx0.16$ a Hopf bifurcation occurs. Time simulations at slightly larger $v_0$ with $v_\mathrm{Hopf}<v_0<v_\mathrm{c}$ exhibit oscillating LS [dark red line, e.g., state~II, corresponding space-time plot in Fig.~\ref{fig:oscillation_spacetime}(a)]. Note that the oscillations are not symmetrical in space similar to the case of wiggling LSs found in reaction-diffusion systems~\cite{SOMS_PRE95}. At $v_\mathrm{c}\approx0.167$, closely above $v_\mathrm{Hopf}$, a drift bifurcation occurs on the branch of resting LS, cf.\ inset of Fig.~\ref{fig:hopf_bif}(a). There, an unstable branch of steadily traveling four-peak LS emerges shown as dotted gray line in the inset. It ends at the branch of resting three-peak LS (not shown).

The time simulations indicate that also the branch of oscillating states undergoes a drift instability. Hence, the oscillatory LS without a net drift only exist in a very narrow range of activity. The onset of motion is close to $v_\mathrm{c}$ indicated by a sudden increase in the time-averaged norm related to the jump onto another branch. The corresponding modulated traveling states (e.g., state~III) show an intricate combination of the destabilized monotonic and oscillatory modes. A simple superposition would lead to a drifting oscillating state. Instead we find a steadily traveling envelope that moves with group velocity $c_\mathrm{group}$ while the individual density peaks travel at larger velocity $c_\mathrm{phase}$ into the same direction. The difference in the two speeds results in a periodic change of the localized pattern moving within the envelope. Figure~\ref{fig:oscillation_spacetime}(b) presents a corresponding space-time plot. Increasing $v_0$ results in a larger difference between $c_\mathrm{group}$ and $c_\mathrm{phase}$. The inset of Fig.~\ref{fig:hopf_bif}~(a) also shows a hysteresis between oscillating and modulated traveling LS (light red branch) revealing the subcritical nature of the drift bifurcation of oscillatory states. Note that due to restrictions of the numerical techniques used our picture of the  transition is incomplete as it is likely that more states are involved.

At another critical activity of $v_0\approx0.28$ the branch of oscillatory modulated traveling four-peak LS seems to fold back as the trace of time simulations in Fig.~\ref{fig:hopf_bif}(a) approaches a vertical. At larger $v_0$ only traveling one-peak LS (e.g., state IV) emerge in the time simulations. For clarity of the diagram, we abstain from including further one-peak branches. 

\begin{figure}[t]
 \centering
\includegraphics{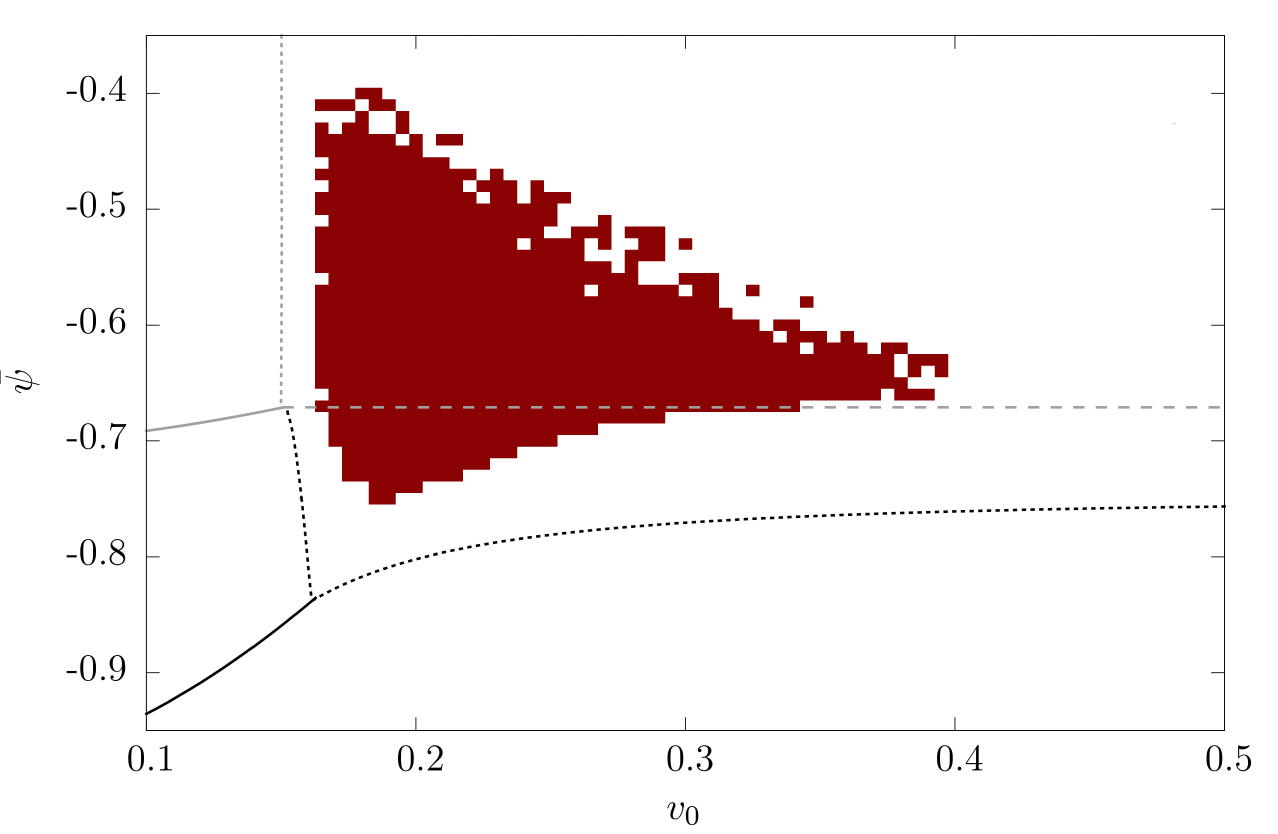}
\caption{\label{fig:oscillating_states} Shaded region of existence of modulated traveling LS in the plane spanned by activity $v_0$ and mean density $\bar{\psi}$. Line styles and other details are as in Fig.~\ref{fig:phasediagram_1dLSA1d}. Oscillating LS (without net drift) only appear in a very narrow $v_0$-range that is not resolved here.}
\end{figure}

The existing region of modulated traveling LS of different peak numbers is shown in Figure~\ref{fig:oscillating_states} extracting the pertinent data from the parameter scan performed for Fig.~\ref{fig:phasediagram_1dLSA1d}.  It is constructed by detecting oscillations in the detected number of density peaks.

\section{Interaction of traveling LS}
\label{sec:scattering}

\subsection{Two colliding LS}
\label{sec:twocoll}

Having obtained an extensive description of individual localized states in terms of their bifurcation structure and existence regions in parameter space, we next explore the interaction between two and more LSs. A simple way to study such interactions is to look at collision kinematics. First, we examine two colliding one-peak LS. Due to the periodic boundaries of the considered domain, collisions will repeatedly occur. By studying the outcome of collisions, we can distinguish different types of interaction between traveling crystallites. As before, our main control parameters are the activity $v_0$ setting mainly the velocity of the colliding crystallites and the mean density $\bar{\psi}$.

Our typical setup of a numerical collision experiment is as follows. We place two identical traveling one-peak LS with opposite drift velocities at sufficient distance into a domain of length $L$ with periodic boundary conditions and perform a direct time simulation. Due to the periodic boundary conditions collisions can occur repeatedly allowing us to probe long-time stability. If the colliding LS are initially regularly spaced, the free path $\lambda$ between two subsequent collisions is $L/2$.

\begin{figure}
\hspace*{0.5cm}
 \includegraphics[width=0.7\textwidth]{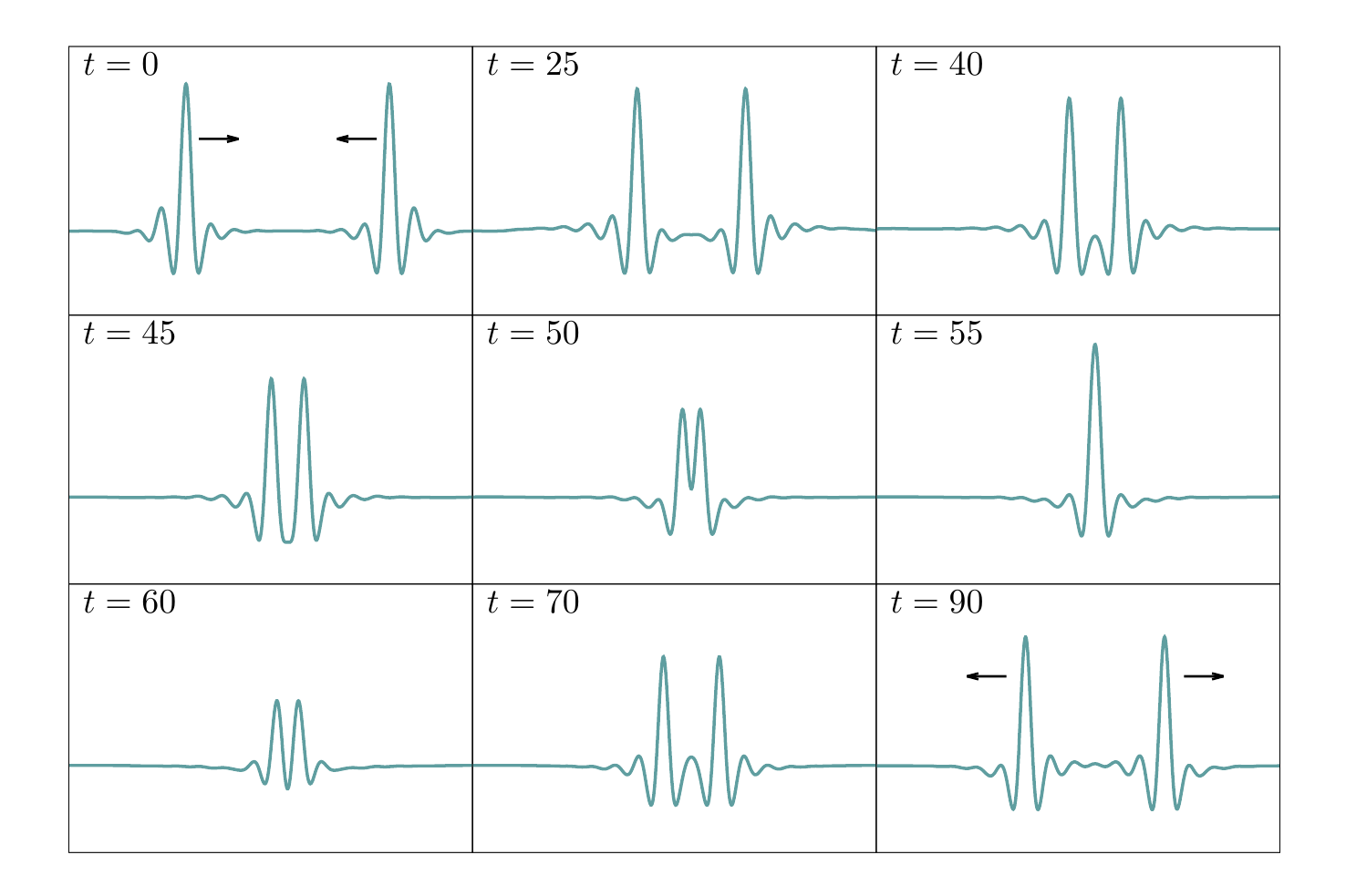}
 \caption{Snapshots of the density profile $\psi(x,t)$ of two colliding one-peak LS at $v_0=0.5$ and $\bar{\psi}=-0.7$, taken at different times $t$ as indicated in the panels. Arrows in the first and final panel indicate the direction of motion. The LS approach each other with a constant velocity, interact till complete fusion at $t\approx55$. Then they separate again, i.e., the LS get reflected. Finally, the LS completely recover from the collision and regain their original height and velocity. A space-time plot of this collision is given in Fig.~\ref{fig:collision_spacetime}~(a). The domain size is $L=200$ and the remaining parameters are as in Fig.~\ref{fig:ubar_0_83}. }
 \label{fig:1dsnaps}
\end{figure}

Figure~\ref{fig:1dsnaps} illustrates a typical ``elastic'' collision of two traveling one-peak LSs.  After an intense phase of interaction the two LSs get completely reflected and fully recover their initial profile and speed. Note that ``elastic'' is used to indicate that the states before and after the collision are identical up to reversed velocities. This is in contrast to an elastic collision in a mechanical system where the kinetic energies before and after the collision are identical. The present nonequilibrium system is overdamped and results from a balance of energy gain (related to self-propulsion mechanism) and dissipation. No kinetic energy can be assigned to a moving LS.

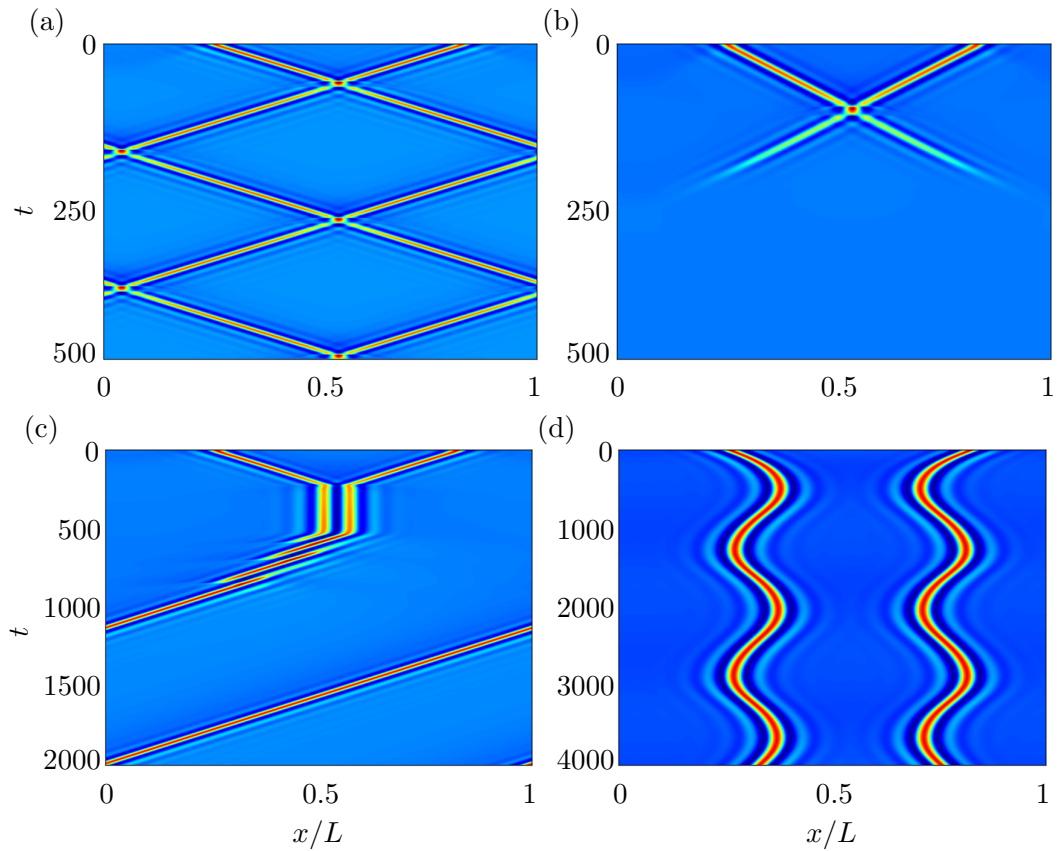
\begin{figure}
\hspace*{1cm}
\begin{overpic}[width=6.5cm]{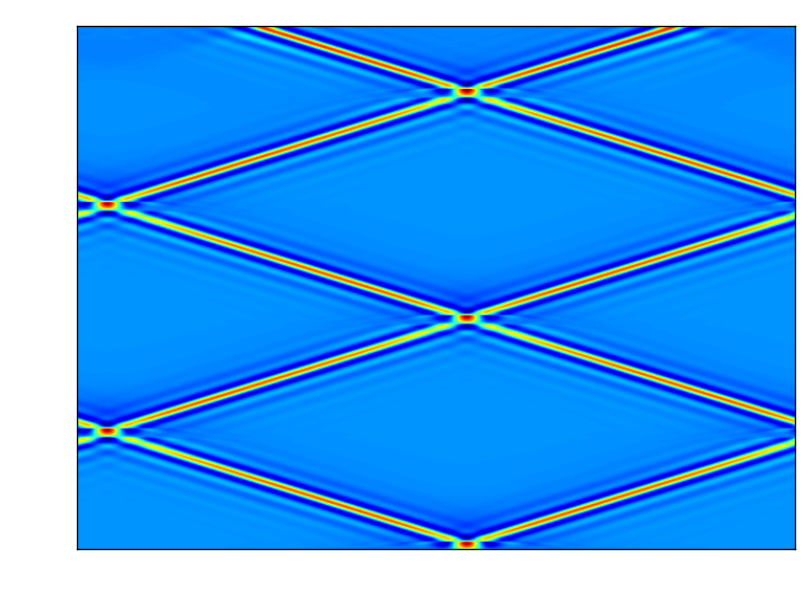}
  \put(-6,75){(a)}
  \put(-7,38){\rotatebox{90}{\makebox(0,0){$t$}}}
  \put(5,70){0}
   \put(-1,36){250}
   \put(-1,7){500}
   \put(8,0){0}
   \put(51,0){0.5}
   \put(96,0){1}
\end{overpic}\hspace*{0.2cm}
\begin{overpic}[width=6.5cm]{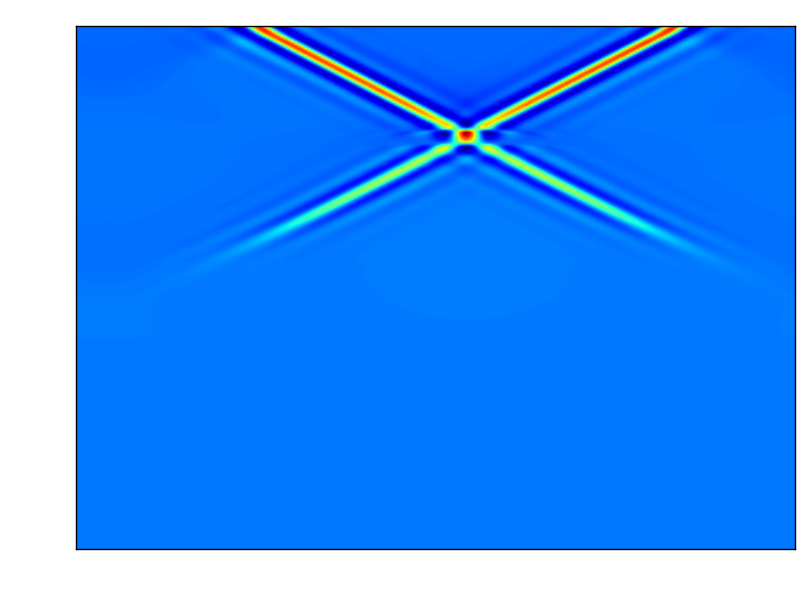}
  \put(-6,75){(b)}
  \put(5,70){0}
   \put(-1,36){250}
   \put(-1,7){500}
   \put(8,0){0}
   \put(51,0){0.5}
   \put(96,0){1}
\end{overpic}\vspace*{0.5cm}

\hspace*{1cm} \begin{overpic}[width=6.5cm]{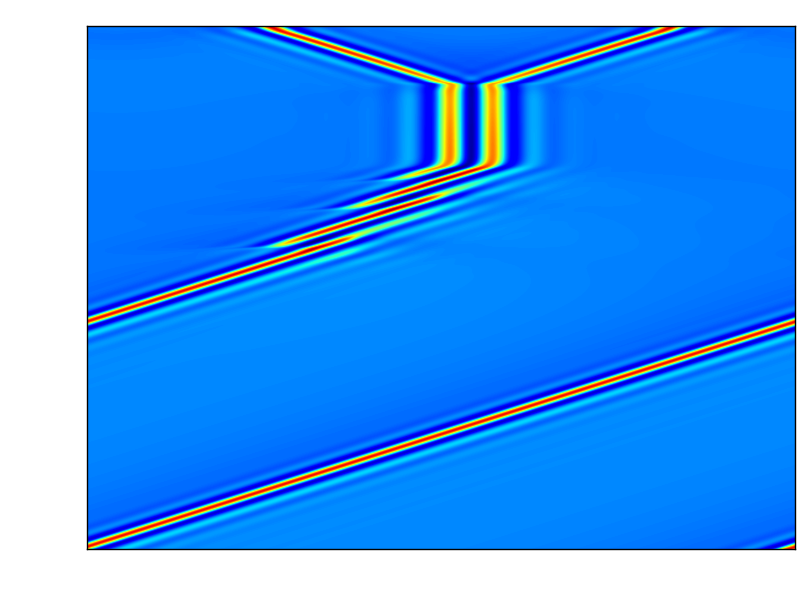}
  \put(-6,75){(c)}
  \put(-7,35){\rotatebox{90}{\makebox(0,0){$t$}}}
  \put(6.5,70){0}
   \put(1,54){500}
   \put(-2,38){1000}
   \put(-2,22){1500}
   \put(-2.,7){2000}
   \put(9.5,0){0}
   \put(51,0){0.5}
   \put(96,0){1}
   \put(49,-8){$x/L$}
\end{overpic}\hspace*{0.2cm}
\begin{overpic}[width=6.5cm]{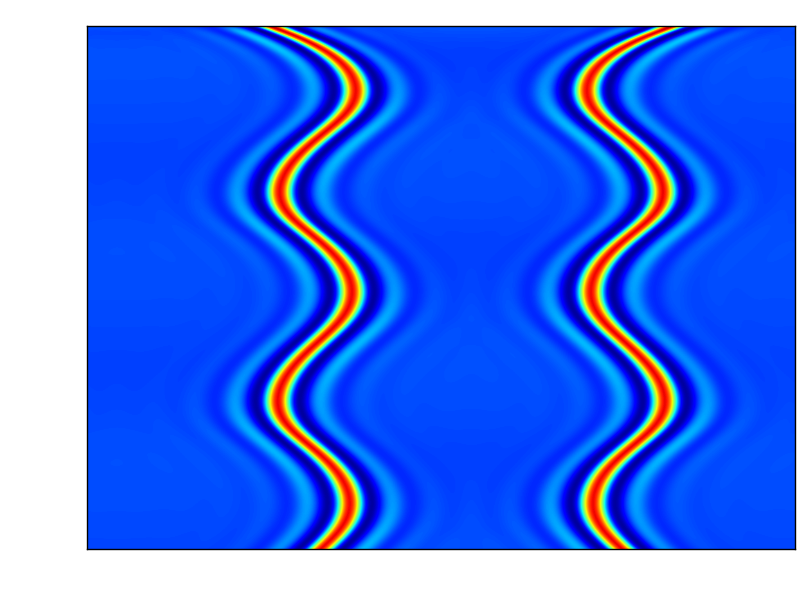}
  \put(-6,75){(d)}
  \put(6.5,70){0}
   \put(-2,54){1000}
   \put(-2,38){2000}
   \put(-2,22){3000}
   \put(-2.,7){4000}
   \put(9.5,0){0}
   \put(51,0){0.5}
   \put(96,0){1}
   \put(49,-8){$x/L$}
\end{overpic}
\vspace*{1cm}
\caption{Shown are space-time plots of the density field $\psi(x)$ for qualitativly different collisions of two one-peak LS. Panel (a) represents the elastic collision at $v_0=0.5$ and $\bar{\psi}=-0.7$ corresponding to Fig.~\ref{fig:1dsnaps}; (b) gives an inelastic collision at $v_0=0.5$ and $\bar{\psi}=-0.75$ resulting in a complete annihilation of both LS, (c) shows an inelastic collision at $v_0=0.2$, $\bar{\psi}=-0.75$ resulting in a sequence of unstable resting and modulated traveling two-peak LS and ultimately in a traveling one-peak LS; and (d) gives an 'avoided collision' at $v_0=0.163$, $\bar{\psi}=-0.71$ where the initial state directly transforms into a state of anti-phase oscillation. In all cases time increases from top to bottom. The domain size is $L=100$ and the remaining parameters as in Fig.~\ref{fig:1dsnaps}. }
\label{fig:collision_spacetime}
\end{figure}

The snapshots in Figure~\ref{fig:1dsnaps} show that for $t\lesssim40$ the LS approach each other at constant velocity, then interact and undergo a deformation that includes a fully fused one-peak intermediate at $t\approx55$. This state we call ``full impact''. After this momentary fusion the LSs reverse velocity and separate again. Overall they are eventually reflected and move steadily away from each other from about $t=60$. Finally, the LSs completely recover from the collision and regain their original profile and velocity. This is most clearly seen in the accompanying space-time plot in Fig.~\ref{fig:collision_spacetime}~(a) where the trajectories of the two LSs before and after the collision show the same absolute value of the slope. Alternatively, one may describe the interaction as the encounter of two solitary pulses that pass through each other.
 
\begin{figure}
\centering
  \includegraphics{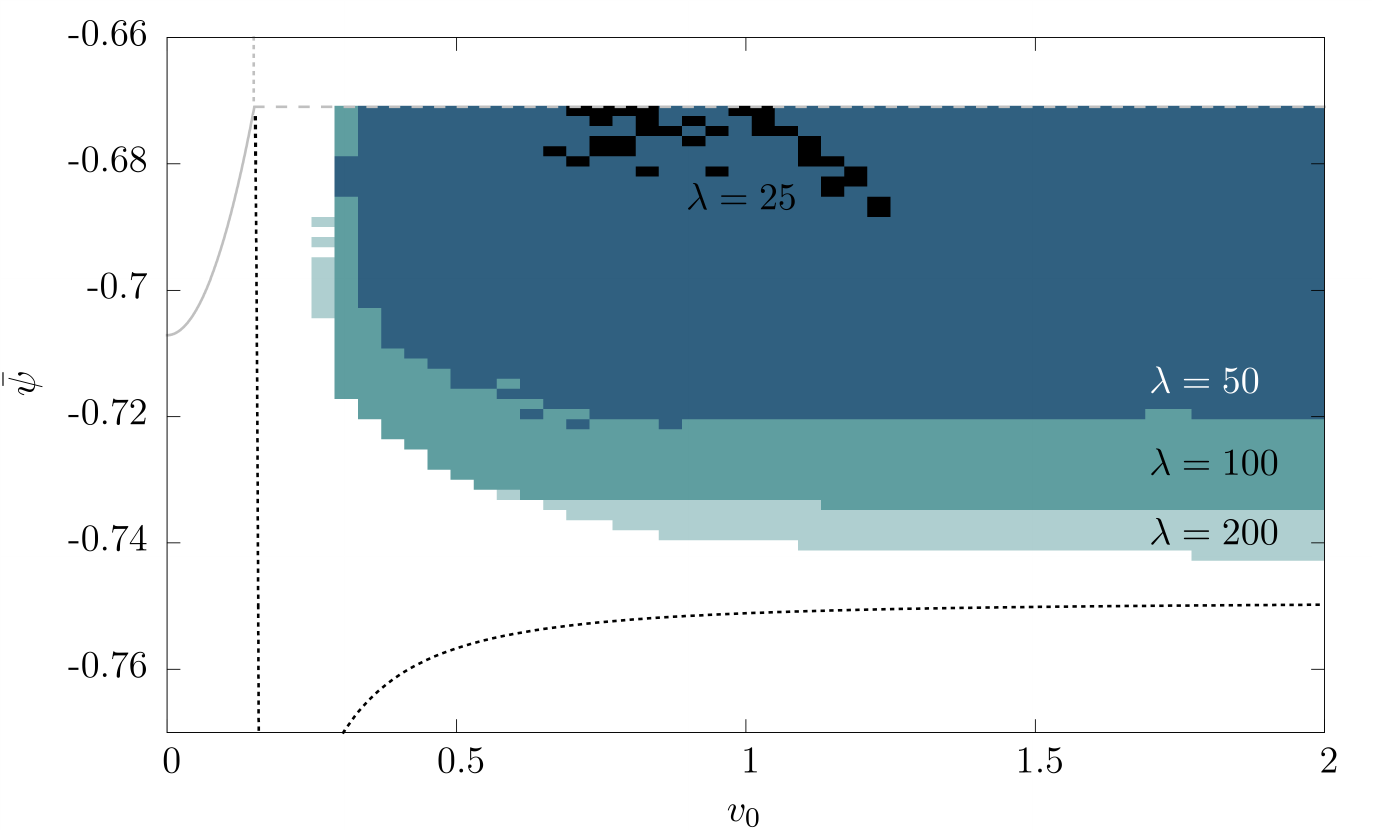}
  \caption{The occurrence of elastic collisions of one-peak LS in the parameter plane spanned by $v_0$ and $\bar{\psi}$ is indicated by the shaded areas for different free path lengths $\lambda$. In the white area collisions are inelastic (or no traveling LS exist) and the number of colliding LS is not conserved. Results for different $\lambda$ are given as black ($\lambda=25$), blue ($\lambda=50$), greenish ($\lambda=100$) and light gray ($\lambda=200$) shading. Parameter increments between simulations are $\Delta v_0=0.006$ and $\Delta \bar{\psi}=0.00125$.  Remaining line styles and parameters are as in Fig.~\ref{fig:phasediagram_1dLSA1d}.}
\label{fig:freepath}
\end{figure}

Due to the periodic boundary conditions, a sequence of  collisions continues ad infinitum. For this to occur the employed domain has to be sufficiently large for the colliding LS to fully recover after a collision before the next one occurs. In consequence, in a systems with several LS each LS needs enough time to recover after each collision to keep a constant number of LS. The recovery time corresponds to a critical free path that needs to be exceeded. This is further discussed in the context of Fig.~\ref{fig:freepath}.

Our main control parameters $v_0$ and $\bar{\psi}$ strongly influence the outcome of a collision. Figure~\ref{fig:collision_spacetime} shows space-time plots of four selected examples that illustrate qualitatively different behavior. Figure~\ref{fig:collision_spacetime}~(a) presents the previously discussed case of two elastically colliding LSs. Decreasing $\bar{\psi}$ from $-0.7$ to $-0.75$ one finds the inelastic collision shown in Fig.~\ref{fig:collision_spacetime}~(b). The traveling LSs interact till they fully merge, then start to move away from the point of full impact. However, instead of recovering their original properties they fade away into the linearly stable homogeneous liquid state that forms the background. The overall result of this inelastic collision is a complete annihilation of the two LSs.

Keeping $\bar{\psi}=-0.75$, a decrease in activity from $v_0=0.5$ to the value $v_0=0.2$ (already close to the onset of motion at $v_\mathrm{c}\approx0.16$) gives the more intriguing inelastic behavior shown in Fig.~\ref{fig:collision_spacetime}~(b). There, the two traveling LSs approach each other, interact but do not merge. Instead, at $t\approx250$ they come to rest in a bound two-peak state. At first it seems to persist (not the time scale of the simulation), however, at $t\approx500$ the state deforms and starts to move to the left as a modulated traveling two-peak state with $c_\mathrm{group}<c_\mathrm{phase}$. Again, the state turns out to be an unstable transient and at  $t\approx800$ a one-peak LS survives that travels with the same speed as the one of the initial collision partners. In fact, it corresponds to the same one-peak traveling LS.

Finally, Fig.~\ref{fig:collision_spacetime}~(d) presents an example at a $v_0$ very close to $v_\mathrm{c}$. The two LSs start to approach but then repel each other already long before a proper collision takes place. Their motion then becomes a periodic anti-phase oscillation about the positions $x=L/4$ and $x=3L/4$, respectively. Depending on the type of interacting LSs and parameter values several other cases are possible. However, most only occur in very small parameter regions and are not discussed here.

Here we focus on the prevalent elastic collisions and investigate their occurrence in the parameter plane spanned by $v_0$ and $\bar{\psi}$. Again we place two traveling one-peak LSs equally spaced on a domain of size $L$ resulting in a free path of $\lambda=L/2$ between collisions. For each set of ($v_0,\bar{\psi}$) and four different values of $\lambda$, time simulations are performed and parameter maps are created that are presented in Fig.~\ref{fig:freepath}. 

Towards large $\bar{\psi}$ we limit the considered region by onset of the linear instability of the homogeneous liquid state (horizontal dashed line). Above the line, the uniform background state is unstable and spatial modulations grow everywhere. Further details are give below.
As before, the nearly vertical dotted line on the left of Fig.~\ref{fig:freepath} marks the onset of motion for one-peak LS and thus represents the lower $v_0$ value of the investigated region. At low $\bar{\psi}$ the region is limited by the end of the existence region of one-peak LS.

The respective shaded regions indicate where the collision is elastic for the different free path lengths $\lambda$. One notices at $v_0\approx0.28$ a rather sharp transition from inelastic to elastic collisions that is nearly independent of $\lambda$. At lower values of $v_0$, the LS spend too much time in the collision process and loose some of their mass before separating again. If this loss is too large complete recovery does not occur. At larger $v_0$ the LSs are faster and spend less time in the actual collision process resulting in a fast complete recovery and long-time conservation of the number of collision partners. One also notices that for elastic collisions to occur, for $\lambda=50$ the mean density has to exceed values about $\bar{\psi}\approx-0.71$ to $\bar{\psi}\approx-0.73$ depending on the exact value of $v_0\gtrsim0.28$.

Next we discuss the influence of the mean free path $\lambda$ on the collision behavior by comparing the differently shaded regions in Fig.~\ref{fig:freepath}. The largest considered free path is $\lambda=200$ and results in the largest region of elastic collisions. At large $v_0$ it nearly coincides with the existence region of traveling one-peak LS.
Decreasing the free path to $\lambda=100$, $\lambda=50$ and finally to $\lambda=25$ results in a decrease and finally disappearance of the region of elastic collisions. Interestingly, the above discussed relatively sharp boundary at $v_0\approx0.3$ does only weakly depend on $\lambda=50\dots200$. In contrast the nearly horizontal low-$\bar{\psi}$ limit at large $v_0$ continuously moves towards larger $\bar{\psi}$ with decreasing $\lambda$. For $\lambda<50$ the change is more dramatic: for $\lambda=25$ most of the region of elastic collisions has vanished. This implies that overall, a free path of at least $\lambda\approx 50$ seems necessary for elastic collisions to occur.

\begin{figure}
\centering
  \begin{overpic}{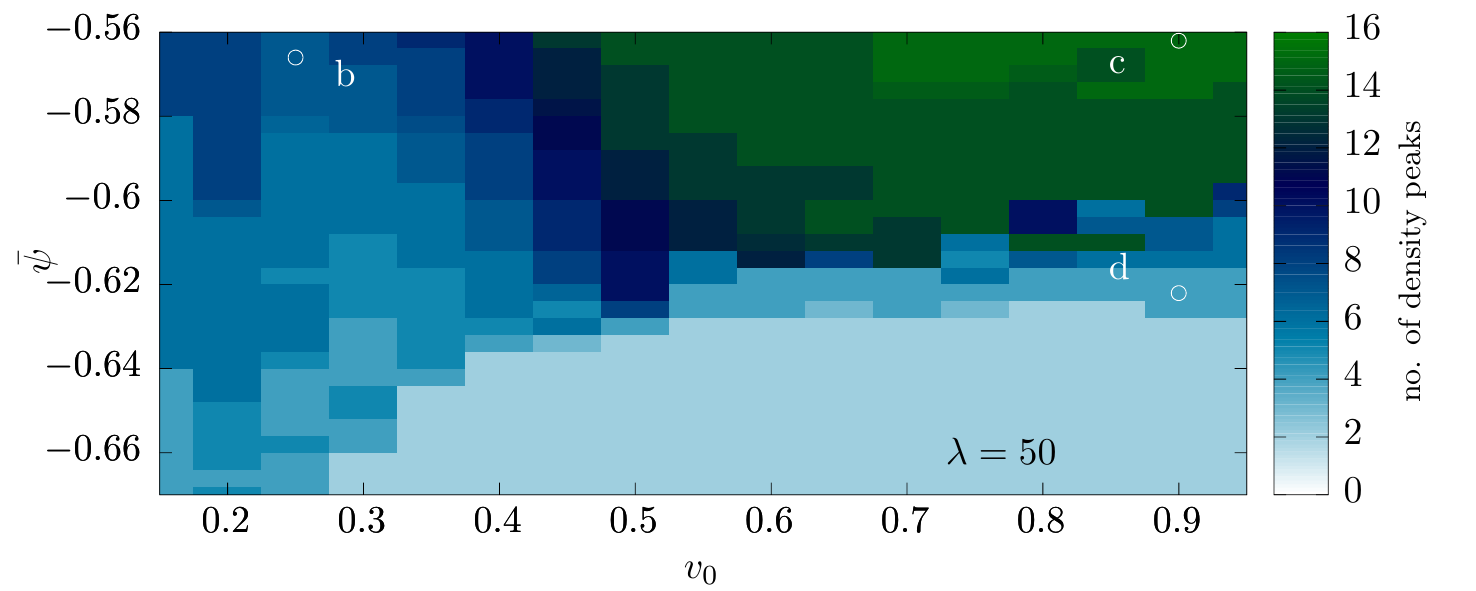}
    \put(4,41){(a)}
  \end{overpic}\vspace*{0.3cm}
\begin{overpic}[width=4cm]{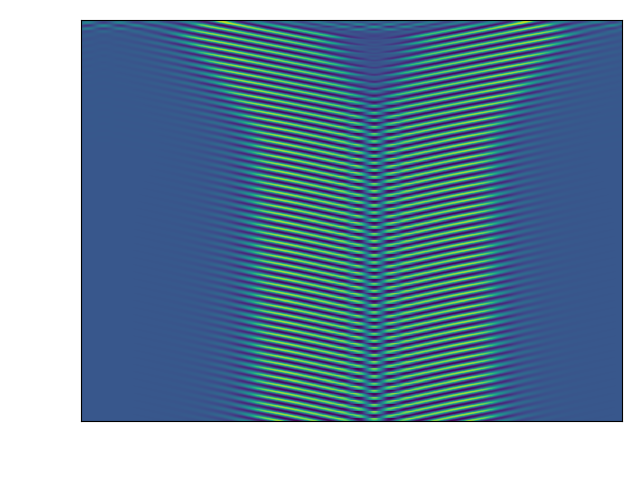}
  \put(-9,80){(b)}
   \put(-14,35){\rotatebox{90}{\makebox(0,0){$t$}}}
  \put(6.5,70){0}
   \put(-9,36){1000}
   \put(-9,7){2000}
   \put(8,0){0}
   \put(51,0){0.5}
   \put(96,0){1}
    \put(46,-10){$x/L$}
\end{overpic}\hspace*{0.4cm}
\begin{overpic}[width=4cm]{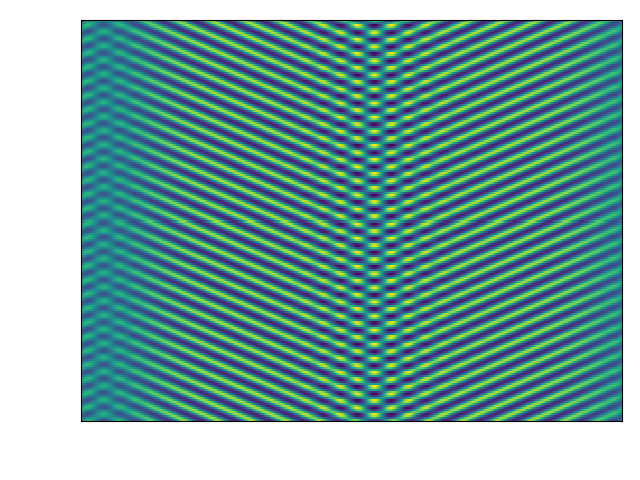}
  \put(-9,80){(c)}
  \put(-9,70){1800}
   \put(-9,38){1900}
   \put(-9.,7){2000}
   \put(9.5,0){0}
   \put(51,0){0.5}
   \put(96,0){1}
   \put(49,-10){$x/L$}
\end{overpic}\hspace*{0.4cm}
\begin{overpic}[width=4cm]{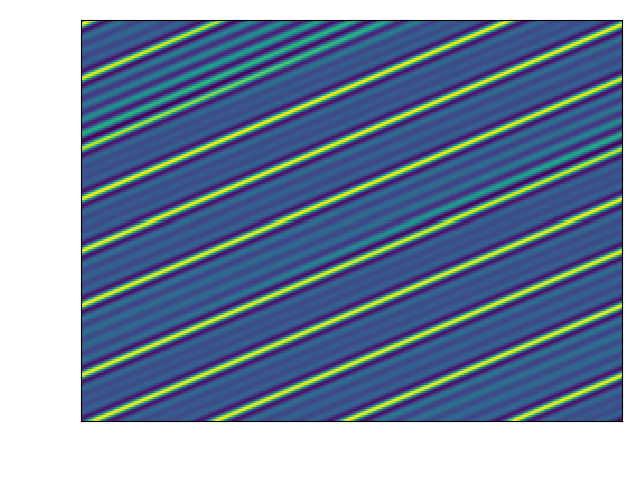}
  \put(-9,80){(d)}
  \put(-9,70){1800}
   \put(-9,38){1900}
   \put(-9.,7){2000}
   \put(9.5,0){0}
   \put(51,0){0.5}
   \put(96,0){1}
   \put(49,-10){$x/L$}
\end{overpic}
  \caption{(a) Morphological phase diagram in the ($v_0$, $\bar{\psi}$)-plane for an initial free path of $\lambda=50$. Shown is the region above the linear stability border of the liquid phase at $\bar{\psi}\approx-0.67$. The initial state corresponds to two traveling one-peak LS on collision course. The long-time outcome is presented in terms of the color-coded number of density peaks. Light blue indicates two peaks what corresponds to an elastic collision. Increasing $\bar{\psi}$, the homogenous background of the collision partners becomes more unstable resulting in the formation of increasingly many peaks.  The remaining parameters are as in Fig.~\ref{fig:freepath}. Panels (b) to (d) give selected space-time plots of long-time behavior at parameters indicated in panel (a) by letters ``b'' to ``d'', respectively.}
\label{fig:aboveLS}
\end{figure}
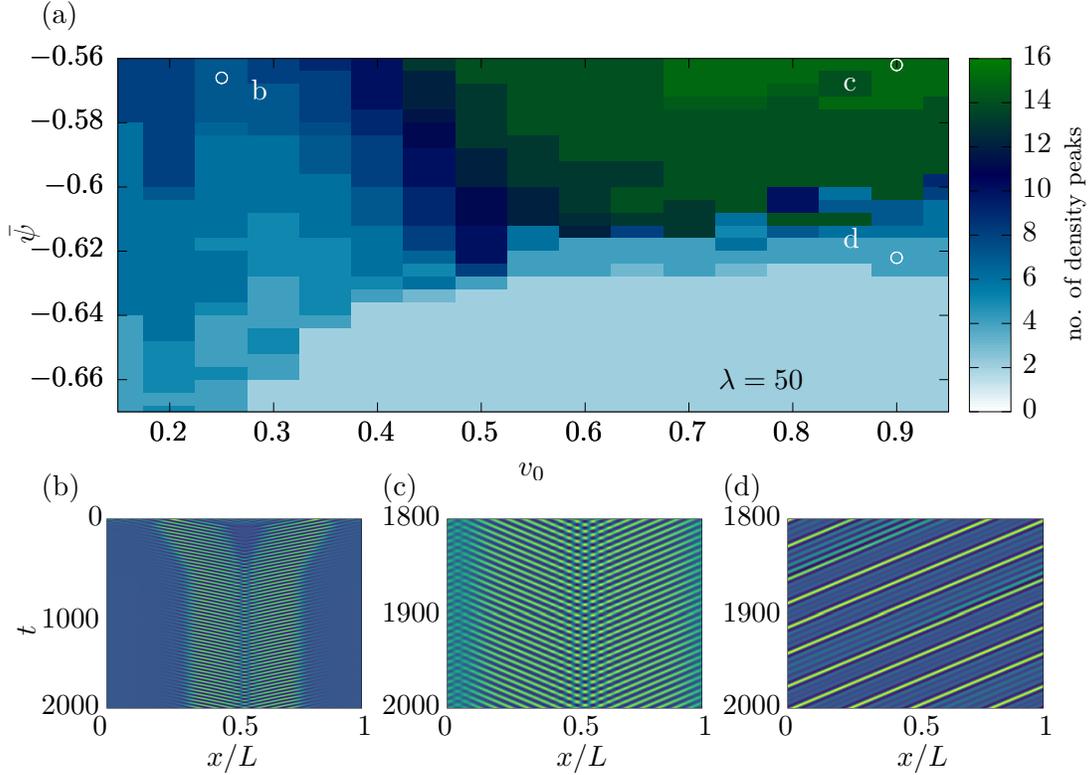

Finally, we shed some light on the behavior in the region above the limit of linear stability of the homogeneous phase. Figure~\ref{fig:aboveLS}~(a) shows that the area of elastic collisions as determined in Fig.~\ref{fig:freepath} with $\lambda=50$ reaches well into the linearly unstable region. This is possible because due to mass conservation the presence of the LS lowers the liquid background density practically shifting the stability border. However, this only holds for a small range of $\bar{\psi}$. The transition from light blue indicating two peaks to darker shading indicates the upper limit of the region of elastic collisions. Increasing $\bar{\psi}$ further leads to the creation of additional spatial modulations in the vicinity of the original LS. Figures~\ref{fig:aboveLS}~(b) to (d) demonstrate selected time evolutions of the colliding density bumps. In panel (b), the initial traveling LSs very fast become much broader by developing further peaks. The enlarged structures collide, their envelops come to rest while the individual peaks continue to travel with opposing phase velocities. Overall a localized source-sink structure of peaks is formed.  Panel (c) shows a domain-filling traveling crystal with a source of traveling peaks on the boundary and a sink at the center. In panel (d) a state with four density peaks has emerged all traveling into the same direction.

\subsection{Gas of colliding LSs in 1d}

\begin{figure}
 \hspace*{1.cm}
  \begin{overpic}[width=6.9cm]{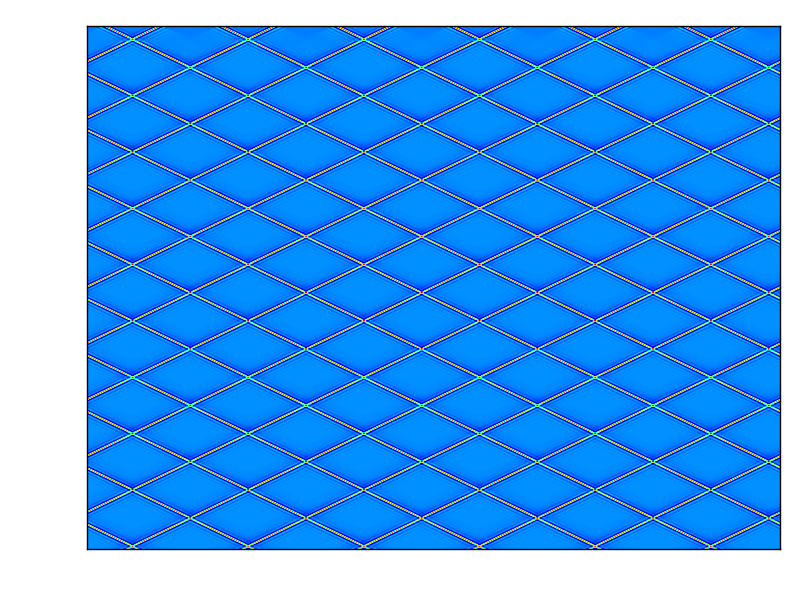}
    \put(-6,77){(a)}
  \put(-7,35){\rotatebox{90}{\makebox(0,0){$t$}}}
  \put(6.5,70){0}
   \put(1,54){500}
   \put(-2,38){1000}
   \put(-2,22){1500}
   \put(-2.,7){2000}
   \put(5,0){-300}
   \put(51.5,0){0}
   \put(91,0){300}
\end{overpic}\hspace*{0.2cm}
\vspace*{0.7cm}

\hspace*{1.cm}  \begin{overpic}[width=6.9cm]{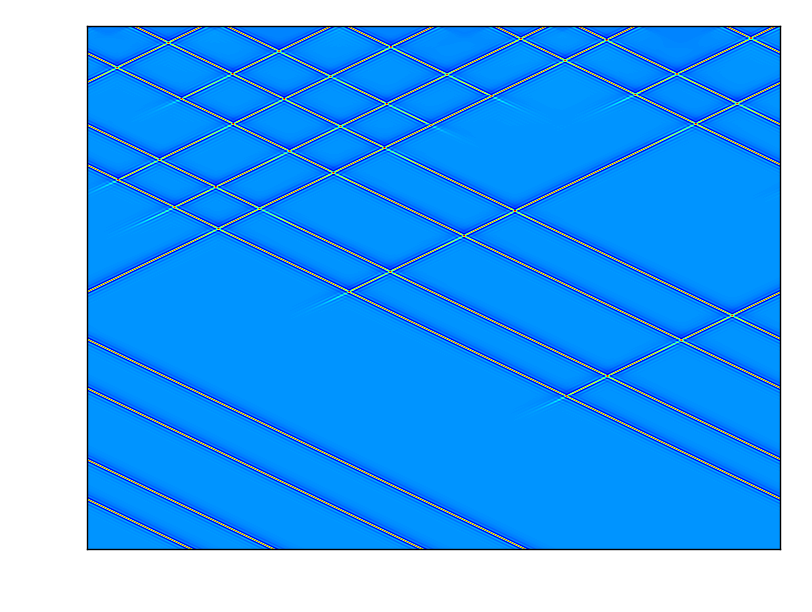}
    \put(-6,77){(b)}
  \put(-7,35){\rotatebox{90}{\makebox(0,0){$t$}}}
  \put(6.5,70){0}
   \put(1,54){500}
   \put(-2,38){1000}
   \put(-2,22){1500}
   \put(-2.,7){2000}
   \put(9.5,0){0}
   \put(5,0){-300}
   \put(51.5,0){0}
   \put(91,0){300}
   \put(51,-8){$x$}
\end{overpic}\hspace*{0.2cm}
\vspace*{1cm}
  \caption{(a) Space-time plot of a ``stable gas'' of traveling one-peak LS on a domain with $L=600$ at $v_0=0.5$ and $\bar{\psi}=-0.7$. Twelve one-peak LS are equally spaced on the domain resulting in a uniform and constant free path of $\lambda=50$. Panel (b) shows a case where the LS are initially randomly shifted from the regular spacing. As a result, most LS decays after a number of collisions and, eventually, only four LS survive. The remaining parameters are as in Fig.~\ref{fig:1dsnaps}.}
  \label{fig:1dgas}
\end{figure}
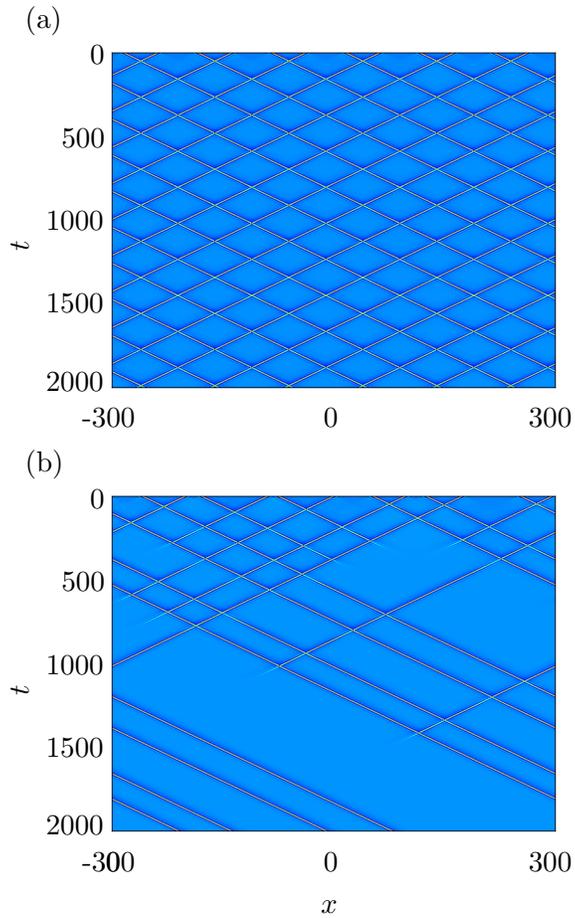

Based on the obtained results for the elastic collisions of two traveling one-peak LSs we now investigate the behavior of several such LSs confined in a large domain. If the number of colliding ``particles'' is conserved such a system could be termed a ``gas''. Figure~\ref{fig:1dgas} demonstrates that the minimal free path is crucial for the conservation of the number of LS. Figure~\ref{fig:1dgas}~(a) shows that for initially twelve equally spaced one-peak LSs with alternating velocities, a domain of size $L=600$ provides sufficient space. The corresponding free path of $\lambda=50$ allows all LSs to recover their original properties and to undergo consecutive elastic collisions.

However, this only works if the initial spacing of the LS is regular, as then all trajectories of the ensemble of LS form a regular pattern in the space-time plot, i.e., all collisions are synchronous and all free path lengths are identical and constant across collisions. This is highly unrealistic. In the more realistic situation of LSs of varying distances but the same mean free path one finds the behavior in Fig.~\ref{fig:1dgas}~(b). There, the twelve LSs are initially randomly shifted away from the regular spacing. Then, the free path is identical to the previous case and above the critical one, however, not all of the individual free paths are larger than the critical one at all times. This implies that time spans between some of the collisions is too short. The involved LSs do not recover their original properties and decay into the homogeneous background. After several collisions the gas is thinned out and only consists of LSs traveling with identical speed into the same direction. In the shown example only four LSs survive.

This implies that a stable gas of LS can not be achieved as in the parameter region where elastic collisions occur there is no mechanism for the creation of individual peaks out of the uniform background. A possible mechanism could exist beyond the onset of linear instability of the homogeneous background state. However, our investigation in section~\ref{sec:twocoll} has shown that no individual 'free' LS are created in this way but only multi-peak structures showing more complicated collective behavior (see Fig.~\ref{fig:aboveLS}).

\section{Conclusion}
\label{sec:conc}

\begin{figure}
\centering
  \begin{overpic}[width=4.2cm]{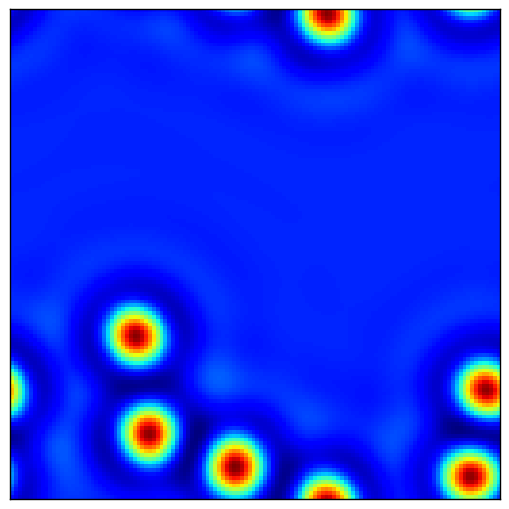} 
  \put(64,90){\color{white}$t=0$}
  \end{overpic}
 \begin{overpic}[width=4.2cm]{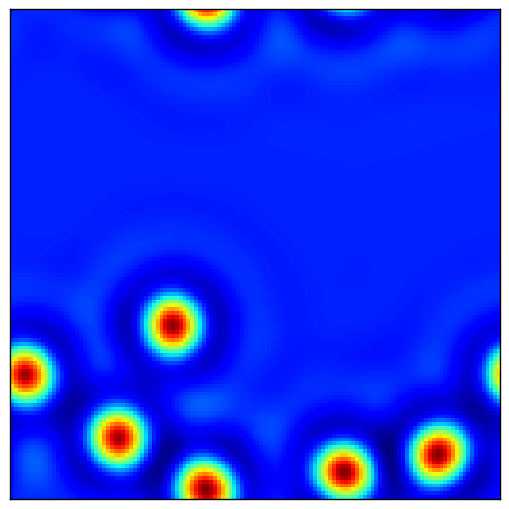}
    \put(64,90){\color{white}$t=20$}
  \end{overpic}
 \begin{overpic}[width=4.2cm]{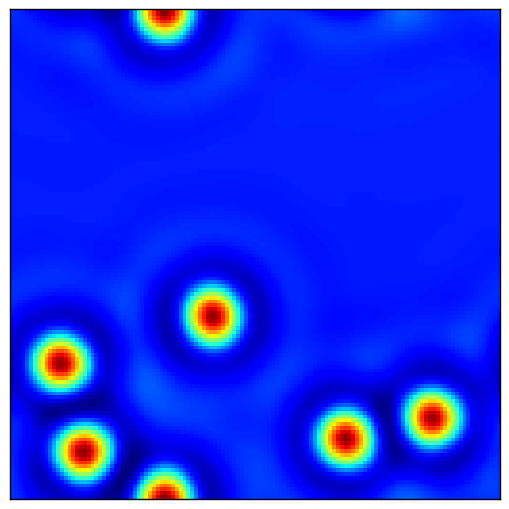}
    \put(64,90){\color{white}$t$\color{white}$=40$}
    \end{overpic}
  \begin{overpic}[width=4.2cm]{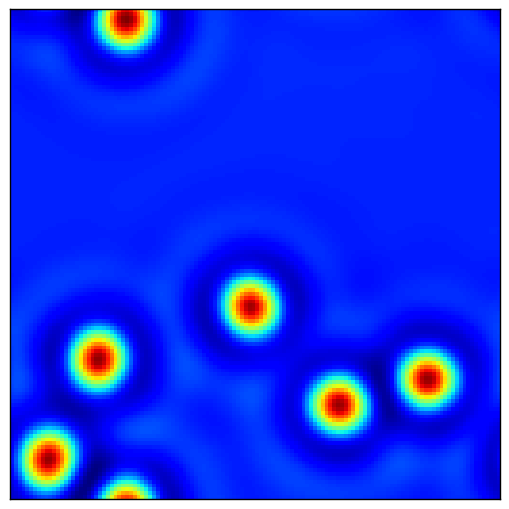} 
  \put(64,90){\color{white}$t=$\color{white}$60$}
  \end{overpic}
 \begin{overpic}[width=4.2cm]{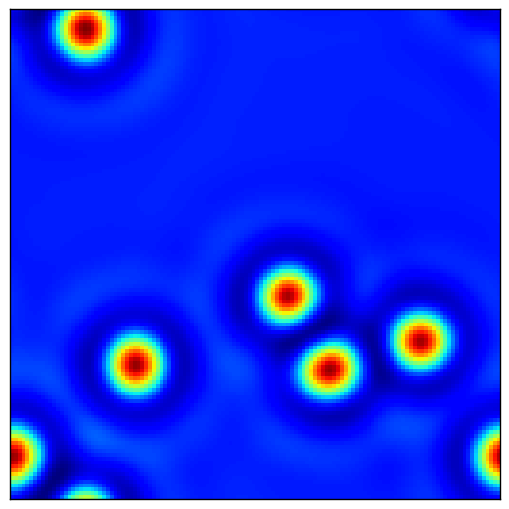}
    \put(64,90){\color{white}$t$\color{white}$=80$}
  \end{overpic}
 \begin{overpic}[width=4.2cm]{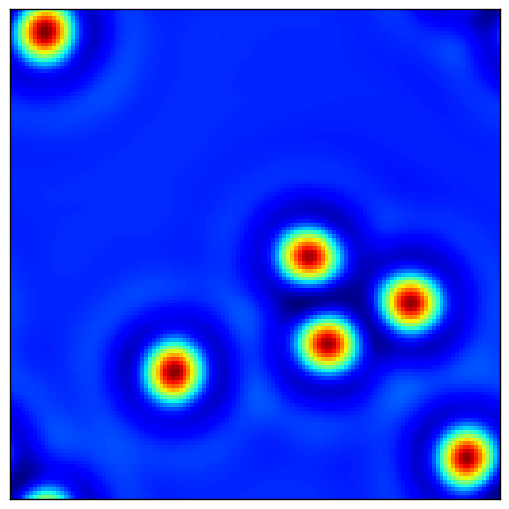}
    \put(64,90){\color{white}$t=100$}
  \end{overpic}
 \begin{overpic}[width=4.2cm]{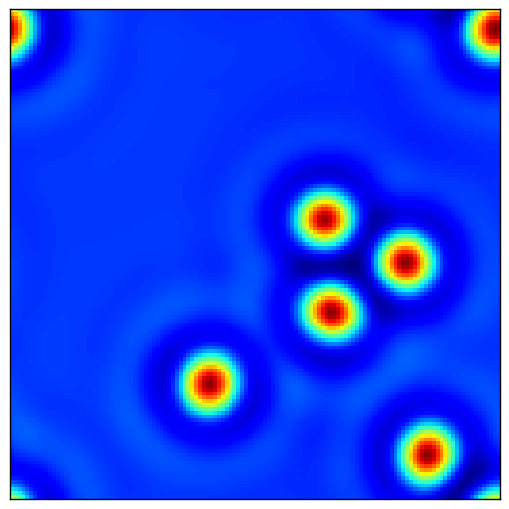}
    \put(64,90){\color{white}$t=120$}
  \end{overpic}
 \begin{overpic}[width=4.2cm]{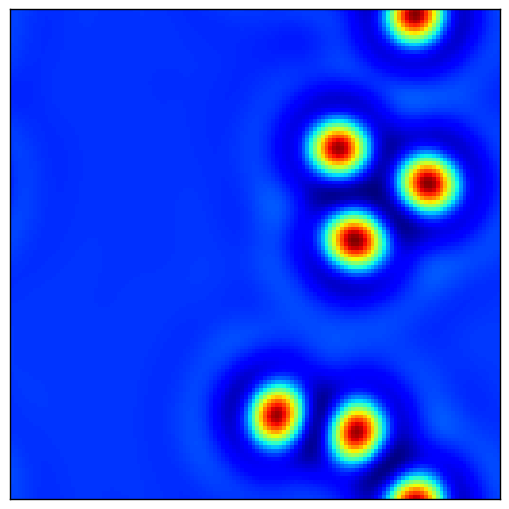}
    \put(64,90){\color{white}$t=160$}
  \end{overpic}
 \begin{overpic}[width=4.2cm]{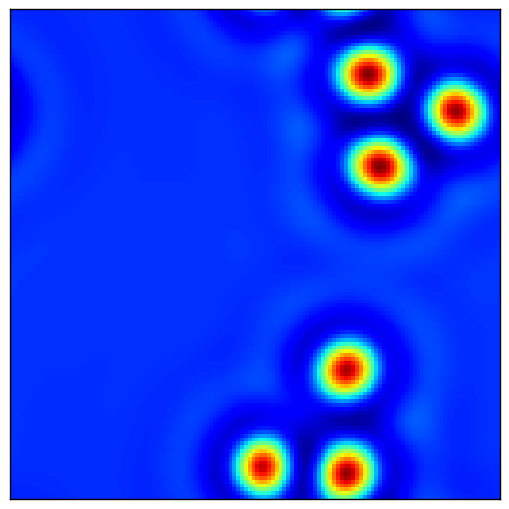}
    \put(64,90){\color{white}$t=200$}
  \end{overpic}
 \begin{overpic}[width=4.2cm]{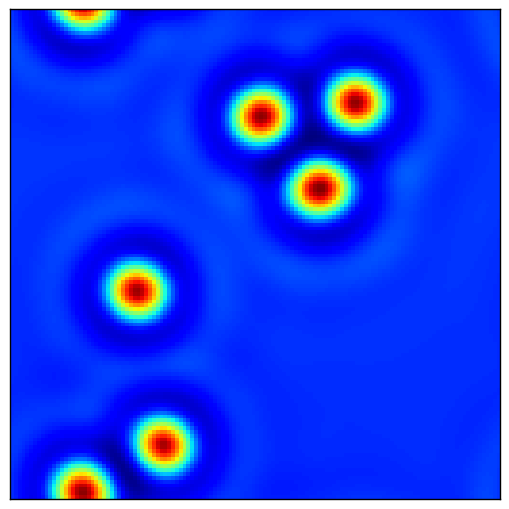}
    \put(64,90){\color{white}$t=720$}
  \end{overpic}
 \begin{overpic}[width=4.2cm]{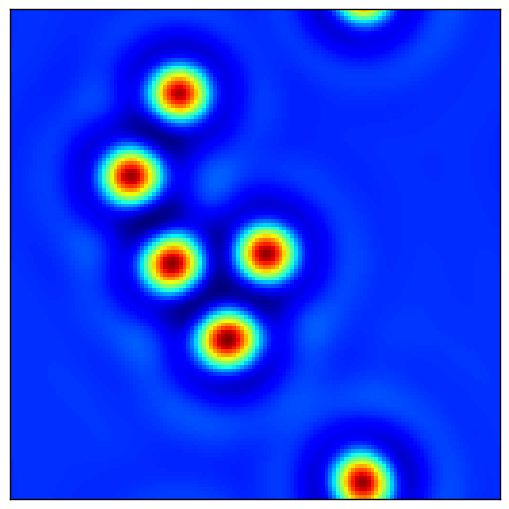}
    \put(64,90){\color{white}$t=900$}
  \end{overpic}
 \begin{overpic}[width=4.2cm]{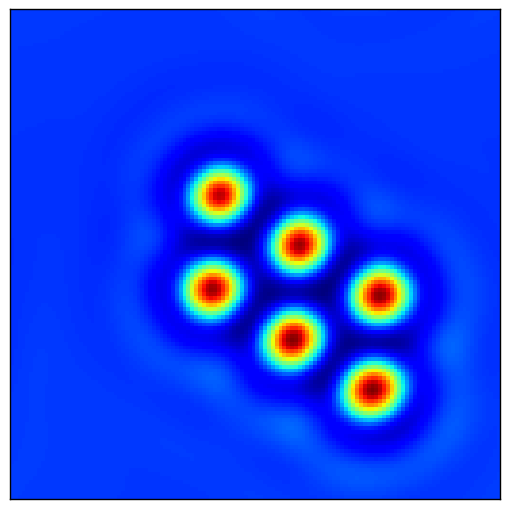}
    \put(64,90){\color{white}$t=4000$}
    \put(42,68){\rotatebox{118}{\makebox(0,0){\strut{}\textcolor{white}{\Huge{$\Longrightarrow$}}}}}
  \end{overpic}
  \caption{Exemplary time series of snapshots showing the collision and condensation dynamics of six traveling LS at $v_0=0.2$ and $\bar{\psi}=-0.8$. Shown is the density profile $\psi(\mathbf{x},t)$ at times $t$ as given in the panels. The domain size is $L_x\times L_y=32\pi / \sqrt{3} \times 14 \pi$ and has periodic boundaries. Note that all LS are in motion at all times. 
Remaining parameters as in Fig.~\ref{fig:1dsnaps}. } 
  \label{fig:2d_scattering}

\end{figure}

We have investigated the collision behavior of localized states of the active Phase-Field-Crystal (aPFC) model that combines elements of the Toner-Tu theory for self-propelled particles and the classical Phase-Field-Crystal (PFC) model for the liquid to crystalline phase transition. Based on results of \cite{OphausPRE18} on the linear stability of the homogeneous liquid state, the existence and onset of motion of space-filling crystals and selected crystallites, i.e., localized states, we have first studied the bifurcation structure and existence regions of various steadily traveling localized states in detail. This has allowed us to identify oscillating and modulated traveling localized states that have as the steadily traveling localized states no counterpart in the classical PFC model that only allows for resting states.

Results of linear stability analyses, two-parameter continuation and direct time simulations have been combined to obtain morphological phase diagrams in the parameter plane spanned by activity and mean concentration. They indicate where the various resting and traveling localized states exist. This has allowed us to identify the parameter region where subsequently we have analyzed the collision behavior of traveling one-peak  localized states. As a result we have distinguished elastic and inelastic collisions. In the elastic case the localized states eventually recover their properties after a collision. In contrast, in the inelastic case the localized states may annihilate or form resting or traveling bound states. This occurs at all studied parameter values if a certain minimal free path is not be guaranteed any more. We have shown that this rules out the possibility of a stable ``gas'' of localized states as in such a gas there is always a spectrum of different free  paths. This always results in the annihilation of localized states if one does not use highly artificial regular initial conditions.

There remains a number of questions that merit further investigation. First, the bifurcation behavior related to the emergence of the newly identified oscillating and modulated traveling localized states should be clarified, e.g., by employing continuation techniques for time-periodic states. This has been beyond the scope of the present work, but has recently been done for several time-periodic spatial structures in reaction-diffusion systems~\cite{SOMS_PRE95} as well as (simpler) systems of driven \cite{LRTT2016pf,TWGT2019prf} or active matter \cite{StGT2020c}. Ultimately, this should contribute to a more systematic understanding of mechanisms that result in the onset of motion in systems of active matter described by nonvariational (or nonreciprocal) models \cite{praetorius2018active,doostmohammadi2017onset}. 


Second, it will be very instructive to expand the present study towards two- and three-dimensional systems as done for the classical passive PFC model in \cite{TARG2013pre, EGUW2018springer,TFEK2019njp} or for reaction-diffusion systems~\cite{PuBA2010ap, liehr2013dissipative,NTU_Chaos05}. However, an increase in dimensionality will result in a much richer bifurcation behavior as there are more symmetries that can be broken. This may, for instance, result in the occurence of drift bifurcations for the onset of translation (as in the present one-dimensional case) and also for rotation similar to transitions observed for LSs in reaction-diffusion systems~ \cite{MoskalenkoEPL2003}. 

As an outlook we consider an example for the collision behavior in two spatial dimensions that allows us to point out certain features and may serve as a starting point for future studies. Figure~\ref{fig:2d_scattering} shows snapshots of a simulation initialized with six randomly placed one-peak LSs that travel with a constant speed of $c\approx0.15$ in various directions (according to an activity of $v_0=0.2$). Interestingly, the collision behavior is much more robust than in the 1d case. We find that the number of peaks remains constant through the various collisions. In contrast to 1d, after the first collisions (from $t\approx50$), the one-peak LSs ``condense`` into larger clusters, first into into two- and three-peak LS (the latter emerges at $t\approx100$) and finally into a six-peak LS ($t=4000$). Within the cluster, the density peaks form a hexagonal pattern. The white arrow indicates the direction of motion and is oriented perpendicularly to one side of the equilateral triangle at the tip of the cluster. Before this direction of translation emerges, rotational modes are visible and the direction of the collective drift keeps changing. 

Such a dynamical coalescence of active particles into extended clusters that move as one entity is also observed in experiments. For instance, in Ref.~\cite{speckPRL110} carbon-coated colloidal Janus particles are dispersed in a mixture of water and lutidine and move with a speed depending on the intensity of illumination. There, the formation of larger clusters is a dynamical process and is counteracted by the break up of clusters. It will be interesting to investigate whether such break up processes can also occur in the dynamics of ensembles of LS in the active PFC model.

\section*{Acknowledgments}

We acknowledge support through the doctoral school ``Active living fluids'' funded by the German French University  (Grant No. CDFA-01-14). UT and LO thank Edgar Knobloch for hosting them during a stay at UC Berkeley, for enlightening discussions, and for the opportunity to tutor Lea (UT). LO wishes to thank ``Studienstiftung des deutschen Volkes'' and ``IP@WWU'' for financial support, Fenna Stegemerten, Max Holl and Tobias Frohoff-H\"ulsmann for frequent discussions.


\begin{thebibliography}{85}%
\makeatletter
\providecommand \@ifxundefined [1]{%
 \@ifx{#1\undefined}
}%
\providecommand \@ifnum [1]{%
 \ifnum #1\expandafter \@firstoftwo
 \else \expandafter \@secondoftwo
 \fi
}%
\providecommand \@ifx [1]{%
 \ifx #1\expandafter \@firstoftwo
 \else \expandafter \@secondoftwo
 \fi
}%
\providecommand \natexlab [1]{#1}%
\providecommand \enquote  [1]{``#1''}%
\providecommand \bibnamefont  [1]{#1}%
\providecommand \bibfnamefont [1]{#1}%
\providecommand \citenamefont [1]{#1}%
\providecommand \href@noop [0]{\@secondoftwo}%
\providecommand \href [0]{\begingroup \@sanitize@url \@href}%
\providecommand \@href[1]{\@@startlink{#1}\@@href}%
\providecommand \@@href[1]{\endgroup#1\@@endlink}%
\providecommand \@sanitize@url [0]{\catcode `\\12\catcode `\$12\catcode
  `\&12\catcode `\#12\catcode `\^12\catcode `\_12\catcode `\%12\relax}%
\providecommand \@@startlink[1]{}%
\providecommand \@@endlink[0]{}%
\providecommand \url  [0]{\begingroup\@sanitize@url \@url }%
\providecommand \@url [1]{\endgroup\@href {#1}{\urlprefix }}%
\providecommand \urlprefix  [0]{URL }%
\providecommand \Eprint [0]{\href }%
\providecommand \doibase [0]{http://dx.doi.org/}%
\providecommand \selectlanguage [0]{\@gobble}%
\providecommand \bibinfo  [0]{\@secondoftwo}%
\providecommand \bibfield  [0]{\@secondoftwo}%
\providecommand \translation [1]{[#1]}%
\providecommand \BibitemOpen [0]{}%
\providecommand \bibitemStop [0]{}%
\providecommand \bibitemNoStop [0]{.\EOS\space}%
\providecommand \EOS [0]{\spacefactor3000\relax}%
\providecommand \BibitemShut  [1]{\csname bibitem#1\endcsname}%
\let\auto@bib@innerbib\@empty
\bibitem [{\citenamefont {Ball}(1999)}]{Ball1999}%
  \BibitemOpen
  \bibfield  {author} {\bibinfo {author} {\bibfnamefont {P.}~\bibnamefont
  {Ball}},\ }\href {https://books.google.de/books?id=kmVV8x-5y-wC} {\emph
  {\bibinfo {title} {The Self-made Tapestry: Pattern Formation in Nature}}}\
  (\bibinfo  {publisher} {Oxford University Press},\ \bibinfo {year}
  {1999})\BibitemShut {NoStop}%
\bibitem [{\citenamefont {Lam}(1998)}]{Lam1998}%
  \BibitemOpen
  \bibfield  {author} {\bibinfo {author} {\bibfnamefont {L.}~\bibnamefont
  {Lam}},\ }\href@noop {} {\emph {\bibinfo {title} {Nonlinear Physics for
  Beginners: Fractals, Chaos, Solitons, Pattern Formation, Cellular Automata,
  Complex Systems}}}\ (\bibinfo  {publisher} {World Scientific},\ \bibinfo
  {year} {1998})\BibitemShut {NoStop}%
\bibitem [{\citenamefont {Pismen}(2006)}]{Pismen2006}%
  \BibitemOpen
  \bibfield  {author} {\bibinfo {author} {\bibfnamefont {L.}~\bibnamefont
  {Pismen}},\ }\href@noop {} {\emph {\bibinfo {title} {Patterns and Interfaces
  in Dissipative Dynamics (Springer Series in Synergetics)}}}\ (\bibinfo
  {publisher} {Springer},\ \bibinfo {year} {2006})\BibitemShut {NoStop}%
\bibitem [{\citenamefont {Misbah}(2017)}]{Misbah2017}%
  \BibitemOpen
  \bibfield  {author} {\bibinfo {author} {\bibfnamefont {C.}~\bibnamefont
  {Misbah}},\ }\href@noop {} {\emph {\bibinfo {title} {Complex dynamics and
  morphogenesis: an introduction to nonlinear science}}}\ (\bibinfo
  {publisher} {Springer Netherlands},\ \bibinfo {year} {2017})\BibitemShut
  {NoStop}%
\bibitem [{\citenamefont {Cross}\ and\ \citenamefont
  {Hohenberg}(1993)}]{CrHo1993rmp}%
  \BibitemOpen
  \bibfield  {author} {\bibinfo {author} {\bibfnamefont {M.~C.}\ \bibnamefont
  {Cross}}\ and\ \bibinfo {author} {\bibfnamefont {P.~C.}\ \bibnamefont
  {Hohenberg}},\ }\href {\doibase 10.1103/RevModPhys.65.851} {\bibfield
  {journal} {\bibinfo  {journal} {Rev. Mod. Phys.}\ }\textbf {\bibinfo {volume}
  {65}},\ \bibinfo {pages} {851} (\bibinfo {year} {1993})}\BibitemShut
  {NoStop}%
\bibitem [{\citenamefont {Wada}\ and\ \citenamefont {Netz}(2007)}]{wadaPRL99}%
  \BibitemOpen
  \bibfield  {author} {\bibinfo {author} {\bibfnamefont {H.}~\bibnamefont
  {Wada}}\ and\ \bibinfo {author} {\bibfnamefont {R.~R.}\ \bibnamefont
  {Netz}},\ }\href {\doibase 10.1103/PhysRevLett.99.108102} {\bibfield
  {journal} {\bibinfo  {journal} {Phys. Rev. Lett.}\ }\textbf {\bibinfo
  {volume} {99}},\ \bibinfo {pages} {108102} (\bibinfo {year}
  {2007})}\BibitemShut {NoStop}%
\bibitem [{\citenamefont {Rappel}\ \emph {et~al.}(1999)\citenamefont {Rappel},
  \citenamefont {Nicol}, \citenamefont {Sarkissian}, \citenamefont {Levine},\
  and\ \citenamefont {Loomis}}]{rappelPRL83}%
  \BibitemOpen
  \bibfield  {author} {\bibinfo {author} {\bibfnamefont {W.-J.}\ \bibnamefont
  {Rappel}}, \bibinfo {author} {\bibfnamefont {A.}~\bibnamefont {Nicol}},
  \bibinfo {author} {\bibfnamefont {A.}~\bibnamefont {Sarkissian}}, \bibinfo
  {author} {\bibfnamefont {H.}~\bibnamefont {Levine}}, \ and\ \bibinfo {author}
  {\bibfnamefont {W.~F.}\ \bibnamefont {Loomis}},\ }\href {\doibase
  10.1103/PhysRevLett.83.1247} {\bibfield  {journal} {\bibinfo  {journal}
  {Phys. Rev. Lett.}\ }\textbf {\bibinfo {volume} {83}},\ \bibinfo {pages}
  {1247} (\bibinfo {year} {1999})}\BibitemShut {NoStop}%
\bibitem [{\citenamefont {Szab\'o}\ \emph {et~al.}(2006)\citenamefont
  {Szab\'o}, \citenamefont {Sz\"oll\"osi}, \citenamefont {G\"onci},
  \citenamefont {Jur\'anyi}, \citenamefont {Selmeczi},\ and\ \citenamefont
  {Vicsek}}]{vicsekPRE74}%
  \BibitemOpen
  \bibfield  {author} {\bibinfo {author} {\bibfnamefont {B.}~\bibnamefont
  {Szab\'o}}, \bibinfo {author} {\bibfnamefont {G.~J.}\ \bibnamefont
  {Sz\"oll\"osi}}, \bibinfo {author} {\bibfnamefont {B.}~\bibnamefont
  {G\"onci}}, \bibinfo {author} {\bibfnamefont {Z.}~\bibnamefont {Jur\'anyi}},
  \bibinfo {author} {\bibfnamefont {D.}~\bibnamefont {Selmeczi}}, \ and\
  \bibinfo {author} {\bibfnamefont {T.}~\bibnamefont {Vicsek}},\ }\href
  {\doibase 10.1103/PhysRevE.74.061908} {\bibfield  {journal} {\bibinfo
  {journal} {Phys. Rev. E}\ }\textbf {\bibinfo {volume} {74}},\ \bibinfo
  {pages} {061908} (\bibinfo {year} {2006})}\BibitemShut {NoStop}%
\bibitem [{\citenamefont {Sumpter}(2010)}]{sumpter}%
  \BibitemOpen
  \bibfield  {author} {\bibinfo {author} {\bibfnamefont {D.~J.}\ \bibnamefont
  {Sumpter}},\ }\href@noop {} {\emph {\bibinfo {title} {Collective animal
  behavior}}}\ (\bibinfo  {publisher} {Princeton University Press},\ \bibinfo
  {address} {Princeton},\ \bibinfo {year} {2010})\BibitemShut {NoStop}%
\bibitem [{\citenamefont {Marchetti}\ \emph {et~al.}(2013)\citenamefont
  {Marchetti}, \citenamefont {Joanny}, \citenamefont {Ramaswamy}, \citenamefont
  {Liverpool}, \citenamefont {Prost}, \citenamefont {Rao},\ and\ \citenamefont
  {Simha}}]{Marchetti.RevModPhys.85}%
  \BibitemOpen
  \bibfield  {author} {\bibinfo {author} {\bibfnamefont {M.~C.}\ \bibnamefont
  {Marchetti}}, \bibinfo {author} {\bibfnamefont {J.~F.}\ \bibnamefont
  {Joanny}}, \bibinfo {author} {\bibfnamefont {S.}~\bibnamefont {Ramaswamy}},
  \bibinfo {author} {\bibfnamefont {T.~B.}\ \bibnamefont {Liverpool}}, \bibinfo
  {author} {\bibfnamefont {J.}~\bibnamefont {Prost}}, \bibinfo {author}
  {\bibfnamefont {M.}~\bibnamefont {Rao}}, \ and\ \bibinfo {author}
  {\bibfnamefont {R.~A.}\ \bibnamefont {Simha}},\ }\href {\doibase
  10.1103/RevModPhys.85.1143} {\bibfield  {journal} {\bibinfo  {journal} {Rev.
  Mod. Phys.}\ }\textbf {\bibinfo {volume} {85}},\ \bibinfo {pages} {1143}
  (\bibinfo {year} {2013})}\BibitemShut {NoStop}%
\bibitem [{\citenamefont {Bechinger}\ \emph {et~al.}(2016)\citenamefont
  {Bechinger}, \citenamefont {Di~Leonardo}, \citenamefont {Lowen},
  \citenamefont {Reichhardt}, \citenamefont {Volpe},\ and\ \citenamefont
  {Volpe}}]{BDLR2016rmp}%
  \BibitemOpen
  \bibfield  {author} {\bibinfo {author} {\bibfnamefont {C.}~\bibnamefont
  {Bechinger}}, \bibinfo {author} {\bibfnamefont {R.}~\bibnamefont
  {Di~Leonardo}}, \bibinfo {author} {\bibfnamefont {H.}~\bibnamefont {Lowen}},
  \bibinfo {author} {\bibfnamefont {C.}~\bibnamefont {Reichhardt}}, \bibinfo
  {author} {\bibfnamefont {G.}~\bibnamefont {Volpe}}, \ and\ \bibinfo {author}
  {\bibfnamefont {G.}~\bibnamefont {Volpe}},\ }\href {\doibase
  10.1103/RevModPhys.88.045006} {\bibfield  {journal} {\bibinfo  {journal}
  {Rev. Mod. Phys.}\ }\textbf {\bibinfo {volume} {88}},\ \bibinfo {pages} {UNSP
  045006} (\bibinfo {year} {2016})}\BibitemShut {NoStop}%
\bibitem [{\citenamefont {Uchida}\ and\ \citenamefont
  {Golestanian}(2011)}]{uchidaPRL106}%
  \BibitemOpen
  \bibfield  {author} {\bibinfo {author} {\bibfnamefont {N.}~\bibnamefont
  {Uchida}}\ and\ \bibinfo {author} {\bibfnamefont {R.}~\bibnamefont
  {Golestanian}},\ }\href {\doibase 10.1103/PhysRevLett.106.058104} {\bibfield
  {journal} {\bibinfo  {journal} {Phys. Rev. Lett.}\ }\textbf {\bibinfo
  {volume} {106}},\ \bibinfo {pages} {058104} (\bibinfo {year}
  {2011})}\BibitemShut {NoStop}%
\bibitem [{\citenamefont {Golestanian}\ \emph {et~al.}(2011)\citenamefont
  {Golestanian}, \citenamefont {Yeomans},\ and\ \citenamefont
  {Uchida}}]{golestanianSoftMatter7}%
  \BibitemOpen
  \bibfield  {author} {\bibinfo {author} {\bibfnamefont {R.}~\bibnamefont
  {Golestanian}}, \bibinfo {author} {\bibfnamefont {J.~M.}\ \bibnamefont
  {Yeomans}}, \ and\ \bibinfo {author} {\bibfnamefont {N.}~\bibnamefont
  {Uchida}},\ }\href {\doibase 10.1039/C0SM01121E} {\bibfield  {journal}
  {\bibinfo  {journal} {Soft Matter}\ }\textbf {\bibinfo {volume} {7}},\
  \bibinfo {pages} {3074} (\bibinfo {year} {2011})}\BibitemShut {NoStop}%
\bibitem [{\citenamefont {Palacci}\ \emph {et~al.}(2013)\citenamefont
  {Palacci}, \citenamefont {Sacanna}, \citenamefont {Steinberg}, \citenamefont
  {Pine},\ and\ \citenamefont {Chaikin}}]{PalacciScience}%
  \BibitemOpen
  \bibfield  {author} {\bibinfo {author} {\bibfnamefont {J.}~\bibnamefont
  {Palacci}}, \bibinfo {author} {\bibfnamefont {S.}~\bibnamefont {Sacanna}},
  \bibinfo {author} {\bibfnamefont {A.~P.}\ \bibnamefont {Steinberg}}, \bibinfo
  {author} {\bibfnamefont {D.~J.}\ \bibnamefont {Pine}}, \ and\ \bibinfo
  {author} {\bibfnamefont {P.~M.}\ \bibnamefont {Chaikin}},\ }\href {\doibase
  10.1126/science.1230020} {\bibfield  {journal} {\bibinfo  {journal}
  {Science}\ }\textbf {\bibinfo {volume} {339}},\ \bibinfo {pages} {936}
  (\bibinfo {year} {2013})}\BibitemShut {NoStop}%
\bibitem [{\citenamefont {Jiang}\ \emph {et~al.}(2010)\citenamefont {Jiang},
  \citenamefont {Yoshinaga},\ and\ \citenamefont {Sano}}]{jiangPRL105}%
  \BibitemOpen
  \bibfield  {author} {\bibinfo {author} {\bibfnamefont {H.-R.}\ \bibnamefont
  {Jiang}}, \bibinfo {author} {\bibfnamefont {N.}~\bibnamefont {Yoshinaga}}, \
  and\ \bibinfo {author} {\bibfnamefont {M.}~\bibnamefont {Sano}},\ }\href
  {\doibase 10.1103/PhysRevLett.105.268302} {\bibfield  {journal} {\bibinfo
  {journal} {Phys. Rev. Lett.}\ }\textbf {\bibinfo {volume} {105}},\ \bibinfo
  {pages} {268302} (\bibinfo {year} {2010})}\BibitemShut {NoStop}%
\bibitem [{\citenamefont {Wang}\ \emph {et~al.}(2012)\citenamefont {Wang},
  \citenamefont {Castro}, \citenamefont {Hoyos},\ and\ \citenamefont
  {Mallouk}}]{WangACSNano2012}%
  \BibitemOpen
  \bibfield  {author} {\bibinfo {author} {\bibfnamefont {W.}~\bibnamefont
  {Wang}}, \bibinfo {author} {\bibfnamefont {L.~A.}\ \bibnamefont {Castro}},
  \bibinfo {author} {\bibfnamefont {M.}~\bibnamefont {Hoyos}}, \ and\ \bibinfo
  {author} {\bibfnamefont {T.~E.}\ \bibnamefont {Mallouk}},\ }\href {\doibase
  10.1021/nn301312z} {\bibfield  {journal} {\bibinfo  {journal} {ACS Nano}\
  }\textbf {\bibinfo {volume} {6}},\ \bibinfo {pages} {6122} (\bibinfo {year}
  {2012})},\ \bibinfo {note} {pMID: 22631222}\BibitemShut {NoStop}%
\bibitem [{\citenamefont {Vo{\ss}}\ and\ \citenamefont
  {Wittkowski}(2020)}]{VoWi2020preprint}%
  \BibitemOpen
  \bibfield  {author} {\bibinfo {author} {\bibfnamefont {J.}~\bibnamefont
  {Vo{\ss}}}\ and\ \bibinfo {author} {\bibfnamefont {R.}~\bibnamefont
  {Wittkowski}},\ }\href@noop {} {\  (\bibinfo {year} {2020})},\ \Eprint
  {http://arxiv.org/abs/http://arxiv.org/abs/2002.02048}
  {http://arxiv.org/abs/2002.02048} \BibitemShut {NoStop}%
\bibitem [{\citenamefont {Howse}\ \emph {et~al.}(2007)\citenamefont {Howse},
  \citenamefont {Jones}, \citenamefont {Ryan}, \citenamefont {Gough},
  \citenamefont {Vafabakhsh},\ and\ \citenamefont {Golestanian}}]{howsePRL99}%
  \BibitemOpen
  \bibfield  {author} {\bibinfo {author} {\bibfnamefont {J.~R.}\ \bibnamefont
  {Howse}}, \bibinfo {author} {\bibfnamefont {R.~A.~L.}\ \bibnamefont {Jones}},
  \bibinfo {author} {\bibfnamefont {A.~J.}\ \bibnamefont {Ryan}}, \bibinfo
  {author} {\bibfnamefont {T.}~\bibnamefont {Gough}}, \bibinfo {author}
  {\bibfnamefont {R.}~\bibnamefont {Vafabakhsh}}, \ and\ \bibinfo {author}
  {\bibfnamefont {R.}~\bibnamefont {Golestanian}},\ }\href {\doibase
  10.1103/PhysRevLett.99.048102} {\bibfield  {journal} {\bibinfo  {journal}
  {Phys. Rev. Lett.}\ }\textbf {\bibinfo {volume} {99}},\ \bibinfo {pages}
  {048102} (\bibinfo {year} {2007})}\BibitemShut {NoStop}%
\bibitem [{\citenamefont {Mallory}\ \emph {et~al.}(2018)\citenamefont
  {Mallory}, \citenamefont {Valeriani},\ and\ \citenamefont
  {Cacciuto}}]{MaVC2018arpc}%
  \BibitemOpen
  \bibfield  {author} {\bibinfo {author} {\bibfnamefont {S.}~\bibnamefont
  {Mallory}}, \bibinfo {author} {\bibfnamefont {C.}~\bibnamefont {Valeriani}},
  \ and\ \bibinfo {author} {\bibfnamefont {A.}~\bibnamefont {Cacciuto}},\
  }\href {\doibase 10.1146/annurev-physchem-050317-021237} {\bibfield
  {journal} {\bibinfo  {journal} {Annu. Rev. Phys. Chem.}\ }\textbf {\bibinfo
  {volume} {69}},\ \bibinfo {pages} {59} (\bibinfo {year} {2018})}\BibitemShut
  {NoStop}%
\bibitem [{\citenamefont {Ginot}\ \emph {et~al.}(2015)\citenamefont {Ginot},
  \citenamefont {Theurkauff}, \citenamefont {Levis}, \citenamefont {Ybert},
  \citenamefont {Bocquet}, \citenamefont {Berthier},\ and\ \citenamefont
  {Cottin-Bizonne}}]{Ginot2015prx}%
  \BibitemOpen
  \bibfield  {author} {\bibinfo {author} {\bibfnamefont {F.}~\bibnamefont
  {Ginot}}, \bibinfo {author} {\bibfnamefont {I.}~\bibnamefont {Theurkauff}},
  \bibinfo {author} {\bibfnamefont {D.}~\bibnamefont {Levis}}, \bibinfo
  {author} {\bibfnamefont {C.}~\bibnamefont {Ybert}}, \bibinfo {author}
  {\bibfnamefont {L.}~\bibnamefont {Bocquet}}, \bibinfo {author} {\bibfnamefont
  {L.}~\bibnamefont {Berthier}}, \ and\ \bibinfo {author} {\bibfnamefont
  {C.}~\bibnamefont {Cottin-Bizonne}},\ }\href {\doibase
  10.1103/PhysRevX.5.011004} {\bibfield  {journal} {\bibinfo  {journal} {Phys.
  Rev. X}\ }\textbf {\bibinfo {volume} {5}},\ \bibinfo {pages} {011004}
  (\bibinfo {year} {2015})}\BibitemShut {NoStop}%
\bibitem [{\citenamefont {Solon}\ \emph {et~al.}(2015)\citenamefont {Solon},
  \citenamefont {Stenhammar}, \citenamefont {Wittkowski}, \citenamefont
  {Kardar}, \citenamefont {Kafri}, \citenamefont {Cates},\ and\ \citenamefont
  {Tailleur}}]{SSWK2015prl}%
  \BibitemOpen
  \bibfield  {author} {\bibinfo {author} {\bibfnamefont {A.~P.}\ \bibnamefont
  {Solon}}, \bibinfo {author} {\bibfnamefont {J.}~\bibnamefont {Stenhammar}},
  \bibinfo {author} {\bibfnamefont {R.}~\bibnamefont {Wittkowski}}, \bibinfo
  {author} {\bibfnamefont {M.}~\bibnamefont {Kardar}}, \bibinfo {author}
  {\bibfnamefont {Y.}~\bibnamefont {Kafri}}, \bibinfo {author} {\bibfnamefont
  {M.~E.}\ \bibnamefont {Cates}}, \ and\ \bibinfo {author} {\bibfnamefont
  {J.}~\bibnamefont {Tailleur}},\ }\href {\doibase
  10.1103/PhysRevLett.114.198301} {\bibfield  {journal} {\bibinfo  {journal}
  {Phys. Rev. Lett.}\ }\textbf {\bibinfo {volume} {114}},\ \bibinfo {pages}
  {198301} (\bibinfo {year} {2015})}\BibitemShut {NoStop}%
\bibitem [{\citenamefont {Cates}\ and\ \citenamefont
  {Tailleur}(2015)}]{CaTa2015arcmp}%
  \BibitemOpen
  \bibfield  {author} {\bibinfo {author} {\bibfnamefont {M.}~\bibnamefont
  {Cates}}\ and\ \bibinfo {author} {\bibfnamefont {J.}~\bibnamefont
  {Tailleur}},\ }\href {\doibase 10.1146/annurev-conmatphys-031214-014710}
  {\bibfield  {journal} {\bibinfo  {journal} {Annu. Rev. Condens. Matter
  Phys.}\ }\textbf {\bibinfo {volume} {6}},\ \bibinfo {pages} {219} (\bibinfo
  {year} {2015})}\BibitemShut {NoStop}%
\bibitem [{\citenamefont {Thar}\ and\ \citenamefont {K\"uhl}(2002)}]{Thar2002}%
  \BibitemOpen
  \bibfield  {author} {\bibinfo {author} {\bibfnamefont {R.}~\bibnamefont
  {Thar}}\ and\ \bibinfo {author} {\bibfnamefont {M.}~\bibnamefont {K\"uhl}},\
  }\href@noop {} {\bibfield  {journal} {\bibinfo  {journal} {Appl. Environ.
  Microbiol.}\ }\textbf {\bibinfo {volume} {68}},\ \bibinfo {pages} {6310}
  (\bibinfo {year} {2002})}\BibitemShut {NoStop}%
\bibitem [{\citenamefont {Thar}\ and\ \citenamefont {K\"uhl}(2005)}]{Thar2005}%
  \BibitemOpen
  \bibfield  {author} {\bibinfo {author} {\bibfnamefont {R.}~\bibnamefont
  {Thar}}\ and\ \bibinfo {author} {\bibfnamefont {M.}~\bibnamefont {K\"uhl}},\
  }\href@noop {} {\bibfield  {journal} {\bibinfo  {journal} {FEMS Microbiol.
  Lett.}\ }\textbf {\bibinfo {volume} {246}},\ \bibinfo {pages} {75} (\bibinfo
  {year} {2005})}\BibitemShut {NoStop}%
\bibitem [{\citenamefont {Theurkauff}\ \emph {et~al.}(2012)\citenamefont
  {Theurkauff}, \citenamefont {Cottin-Bizonne}, \citenamefont {Palacci},
  \citenamefont {Ybert},\ and\ \citenamefont {Bocquet}}]{theurkauff2012prl}%
  \BibitemOpen
  \bibfield  {author} {\bibinfo {author} {\bibfnamefont {I.}~\bibnamefont
  {Theurkauff}}, \bibinfo {author} {\bibfnamefont {C.}~\bibnamefont
  {Cottin-Bizonne}}, \bibinfo {author} {\bibfnamefont {J.}~\bibnamefont
  {Palacci}}, \bibinfo {author} {\bibfnamefont {C.}~\bibnamefont {Ybert}}, \
  and\ \bibinfo {author} {\bibfnamefont {L.}~\bibnamefont {Bocquet}},\ }\href
  {\doibase 10.1103/PhysRevLett.108.268303} {\bibfield  {journal} {\bibinfo
  {journal} {Phys. Rev. Lett.}\ }\textbf {\bibinfo {volume} {108}},\ \bibinfo
  {pages} {268303} (\bibinfo {year} {2012})}\BibitemShut {NoStop}%
\bibitem [{\citenamefont {Petroff}\ \emph {et~al.}(2015)\citenamefont
  {Petroff}, \citenamefont {Wu},\ and\ \citenamefont
  {Libchaber}}]{LibchaberPRL2015}%
  \BibitemOpen
  \bibfield  {author} {\bibinfo {author} {\bibfnamefont {A.~P.}\ \bibnamefont
  {Petroff}}, \bibinfo {author} {\bibfnamefont {X.-L.}\ \bibnamefont {Wu}}, \
  and\ \bibinfo {author} {\bibfnamefont {A.}~\bibnamefont {Libchaber}},\ }\href
  {\doibase 10.1103/PhysRevLett.114.158102} {\bibfield  {journal} {\bibinfo
  {journal} {Phys. Rev. Lett.}\ }\textbf {\bibinfo {volume} {114}},\ \bibinfo
  {pages} {158102} (\bibinfo {year} {2015})}\BibitemShut {NoStop}%
\bibitem [{\citenamefont {Ginot}\ \emph {et~al.}(2018)\citenamefont {Ginot},
  \citenamefont {Theurkauff}, \citenamefont {Detcheverry}, \citenamefont
  {Ybert},\ and\ \citenamefont {Cottin-Bizonne}}]{ginot2018aggregation}%
  \BibitemOpen
  \bibfield  {author} {\bibinfo {author} {\bibfnamefont {F.}~\bibnamefont
  {Ginot}}, \bibinfo {author} {\bibfnamefont {I.}~\bibnamefont {Theurkauff}},
  \bibinfo {author} {\bibfnamefont {F.}~\bibnamefont {Detcheverry}}, \bibinfo
  {author} {\bibfnamefont {C.}~\bibnamefont {Ybert}}, \ and\ \bibinfo {author}
  {\bibfnamefont {C.}~\bibnamefont {Cottin-Bizonne}},\ }\href@noop {}
  {\bibfield  {journal} {\bibinfo  {journal} {Nature communications}\ }\textbf
  {\bibinfo {volume} {9}},\ \bibinfo {pages} {696} (\bibinfo {year}
  {2018})}\BibitemShut {NoStop}%
\bibitem [{\citenamefont {Toner}\ \emph {et~al.}(2005)\citenamefont {Toner},
  \citenamefont {Tu},\ and\ \citenamefont {Ramaswamy}}]{ToTR2005ap}%
  \BibitemOpen
  \bibfield  {author} {\bibinfo {author} {\bibfnamefont {J.}~\bibnamefont
  {Toner}}, \bibinfo {author} {\bibfnamefont {Y.~H.}\ \bibnamefont {Tu}}, \
  and\ \bibinfo {author} {\bibfnamefont {S.}~\bibnamefont {Ramaswamy}},\ }\href
  {\doibase 10.1016/j.aop.2005.04.011} {\bibfield  {journal} {\bibinfo
  {journal} {Ann. Phys.}\ }\textbf {\bibinfo {volume} {318}},\ \bibinfo {pages}
  {170} (\bibinfo {year} {2005})}\BibitemShut {NoStop}%
\bibitem [{\citenamefont {Mognetti}\ \emph {et~al.}(2013)\citenamefont
  {Mognetti}, \citenamefont {Saric}, \citenamefont {Angioletti-Uberti},
  \citenamefont {Cacciuto}, \citenamefont {Valeriani},\ and\ \citenamefont
  {Frenkel}}]{MSAC2013prl}%
  \BibitemOpen
  \bibfield  {author} {\bibinfo {author} {\bibfnamefont {B.~M.}\ \bibnamefont
  {Mognetti}}, \bibinfo {author} {\bibfnamefont {A.}~\bibnamefont {Saric}},
  \bibinfo {author} {\bibfnamefont {S.}~\bibnamefont {Angioletti-Uberti}},
  \bibinfo {author} {\bibfnamefont {A.}~\bibnamefont {Cacciuto}}, \bibinfo
  {author} {\bibfnamefont {C.}~\bibnamefont {Valeriani}}, \ and\ \bibinfo
  {author} {\bibfnamefont {D.}~\bibnamefont {Frenkel}},\ }\href {\doibase
  10.1103/PhysRevLett.111.245702} {\bibfield  {journal} {\bibinfo  {journal}
  {Phys. Rev. Lett.}\ }\textbf {\bibinfo {volume} {111}},\ \bibinfo {pages}
  {245702} (\bibinfo {year} {2013})}\BibitemShut {NoStop}%
\bibitem [{\citenamefont {Menzel}(2015)}]{Menz2015prspl}%
  \BibitemOpen
  \bibfield  {author} {\bibinfo {author} {\bibfnamefont {A.~M.}\ \bibnamefont
  {Menzel}},\ }\href@noop {} {\bibfield  {journal} {\bibinfo  {journal}
  {Physics reports}\ }\textbf {\bibinfo {volume} {554}},\ \bibinfo {pages} {1}
  (\bibinfo {year} {2015})}\BibitemShut {NoStop}%
\bibitem [{\citenamefont {Reinken}\ \emph {et~al.}(2018)\citenamefont
  {Reinken}, \citenamefont {Klapp}, \citenamefont {B\"ar},\ and\ \citenamefont
  {Heidenreich}}]{RKBH2018pre}%
  \BibitemOpen
  \bibfield  {author} {\bibinfo {author} {\bibfnamefont {H.}~\bibnamefont
  {Reinken}}, \bibinfo {author} {\bibfnamefont {S.~H.~L.}\ \bibnamefont
  {Klapp}}, \bibinfo {author} {\bibfnamefont {M.}~\bibnamefont {B\"ar}}, \ and\
  \bibinfo {author} {\bibfnamefont {S.}~\bibnamefont {Heidenreich}},\ }\href
  {\doibase 10.1103/PhysRevE.97.022613} {\bibfield  {journal} {\bibinfo
  {journal} {Phys. Rev. E}\ }\textbf {\bibinfo {volume} {97}},\ \bibinfo
  {pages} {022613} (\bibinfo {year} {2018})}\BibitemShut {NoStop}%
\bibitem [{\citenamefont {Toner}\ and\ \citenamefont {Tu}(1995)}]{ToTu1995prl}%
  \BibitemOpen
  \bibfield  {author} {\bibinfo {author} {\bibfnamefont {J.}~\bibnamefont
  {Toner}}\ and\ \bibinfo {author} {\bibfnamefont {Y.}~\bibnamefont {Tu}},\
  }\href {\doibase 10.1103/PhysRevLett.75.4326} {\bibfield  {journal} {\bibinfo
   {journal} {Phys. Rev. Lett.}\ }\textbf {\bibinfo {volume} {75}},\ \bibinfo
  {pages} {4326} (\bibinfo {year} {1995})}\BibitemShut {NoStop}%
\bibitem [{\citenamefont {Toner}\ and\ \citenamefont {Tu}(1998)}]{TonerTu}%
  \BibitemOpen
  \bibfield  {author} {\bibinfo {author} {\bibfnamefont {J.}~\bibnamefont
  {Toner}}\ and\ \bibinfo {author} {\bibfnamefont {Y.}~\bibnamefont {Tu}},\
  }\href {\doibase 10.1103/PhysRevE.58.4828} {\bibfield  {journal} {\bibinfo
  {journal} {Phys. Rev. E}\ }\textbf {\bibinfo {volume} {58}},\ \bibinfo
  {pages} {4828} (\bibinfo {year} {1998})}\BibitemShut {NoStop}%
\bibitem [{\citenamefont {Menzel}\ and\ \citenamefont
  {L\"owen}(2013)}]{MenzelLoewen}%
  \BibitemOpen
  \bibfield  {author} {\bibinfo {author} {\bibfnamefont {A.~M.}\ \bibnamefont
  {Menzel}}\ and\ \bibinfo {author} {\bibfnamefont {H.}~\bibnamefont
  {L\"owen}},\ }\href {\doibase 10.1103/PhysRevLett.110.055702} {\bibfield
  {journal} {\bibinfo  {journal} {Phys. Rev. Lett.}\ }\textbf {\bibinfo
  {volume} {110}},\ \bibinfo {pages} {055702} (\bibinfo {year}
  {2013})}\BibitemShut {NoStop}%
\bibitem [{\citenamefont {Elder}\ \emph {et~al.}(2002)\citenamefont {Elder},
  \citenamefont {Katakowski}, \citenamefont {Haataja},\ and\ \citenamefont
  {Grant}}]{ElderGrantPRL88}%
  \BibitemOpen
  \bibfield  {author} {\bibinfo {author} {\bibfnamefont {K.~R.}\ \bibnamefont
  {Elder}}, \bibinfo {author} {\bibfnamefont {M.}~\bibnamefont {Katakowski}},
  \bibinfo {author} {\bibfnamefont {M.}~\bibnamefont {Haataja}}, \ and\
  \bibinfo {author} {\bibfnamefont {M.}~\bibnamefont {Grant}},\ }\href
  {\doibase 10.1103/PhysRevLett.88.245701} {\bibfield  {journal} {\bibinfo
  {journal} {Phys. Rev. Lett.}\ }\textbf {\bibinfo {volume} {88}},\ \bibinfo
  {pages} {245701} (\bibinfo {year} {2002})}\BibitemShut {NoStop}%
\bibitem [{\citenamefont {Emmerich}\ \emph {et~al.}(2012)\citenamefont
  {Emmerich}, \citenamefont {L{\"o}wen}, \citenamefont {Wittkowski},
  \citenamefont {Gruhn}, \citenamefont {T{\'o}th}, \citenamefont {Tegze},\ and\
  \citenamefont {Gr{\'a}n{\'a}sy}}]{EmmerichPFC}%
  \BibitemOpen
  \bibfield  {author} {\bibinfo {author} {\bibfnamefont {H.}~\bibnamefont
  {Emmerich}}, \bibinfo {author} {\bibfnamefont {H.}~\bibnamefont {L{\"o}wen}},
  \bibinfo {author} {\bibfnamefont {R.}~\bibnamefont {Wittkowski}}, \bibinfo
  {author} {\bibfnamefont {T.}~\bibnamefont {Gruhn}}, \bibinfo {author}
  {\bibfnamefont {G.~I.}\ \bibnamefont {T{\'o}th}}, \bibinfo {author}
  {\bibfnamefont {G.}~\bibnamefont {Tegze}}, \ and\ \bibinfo {author}
  {\bibfnamefont {L.}~\bibnamefont {Gr{\'a}n{\'a}sy}},\ }\href@noop {}
  {\bibfield  {journal} {\bibinfo  {journal} {Advances in Physics}\ }\textbf
  {\bibinfo {volume} {61}},\ \bibinfo {pages} {665} (\bibinfo {year}
  {2012})}\BibitemShut {NoStop}%
\bibitem [{\citenamefont {Tegze}\ \emph {et~al.}(2009)\citenamefont {Tegze},
  \citenamefont {Granasy}, \citenamefont {Toth}, \citenamefont {Podmaniczky},
  \citenamefont {Jaatinen}, \citenamefont {Ala-Nissila},\ and\ \citenamefont
  {Pusztai}}]{TGTP2009prl}%
  \BibitemOpen
  \bibfield  {author} {\bibinfo {author} {\bibfnamefont {G.}~\bibnamefont
  {Tegze}}, \bibinfo {author} {\bibfnamefont {L.}~\bibnamefont {Granasy}},
  \bibinfo {author} {\bibfnamefont {G.~I.}\ \bibnamefont {Toth}}, \bibinfo
  {author} {\bibfnamefont {F.}~\bibnamefont {Podmaniczky}}, \bibinfo {author}
  {\bibfnamefont {A.}~\bibnamefont {Jaatinen}}, \bibinfo {author}
  {\bibfnamefont {T.}~\bibnamefont {Ala-Nissila}}, \ and\ \bibinfo {author}
  {\bibfnamefont {T.}~\bibnamefont {Pusztai}},\ }\href {\doibase
  10.1103/PhysRevLett.103.035702} {\bibfield  {journal} {\bibinfo  {journal}
  {Phys. Rev. Lett.}\ }\textbf {\bibinfo {volume} {103}},\ \bibinfo {pages}
  {035702} (\bibinfo {year} {2009})}\BibitemShut {NoStop}%
\bibitem [{\citenamefont {Elder}\ \emph {et~al.}(2012)\citenamefont {Elder},
  \citenamefont {Rossi}, \citenamefont {Kanerva}, \citenamefont {Sanches},
  \citenamefont {Ying}, \citenamefont {Granato}, \citenamefont {Achim},\ and\
  \citenamefont {Ala-Nissila}}]{ERKS2012prl}%
  \BibitemOpen
  \bibfield  {author} {\bibinfo {author} {\bibfnamefont {K.~R.}\ \bibnamefont
  {Elder}}, \bibinfo {author} {\bibfnamefont {G.}~\bibnamefont {Rossi}},
  \bibinfo {author} {\bibfnamefont {P.}~\bibnamefont {Kanerva}}, \bibinfo
  {author} {\bibfnamefont {F.}~\bibnamefont {Sanches}}, \bibinfo {author}
  {\bibfnamefont {S.~C.}\ \bibnamefont {Ying}}, \bibinfo {author}
  {\bibfnamefont {E.}~\bibnamefont {Granato}}, \bibinfo {author} {\bibfnamefont
  {C.~V.}\ \bibnamefont {Achim}}, \ and\ \bibinfo {author} {\bibfnamefont
  {T.}~\bibnamefont {Ala-Nissila}},\ }\href {\doibase
  10.1103/PhysRevLett.108.226102} {\bibfield  {journal} {\bibinfo  {journal}
  {Phys. Rev. Lett.}\ }\textbf {\bibinfo {volume} {108}},\ \bibinfo {pages}
  {226102} (\bibinfo {year} {2012})}\BibitemShut {NoStop}%
\bibitem [{\citenamefont {Thiele}\ \emph {et~al.}(2013)\citenamefont {Thiele},
  \citenamefont {Archer}, \citenamefont {Robbins}, \citenamefont {Gomez},\ and\
  \citenamefont {Knobloch}}]{TARG2013pre}%
  \BibitemOpen
  \bibfield  {author} {\bibinfo {author} {\bibfnamefont {U.}~\bibnamefont
  {Thiele}}, \bibinfo {author} {\bibfnamefont {A.~J.}\ \bibnamefont {Archer}},
  \bibinfo {author} {\bibfnamefont {M.~J.}\ \bibnamefont {Robbins}}, \bibinfo
  {author} {\bibfnamefont {H.}~\bibnamefont {Gomez}}, \ and\ \bibinfo {author}
  {\bibfnamefont {E.}~\bibnamefont {Knobloch}},\ }\href {\doibase
  10.1103/PhysRevE.87.042915} {\bibfield  {journal} {\bibinfo  {journal} {Phys.
  Rev. E}\ }\textbf {\bibinfo {volume} {87}},\ \bibinfo {pages} {042915}
  (\bibinfo {year} {2013})}\BibitemShut {NoStop}%
\bibitem [{\citenamefont {Knobloch}(2016{\natexlab{a}})}]{Knob2016ijam}%
  \BibitemOpen
  \bibfield  {author} {\bibinfo {author} {\bibfnamefont {E.}~\bibnamefont
  {Knobloch}},\ }\href {\doibase 10.1093/imamat/hxw029} {\bibfield  {journal}
  {\bibinfo  {journal} {IMA J. Appl. Math.}\ }\textbf {\bibinfo {volume}
  {81}},\ \bibinfo {pages} {457} (\bibinfo {year}
  {2016}{\natexlab{a}})}\BibitemShut {NoStop}%
\bibitem [{\citenamefont {Menzel}\ \emph {et~al.}(2014)\citenamefont {Menzel},
  \citenamefont {Ohta},\ and\ \citenamefont
  {L\"owen}}]{MenzelOhtaLoewenPhysRevE.89}%
  \BibitemOpen
  \bibfield  {author} {\bibinfo {author} {\bibfnamefont {A.~M.}\ \bibnamefont
  {Menzel}}, \bibinfo {author} {\bibfnamefont {T.}~\bibnamefont {Ohta}}, \ and\
  \bibinfo {author} {\bibfnamefont {H.}~\bibnamefont {L\"owen}},\ }\href
  {\doibase 10.1103/PhysRevE.89.022301} {\bibfield  {journal} {\bibinfo
  {journal} {Phys. Rev. E}\ }\textbf {\bibinfo {volume} {89}},\ \bibinfo
  {pages} {022301} (\bibinfo {year} {2014})}\BibitemShut {NoStop}%
\bibitem [{\citenamefont {Chervanyov}\ \emph {et~al.}(2016)\citenamefont
  {Chervanyov}, \citenamefont {Gomez},\ and\ \citenamefont
  {Thiele}}]{ChGT2016el}%
  \BibitemOpen
  \bibfield  {author} {\bibinfo {author} {\bibfnamefont {A.}~\bibnamefont
  {Chervanyov}}, \bibinfo {author} {\bibfnamefont {H.}~\bibnamefont {Gomez}}, \
  and\ \bibinfo {author} {\bibfnamefont {U.}~\bibnamefont {Thiele}},\ }\href
  {\doibase 10.1209/0295-5075/115/68001} {\bibfield  {journal} {\bibinfo
  {journal} {Europhys. Lett.}\ }\textbf {\bibinfo {volume} {115}},\ \bibinfo
  {pages} {68001} (\bibinfo {year} {2016})}\BibitemShut {NoStop}%
\bibitem [{\citenamefont {Ophaus}\ \emph {et~al.}(2018)\citenamefont {Ophaus},
  \citenamefont {Gurevich},\ and\ \citenamefont {Thiele}}]{OphausPRE18}%
  \BibitemOpen
  \bibfield  {author} {\bibinfo {author} {\bibfnamefont {L.}~\bibnamefont
  {Ophaus}}, \bibinfo {author} {\bibfnamefont {S.~V.}\ \bibnamefont
  {Gurevich}}, \ and\ \bibinfo {author} {\bibfnamefont {U.}~\bibnamefont
  {Thiele}},\ }\href {\doibase 10.1103/PhysRevE.98.022608} {\bibfield
  {journal} {\bibinfo  {journal} {Phys. Rev. E}\ }\textbf {\bibinfo {volume}
  {98}},\ \bibinfo {pages} {022608} (\bibinfo {year} {2018})}\BibitemShut
  {NoStop}%
\bibitem [{\citenamefont {Murray}(1993)}]{MathBio}%
  \BibitemOpen
  \bibfield  {author} {\bibinfo {author} {\bibfnamefont {J.~D.}\ \bibnamefont
  {Murray}},\ }\href@noop {} {\emph {\bibinfo {title} {Mathematical Biology}}}\
  (\bibinfo  {publisher} {Springer},\ \bibinfo {address} {Berlin},\ \bibinfo
  {year} {1993})\BibitemShut {NoStop}%
\bibitem [{\citenamefont {Meinhardt}(1982)}]{BioPatterns}%
  \BibitemOpen
  \bibfield  {author} {\bibinfo {author} {\bibfnamefont {H.}~\bibnamefont
  {Meinhardt}},\ }\href@noop {} {\emph {\bibinfo {title} {Models of Biological
  Pattern Formation}}}\ (\bibinfo  {publisher} {Academic Press},\ \bibinfo
  {address} {London},\ \bibinfo {year} {1982})\BibitemShut {NoStop}%
\bibitem [{\citenamefont {Coullet}\ \emph {et~al.}(2000)\citenamefont
  {Coullet}, \citenamefont {Riera},\ and\ \citenamefont
  {Tresser}}]{coulletPRL84}%
  \BibitemOpen
  \bibfield  {author} {\bibinfo {author} {\bibfnamefont {P.}~\bibnamefont
  {Coullet}}, \bibinfo {author} {\bibfnamefont {C.}~\bibnamefont {Riera}}, \
  and\ \bibinfo {author} {\bibfnamefont {C.}~\bibnamefont {Tresser}},\ }\href
  {\doibase 10.1103/PhysRevLett.84.3069} {\bibfield  {journal} {\bibinfo
  {journal} {Phys. Rev. Lett.}\ }\textbf {\bibinfo {volume} {84}},\ \bibinfo
  {pages} {3069} (\bibinfo {year} {2000})}\BibitemShut {NoStop}%
\bibitem [{\citenamefont {Kapral}(1995)}]{ChemWaves}%
  \BibitemOpen
  \bibfield  {author} {\bibinfo {author} {\bibfnamefont {R.}~\bibnamefont
  {Kapral}},\ }\href@noop {} {\emph {\bibinfo {title} {Chemical Waves and
  Patterns, Understanding Chemical Reactivity}}},\ edited by\ \bibinfo {editor}
  {\bibfnamefont {K.}~\bibnamefont {Showalter}},\ Vol.~\bibinfo {volume} {10}\
  (\bibinfo  {publisher} {Kluwer Academic Publishers},\ \bibinfo {address}
  {Dordrecht},\ \bibinfo {year} {1995})\BibitemShut {NoStop}%
\bibitem [{\citenamefont {Burke}\ and\ \citenamefont
  {Knobloch}(2006)}]{BurkeKnoblochLSgenSHe}%
  \BibitemOpen
  \bibfield  {author} {\bibinfo {author} {\bibfnamefont {J.}~\bibnamefont
  {Burke}}\ and\ \bibinfo {author} {\bibfnamefont {E.}~\bibnamefont
  {Knobloch}},\ }\href {\doibase 10.1103/PhysRevE.73.056211} {\bibfield
  {journal} {\bibinfo  {journal} {Phys. Rev. E}\ }\textbf {\bibinfo {volume}
  {73}},\ \bibinfo {pages} {056211} (\bibinfo {year} {2006})}\BibitemShut
  {NoStop}%
\bibitem [{\citenamefont {Purwins}\ \emph {et~al.}(2010)\citenamefont
  {Purwins}, \citenamefont {Bodeker},\ and\ \citenamefont
  {Amiranashvili}}]{PuBA2010ap}%
  \BibitemOpen
  \bibfield  {author} {\bibinfo {author} {\bibfnamefont {H.}~\bibnamefont
  {Purwins}}, \bibinfo {author} {\bibfnamefont {H.}~\bibnamefont {Bodeker}}, \
  and\ \bibinfo {author} {\bibfnamefont {S.}~\bibnamefont {Amiranashvili}},\
  }\href {\doibase 10.1080/00018732.2010.498228} {\bibfield  {journal}
  {\bibinfo  {journal} {Adv. Phys.}\ }\textbf {\bibinfo {volume} {59}},\
  \bibinfo {pages} {485} (\bibinfo {year} {2010})}\BibitemShut {NoStop}%
\bibitem [{\citenamefont {Akhmediev}\ and\ \citenamefont
  {Ankiewicz}(2008)}]{AA-LNP-08}%
  \BibitemOpen
  \bibfield  {author} {\bibinfo {author} {\bibfnamefont {N.}~\bibnamefont
  {Akhmediev}}\ and\ \bibinfo {author} {\bibfnamefont {A.}~\bibnamefont
  {Ankiewicz}},\ }\href {\doibase 10.1007/978-3-540-78217-9} {\emph {\bibinfo
  {title} {Dissipative Solitons: From Optics to Biology and Medicine
  Series}}},\ \bibinfo {series} {Lecture Notes in Physics}, Vol.\ \bibinfo
  {volume} {751}\ (\bibinfo  {publisher} {Springer Berlin Heidelberg},\
  \bibinfo {year} {2008})\BibitemShut {NoStop}%
\bibitem [{\citenamefont {Tlidi}\ and\ \citenamefont
  {Mandel}(1999)}]{TM_PRL_99}%
  \BibitemOpen
  \bibfield  {author} {\bibinfo {author} {\bibfnamefont {M.}~\bibnamefont
  {Tlidi}}\ and\ \bibinfo {author} {\bibfnamefont {P.}~\bibnamefont {Mandel}},\
  }\href {\doibase 10.1103/PhysRevLett.83.4995} {\bibfield  {journal} {\bibinfo
   {journal} {Phys. Rev. Lett.}\ }\textbf {\bibinfo {volume} {83}},\ \bibinfo
  {pages} {4995} (\bibinfo {year} {1999})}\BibitemShut {NoStop}%
\bibitem [{\citenamefont {Meron}\ \emph {et~al.}(2004)\citenamefont {Meron},
  \citenamefont {Gilad}, \citenamefont {von Hardenberg}, \citenamefont
  {Shachak},\ and\ \citenamefont {Zarmi}}]{MERON2004}%
  \BibitemOpen
  \bibfield  {author} {\bibinfo {author} {\bibfnamefont {E.}~\bibnamefont
  {Meron}}, \bibinfo {author} {\bibfnamefont {E.}~\bibnamefont {Gilad}},
  \bibinfo {author} {\bibfnamefont {J.}~\bibnamefont {von Hardenberg}},
  \bibinfo {author} {\bibfnamefont {M.}~\bibnamefont {Shachak}}, \ and\
  \bibinfo {author} {\bibfnamefont {Y.}~\bibnamefont {Zarmi}},\ }\href
  {\doibase https://doi.org/10.1016/S0960-0779(03)00049-3} {\bibfield
  {journal} {\bibinfo  {journal} {Chaos, Solitons \& Fractals}\ }\textbf
  {\bibinfo {volume} {19}},\ \bibinfo {pages} {367 } (\bibinfo {year}
  {2004})},\ \bibinfo {note} {fractals in Geophysics}\BibitemShut {NoStop}%
\bibitem [{\citenamefont {Richter}\ and\ \citenamefont
  {Barashenkov}(2005)}]{richterPRL94}%
  \BibitemOpen
  \bibfield  {author} {\bibinfo {author} {\bibfnamefont {R.}~\bibnamefont
  {Richter}}\ and\ \bibinfo {author} {\bibfnamefont {I.~V.}\ \bibnamefont
  {Barashenkov}},\ }\href {\doibase 10.1103/PhysRevLett.94.184503} {\bibfield
  {journal} {\bibinfo  {journal} {Phys. Rev. Lett.}\ }\textbf {\bibinfo
  {volume} {94}},\ \bibinfo {pages} {184503} (\bibinfo {year}
  {2005})}\BibitemShut {NoStop}%
\bibitem [{\citenamefont {Sch\"apers}\ \emph {et~al.}(2000)\citenamefont
  {Sch\"apers}, \citenamefont {Feldmann}, \citenamefont {Ackemann},\ and\
  \citenamefont {Lange}}]{SchaepersPRL2000}%
  \BibitemOpen
  \bibfield  {author} {\bibinfo {author} {\bibfnamefont {B.}~\bibnamefont
  {Sch\"apers}}, \bibinfo {author} {\bibfnamefont {M.}~\bibnamefont
  {Feldmann}}, \bibinfo {author} {\bibfnamefont {T.}~\bibnamefont {Ackemann}},
  \ and\ \bibinfo {author} {\bibfnamefont {W.}~\bibnamefont {Lange}},\ }\href
  {\doibase 10.1103/PhysRevLett.85.748} {\bibfield  {journal} {\bibinfo
  {journal} {Phys. Rev. Lett.}\ }\textbf {\bibinfo {volume} {85}},\ \bibinfo
  {pages} {748} (\bibinfo {year} {2000})}\BibitemShut {NoStop}%
\bibitem [{\citenamefont {Lioubashevski}\ \emph {et~al.}(1999)\citenamefont
  {Lioubashevski}, \citenamefont {Hamiel}, \citenamefont {Agnon}, \citenamefont
  {Reches},\ and\ \citenamefont {Fineberg}}]{LiouPRL1999}%
  \BibitemOpen
  \bibfield  {author} {\bibinfo {author} {\bibfnamefont {O.}~\bibnamefont
  {Lioubashevski}}, \bibinfo {author} {\bibfnamefont {Y.}~\bibnamefont
  {Hamiel}}, \bibinfo {author} {\bibfnamefont {A.}~\bibnamefont {Agnon}},
  \bibinfo {author} {\bibfnamefont {Z.}~\bibnamefont {Reches}}, \ and\ \bibinfo
  {author} {\bibfnamefont {J.}~\bibnamefont {Fineberg}},\ }\href {\doibase
  10.1103/PhysRevLett.83.3190} {\bibfield  {journal} {\bibinfo  {journal}
  {Phys. Rev. Lett.}\ }\textbf {\bibinfo {volume} {83}},\ \bibinfo {pages}
  {3190} (\bibinfo {year} {1999})}\BibitemShut {NoStop}%
\bibitem [{\citenamefont {Robbins}\ \emph {et~al.}(2012)\citenamefont
  {Robbins}, \citenamefont {Archer}, \citenamefont {Thiele},\ and\
  \citenamefont {Knobloch}}]{RATK2012pre}%
  \BibitemOpen
  \bibfield  {author} {\bibinfo {author} {\bibfnamefont {M.~J.}\ \bibnamefont
  {Robbins}}, \bibinfo {author} {\bibfnamefont {A.~J.}\ \bibnamefont {Archer}},
  \bibinfo {author} {\bibfnamefont {U.}~\bibnamefont {Thiele}}, \ and\ \bibinfo
  {author} {\bibfnamefont {E.}~\bibnamefont {Knobloch}},\ }\href@noop {}
  {\bibfield  {journal} {\bibinfo  {journal} {Phys. Rev. E}\ }\textbf {\bibinfo
  {volume} {85}},\ \bibinfo {pages} {061408} (\bibinfo {year}
  {2012})}\BibitemShut {NoStop}%
\bibitem [{\citenamefont {Engelnkemper}\ \emph {et~al.}(2018)\citenamefont
  {Engelnkemper}, \citenamefont {Gurevich}, \citenamefont {Uecker},
  \citenamefont {Wetzel},\ and\ \citenamefont {Thiele}}]{EGUW2018springer}%
  \BibitemOpen
  \bibfield  {author} {\bibinfo {author} {\bibfnamefont {S.}~\bibnamefont
  {Engelnkemper}}, \bibinfo {author} {\bibfnamefont {S.}~\bibnamefont
  {Gurevich}}, \bibinfo {author} {\bibfnamefont {H.}~\bibnamefont {Uecker}},
  \bibinfo {author} {\bibfnamefont {D.}~\bibnamefont {Wetzel}}, \ and\ \bibinfo
  {author} {\bibfnamefont {U.}~\bibnamefont {Thiele}},\ }\enquote {\bibinfo
  {title} {Computational modeling of bifurcations and instabilities in fluid
  mechanics},}\ \ (\bibinfo  {publisher} {Springer, Berlin},\ \bibinfo {year}
  {2018})\ Chap.\ \bibinfo {chapter} {Continuation for thin film hydrodynamics
  and related scalar problems}, pp.\ \bibinfo {pages} {459--501}\BibitemShut
  {NoStop}%
\bibitem [{\citenamefont {Burke}\ and\ \citenamefont
  {Knobloch}(2007)}]{BurkeKnoblochSnakingChaos2007}%
  \BibitemOpen
  \bibfield  {author} {\bibinfo {author} {\bibfnamefont {J.}~\bibnamefont
  {Burke}}\ and\ \bibinfo {author} {\bibfnamefont {E.}~\bibnamefont
  {Knobloch}},\ }\href {\doibase 10.1063/1.2746816} {\bibfield  {journal}
  {\bibinfo  {journal} {Chaos: An Interdisciplinary Journal of Nonlinear
  Science}\ }\textbf {\bibinfo {volume} {17}},\ \bibinfo {pages} {037102}
  (\bibinfo {year} {2007})}\BibitemShut {NoStop}%
\bibitem [{\citenamefont {Beck}\ \emph {et~al.}(2009)\citenamefont {Beck},
  \citenamefont {Knobloch}, \citenamefont {Lloyd}, \citenamefont {Sandstede},\
  and\ \citenamefont {Wagenknecht}}]{SandstedeSnakes}%
  \BibitemOpen
  \bibfield  {author} {\bibinfo {author} {\bibfnamefont {M.}~\bibnamefont
  {Beck}}, \bibinfo {author} {\bibfnamefont {J.}~\bibnamefont {Knobloch}},
  \bibinfo {author} {\bibfnamefont {D.~J.~B.}\ \bibnamefont {Lloyd}}, \bibinfo
  {author} {\bibfnamefont {B.}~\bibnamefont {Sandstede}}, \ and\ \bibinfo
  {author} {\bibfnamefont {T.}~\bibnamefont {Wagenknecht}},\ }\href {\doibase
  10.1137/080713306} {\bibfield  {journal} {\bibinfo  {journal} {SIAM Journal
  on Mathematical Analysis}\ }\textbf {\bibinfo {volume} {41}},\ \bibinfo
  {pages} {936} (\bibinfo {year} {2009})}\BibitemShut {NoStop}%
\bibitem [{\citenamefont {Bortolozzo}\ \emph {et~al.}(2008)\citenamefont
  {Bortolozzo}, \citenamefont {Clerc},\ and\ \citenamefont
  {Residori}}]{BoCR2008pre}%
  \BibitemOpen
  \bibfield  {author} {\bibinfo {author} {\bibfnamefont {U.}~\bibnamefont
  {Bortolozzo}}, \bibinfo {author} {\bibfnamefont {M.~G.}\ \bibnamefont
  {Clerc}}, \ and\ \bibinfo {author} {\bibfnamefont {S.}~\bibnamefont
  {Residori}},\ }\href {\doibase 10.1103/PhysRevE.78.036214} {\bibfield
  {journal} {\bibinfo  {journal} {Phys. Rev. E}\ }\textbf {\bibinfo {volume}
  {78}},\ \bibinfo {pages} {036214} (\bibinfo {year} {2008})}\BibitemShut
  {NoStop}%
\bibitem [{\citenamefont {Dawes}(2008)}]{Dawe2008sjads}%
  \BibitemOpen
  \bibfield  {author} {\bibinfo {author} {\bibfnamefont {J.~H.~P.}\
  \bibnamefont {Dawes}},\ }\href {\doibase 10.1137/06067794X} {\bibfield
  {journal} {\bibinfo  {journal} {SIAM J. Appl. Dyn. Syst.}\ }\textbf {\bibinfo
  {volume} {7}},\ \bibinfo {pages} {186} (\bibinfo {year} {2008})}\BibitemShut
  {NoStop}%
\bibitem [{\citenamefont {Lo~Jacono}\ \emph {et~al.}(2011)\citenamefont
  {Lo~Jacono}, \citenamefont {Bergeon},\ and\ \citenamefont
  {Knobloch}}]{LoBK2011jfm}%
  \BibitemOpen
  \bibfield  {author} {\bibinfo {author} {\bibfnamefont {D.}~\bibnamefont
  {Lo~Jacono}}, \bibinfo {author} {\bibfnamefont {A.}~\bibnamefont {Bergeon}},
  \ and\ \bibinfo {author} {\bibfnamefont {E.}~\bibnamefont {Knobloch}},\
  }\href {\doibase 10.1017/jfm.2011.402} {\bibfield  {journal} {\bibinfo
  {journal} {J. Fluid Mech.}\ }\textbf {\bibinfo {volume} {687}},\ \bibinfo
  {pages} {595} (\bibinfo {year} {2011})}\BibitemShut {NoStop}%
\bibitem [{\citenamefont {Pradenas}\ \emph {et~al.}(2017)\citenamefont
  {Pradenas}, \citenamefont {Araya}, \citenamefont {Clerc}, \citenamefont
  {Falcon}, \citenamefont {Gandhi},\ and\ \citenamefont
  {Knobloch}}]{PACF2017prf}%
  \BibitemOpen
  \bibfield  {author} {\bibinfo {author} {\bibfnamefont {B.}~\bibnamefont
  {Pradenas}}, \bibinfo {author} {\bibfnamefont {I.}~\bibnamefont {Araya}},
  \bibinfo {author} {\bibfnamefont {M.~G.}\ \bibnamefont {Clerc}}, \bibinfo
  {author} {\bibfnamefont {C.}~\bibnamefont {Falcon}}, \bibinfo {author}
  {\bibfnamefont {P.}~\bibnamefont {Gandhi}}, \ and\ \bibinfo {author}
  {\bibfnamefont {E.}~\bibnamefont {Knobloch}},\ }\href {\doibase
  10.1103/PhysRevFluids.2.064401} {\bibfield  {journal} {\bibinfo  {journal}
  {Phys. Rev. Fluids}\ }\textbf {\bibinfo {volume} {2}},\ \bibinfo {pages}
  {064401} (\bibinfo {year} {2017})}\BibitemShut {NoStop}%
\bibitem [{\citenamefont {Knobloch}(2016{\natexlab{b}})}]{K_IMA16}%
  \BibitemOpen
  \bibfield  {author} {\bibinfo {author} {\bibfnamefont {E.}~\bibnamefont
  {Knobloch}},\ }\href@noop {} {\bibfield  {journal} {\bibinfo  {journal} {IMA
  J. Appl. Math.}\ }\textbf {\bibinfo {volume} {81}},\ \bibinfo {pages} {457}
  (\bibinfo {year} {2016}{\natexlab{b}})}\BibitemShut {NoStop}%
\bibitem [{\citenamefont {Avitabile}\ \emph {et~al.}(2010)\citenamefont
  {Avitabile}, \citenamefont {Lloyd}, \citenamefont {Burke}, \citenamefont
  {Knobloch},\ and\ \citenamefont {Sandstede}}]{ALBK2010sjads}%
  \BibitemOpen
  \bibfield  {author} {\bibinfo {author} {\bibfnamefont {D.}~\bibnamefont
  {Avitabile}}, \bibinfo {author} {\bibfnamefont {D.~J.~B.}\ \bibnamefont
  {Lloyd}}, \bibinfo {author} {\bibfnamefont {J.}~\bibnamefont {Burke}},
  \bibinfo {author} {\bibfnamefont {E.}~\bibnamefont {Knobloch}}, \ and\
  \bibinfo {author} {\bibfnamefont {B.}~\bibnamefont {Sandstede}},\ }\href
  {\doibase 10.1137/100782747} {\bibfield  {journal} {\bibinfo  {journal} {SIAM
  J. Appl. Dyn. Syst.}\ }\textbf {\bibinfo {volume} {9}},\ \bibinfo {pages}
  {704} (\bibinfo {year} {2010})}\BibitemShut {NoStop}%
\bibitem [{\citenamefont {Lloyd}\ \emph {et~al.}(2008)\citenamefont {Lloyd},
  \citenamefont {Sandstede}, \citenamefont {Avitabile},\ and\ \citenamefont
  {Champneys}}]{LSAC2008sjads}%
  \BibitemOpen
  \bibfield  {author} {\bibinfo {author} {\bibfnamefont {D.~J.~B.}\
  \bibnamefont {Lloyd}}, \bibinfo {author} {\bibfnamefont {B.}~\bibnamefont
  {Sandstede}}, \bibinfo {author} {\bibfnamefont {D.}~\bibnamefont
  {Avitabile}}, \ and\ \bibinfo {author} {\bibfnamefont {A.~R.}\ \bibnamefont
  {Champneys}},\ }\href {\doibase 10.1137/070707622} {\bibfield  {journal}
  {\bibinfo  {journal} {SIAM J. Appl. Dyn. Syst.}\ }\textbf {\bibinfo {volume}
  {7}},\ \bibinfo {pages} {1049} (\bibinfo {year} {2008})}\BibitemShut
  {NoStop}%
\bibitem [{\citenamefont {Elder}\ and\ \citenamefont
  {Grant}(2004)}]{ElderGrantPRE70}%
  \BibitemOpen
  \bibfield  {author} {\bibinfo {author} {\bibfnamefont {K.~R.}\ \bibnamefont
  {Elder}}\ and\ \bibinfo {author} {\bibfnamefont {M.}~\bibnamefont {Grant}},\
  }\href {\doibase 10.1103/PhysRevE.70.051605} {\bibfield  {journal} {\bibinfo
  {journal} {Phys. Rev. E}\ }\textbf {\bibinfo {volume} {70}},\ \bibinfo
  {pages} {051605} (\bibinfo {year} {2004})}\BibitemShut {NoStop}%
\bibitem [{\citenamefont {Krauskopf}\ \emph {et~al.}(2007)\citenamefont
  {Krauskopf}, \citenamefont {Osinga},\ and\ \citenamefont
  {Galan-Vioque}}]{KrauskopfOsingaGalan-Vioque2007}%
  \BibitemOpen
  \bibinfo {editor} {\bibfnamefont {B.}~\bibnamefont {Krauskopf}}, \bibinfo
  {editor} {\bibfnamefont {H.~M.}\ \bibnamefont {Osinga}}, \ and\ \bibinfo
  {editor} {\bibfnamefont {J.}~\bibnamefont {Galan-Vioque}},\ eds.,\ \href@noop
  {} {\emph {\bibinfo {title} {Numerical Continuation Methods for Dynamical
  Systems}}}\ (\bibinfo  {publisher} {Springer},\ \bibinfo {address}
  {Dordrecht},\ \bibinfo {year} {2007})\BibitemShut {NoStop}%
\bibitem [{\citenamefont {Dijkstra}\ \emph {et~al.}(2014)\citenamefont
  {Dijkstra}, \citenamefont {Wubs}, \citenamefont {Cliffe}, \citenamefont
  {Doedel}, \citenamefont {Dragomirescu}, \citenamefont {Eckhardt},
  \citenamefont {Gelfgat}, \citenamefont {Hazel}, \citenamefont {Lucarini},
  \citenamefont {Salinger}, \citenamefont {Phipps}, \citenamefont
  {Sanchez-Umbria}, \citenamefont {Schuttelaars}, \citenamefont {Tuckerman},\
  and\ \citenamefont {Thiele}}]{DWCD2014ccp}%
  \BibitemOpen
  \bibfield  {author} {\bibinfo {author} {\bibfnamefont {H.~A.}\ \bibnamefont
  {Dijkstra}}, \bibinfo {author} {\bibfnamefont {F.~W.}\ \bibnamefont {Wubs}},
  \bibinfo {author} {\bibfnamefont {A.~K.}\ \bibnamefont {Cliffe}}, \bibinfo
  {author} {\bibfnamefont {E.}~\bibnamefont {Doedel}}, \bibinfo {author}
  {\bibfnamefont {I.~F.}\ \bibnamefont {Dragomirescu}}, \bibinfo {author}
  {\bibfnamefont {B.}~\bibnamefont {Eckhardt}}, \bibinfo {author}
  {\bibfnamefont {A.~Y.}\ \bibnamefont {Gelfgat}}, \bibinfo {author}
  {\bibfnamefont {A.}~\bibnamefont {Hazel}}, \bibinfo {author} {\bibfnamefont
  {V.}~\bibnamefont {Lucarini}}, \bibinfo {author} {\bibfnamefont {A.~G.}\
  \bibnamefont {Salinger}}, \bibinfo {author} {\bibfnamefont {E.~T.}\
  \bibnamefont {Phipps}}, \bibinfo {author} {\bibfnamefont {J.}~\bibnamefont
  {Sanchez-Umbria}}, \bibinfo {author} {\bibfnamefont {H.}~\bibnamefont
  {Schuttelaars}}, \bibinfo {author} {\bibfnamefont {L.~S.}\ \bibnamefont
  {Tuckerman}}, \ and\ \bibinfo {author} {\bibfnamefont {U.}~\bibnamefont
  {Thiele}},\ }\href {\doibase 10.4208/cicp.240912.180613a} {\bibfield
  {journal} {\bibinfo  {journal} {Commun. Comput. Phys.}\ }\textbf {\bibinfo
  {volume} {15}},\ \bibinfo {pages} {1} (\bibinfo {year} {2014})}\BibitemShut
  {NoStop}%
\bibitem [{\citenamefont {Engelnkemper}\ \emph {et~al.}(2019)\citenamefont
  {Engelnkemper}, \citenamefont {Gurevich}, \citenamefont {Uecker},
  \citenamefont {Wetzel},\ and\ \citenamefont {Thiele}}]{EGUW2019springer}%
  \BibitemOpen
  \bibfield  {author} {\bibinfo {author} {\bibfnamefont {S.}~\bibnamefont
  {Engelnkemper}}, \bibinfo {author} {\bibfnamefont {S.}~\bibnamefont
  {Gurevich}}, \bibinfo {author} {\bibfnamefont {H.}~\bibnamefont {Uecker}},
  \bibinfo {author} {\bibfnamefont {D.}~\bibnamefont {Wetzel}}, \ and\ \bibinfo
  {author} {\bibfnamefont {U.}~\bibnamefont {Thiele}},\ }in\ \href {\doibase
  10.1007/978-3-319-91494-7_13} {\emph {\bibinfo {booktitle} {Computational
  Modeling of Bifurcations and Instabilities in Fluid Mechanics}}},\ \bibinfo
  {series and number} {Computational Methods in Applied Sciences, vol 50},\
  \bibinfo {editor} {edited by\ \bibinfo {editor} {\bibfnamefont
  {A.}~\bibnamefont {Gelfgat}}}\ (\bibinfo  {publisher} {Springer},\ \bibinfo
  {year} {2019})\ pp.\ \bibinfo {pages} {459--501},\ \Eprint
  {http://arxiv.org/abs/http://arxiv.org/abs/1808.02321}
  {http://arxiv.org/abs/1808.02321} \BibitemShut {NoStop}%
\bibitem [{\citenamefont {Doedel}\ \emph {et~al.}(1991)\citenamefont {Doedel},
  \citenamefont {Keller},\ and\ \citenamefont {Kernevez}}]{DoKK1991ijbc}%
  \BibitemOpen
  \bibfield  {author} {\bibinfo {author} {\bibfnamefont {E.}~\bibnamefont
  {Doedel}}, \bibinfo {author} {\bibfnamefont {H.~B.}\ \bibnamefont {Keller}},
  \ and\ \bibinfo {author} {\bibfnamefont {J.~P.}\ \bibnamefont {Kernevez}},\
  }\href {\doibase 10.1142/S0218127491000397} {\bibfield  {journal} {\bibinfo
  {journal} {Int. J. Bifurcation Chaos}\ }\textbf {\bibinfo {volume} {1}},\
  \bibinfo {pages} {493} (\bibinfo {year} {1991})}\BibitemShut {NoStop}%
\bibitem [{\citenamefont {Doedel}\ and\ \citenamefont
  {Oldeman}(2009)}]{DoedelOldeman2009}%
  \BibitemOpen
  \bibfield  {author} {\bibinfo {author} {\bibfnamefont {E.~J.}\ \bibnamefont
  {Doedel}}\ and\ \bibinfo {author} {\bibfnamefont {B.~E.}\ \bibnamefont
  {Oldeman}},\ }\href@noop {} {\emph {\bibinfo {title} {AUTO07p: Continuation
  and bifurcation software for ordinary differential equations}}}\ (\bibinfo
  {publisher} {Concordia University},\ \bibinfo {address} {Montreal},\ \bibinfo
  {year} {2009})\BibitemShut {NoStop}%
\bibitem [{\citenamefont {Thiele}\ \emph {et~al.}(2014)\citenamefont {Thiele},
  \citenamefont {Kamps},\ and\ \citenamefont {Gurevich}}]{cenosTutorial}%
  \BibitemOpen
  \bibinfo {editor} {\bibfnamefont {U.}~\bibnamefont {Thiele}}, \bibinfo
  {editor} {\bibfnamefont {O.}~\bibnamefont {Kamps}}, \ and\ \bibinfo {editor}
  {\bibfnamefont {S.~V.}\ \bibnamefont {Gurevich}},\ eds.,\ \href@noop {}
  {\emph {\bibinfo {title} {M{\"u}nsteranian Torturials on Nonlinear Science:
  Continuation}}}\ (\bibinfo  {publisher} {CeNoS},\ \bibinfo {address}
  {M{\"u}nster},\ \bibinfo {year} {2014})\ \bibinfo {note}
  {\url{http://www.uni-muenster.de/CeNoS/Lehre/Tutorials}}\BibitemShut
  {NoStop}%
\bibitem [{\citenamefont {Thiele}\ \emph {et~al.}(2019)\citenamefont {Thiele},
  \citenamefont {Frohoff-H\"ulsmann}, \citenamefont {Engelnkemper},
  \citenamefont {Knobloch},\ and\ \citenamefont {Archer}}]{TFEK2019njp}%
  \BibitemOpen
  \bibfield  {author} {\bibinfo {author} {\bibfnamefont {U.}~\bibnamefont
  {Thiele}}, \bibinfo {author} {\bibfnamefont {T.}~\bibnamefont
  {Frohoff-H\"ulsmann}}, \bibinfo {author} {\bibfnamefont {S.}~\bibnamefont
  {Engelnkemper}}, \bibinfo {author} {\bibfnamefont {E.}~\bibnamefont
  {Knobloch}}, \ and\ \bibinfo {author} {\bibfnamefont {A.~J.}\ \bibnamefont
  {Archer}},\ }\href {\doibase 10.1088/1367-2630/ab5caf} {\bibfield  {journal}
  {\bibinfo  {journal} {New J. Phys.}\ }\textbf {\bibinfo {volume} {21}},\
  \bibinfo {pages} {123021} (\bibinfo {year} {2019})}\BibitemShut {NoStop}%
\bibitem [{\citenamefont {Stegemerten}\ \emph {et~al.}(2020)\citenamefont
  {Stegemerten}, \citenamefont {Gurevich},\ and\ \citenamefont
  {Thiele}}]{StGT2020c}%
  \BibitemOpen
  \bibfield  {author} {\bibinfo {author} {\bibfnamefont {F.}~\bibnamefont
  {Stegemerten}}, \bibinfo {author} {\bibfnamefont {S.~V.}\ \bibnamefont
  {Gurevich}}, \ and\ \bibinfo {author} {\bibfnamefont {U.}~\bibnamefont
  {Thiele}},\ }\href {\doibase 10.1063/5.0003271} {\bibfield  {journal}
  {\bibinfo  {journal} {Chaos}\ }\textbf {\bibinfo {volume} {30}},\ \bibinfo
  {pages} {053136} (\bibinfo {year} {2020})},\ \Eprint
  {http://arxiv.org/abs/http://arxiv.org/abs/2002.00777}
  {http://arxiv.org/abs/2002.00777} \BibitemShut {NoStop}%
\bibitem [{\citenamefont {Trinschek}\ \emph {et~al.}(2020)\citenamefont
  {Trinschek}, \citenamefont {Stegemerten}, \citenamefont {John},\ and\
  \citenamefont {Thiele}}]{TSJT2020pre}%
  \BibitemOpen
  \bibfield  {author} {\bibinfo {author} {\bibfnamefont {S.}~\bibnamefont
  {Trinschek}}, \bibinfo {author} {\bibfnamefont {F.}~\bibnamefont
  {Stegemerten}}, \bibinfo {author} {\bibfnamefont {K.}~\bibnamefont {John}}, \
  and\ \bibinfo {author} {\bibfnamefont {U.}~\bibnamefont {Thiele}},\ }\href
  {\doibase x} {\bibfield  {journal} {\bibinfo  {journal} {Phys. Rev. E}\
  }\textbf {\bibinfo {volume} {x}},\ \bibinfo {pages} {x} (\bibinfo {year}
  {2020})},\ \bibinfo {note} {(at press)},\ \Eprint
  {http://arxiv.org/abs/http://arxiv.org/abs/1911.08258}
  {http://arxiv.org/abs/1911.08258} \BibitemShut {NoStop}%
\bibitem [{\citenamefont {Suzuki}\ \emph {et~al.}(1995)\citenamefont {Suzuki},
  \citenamefont {Ohta}, \citenamefont {Mimura},\ and\ \citenamefont
  {Sakaguchi}}]{SOMS_PRE95}%
  \BibitemOpen
  \bibfield  {author} {\bibinfo {author} {\bibfnamefont {M.}~\bibnamefont
  {Suzuki}}, \bibinfo {author} {\bibfnamefont {T.}~\bibnamefont {Ohta}},
  \bibinfo {author} {\bibfnamefont {M.}~\bibnamefont {Mimura}}, \ and\ \bibinfo
  {author} {\bibfnamefont {H.}~\bibnamefont {Sakaguchi}},\ }\href {\doibase
  10.1103/PhysRevE.52.3645} {\bibfield  {journal} {\bibinfo  {journal} {Phys.
  Rev. E}\ }\textbf {\bibinfo {volume} {52}},\ \bibinfo {pages} {3645}
  (\bibinfo {year} {1995})}\BibitemShut {NoStop}%
\bibitem [{\citenamefont {Lin}\ \emph {et~al.}(2016)\citenamefont {Lin},
  \citenamefont {Rogers}, \citenamefont {Tseluiko},\ and\ \citenamefont
  {Thiele}}]{LRTT2016pf}%
  \BibitemOpen
  \bibfield  {author} {\bibinfo {author} {\bibfnamefont {T.-S.}\ \bibnamefont
  {Lin}}, \bibinfo {author} {\bibfnamefont {S.}~\bibnamefont {Rogers}},
  \bibinfo {author} {\bibfnamefont {D.}~\bibnamefont {Tseluiko}}, \ and\
  \bibinfo {author} {\bibfnamefont {U.}~\bibnamefont {Thiele}},\ }\href
  {\doibase 10.1063/1.4959890} {\bibfield  {journal} {\bibinfo  {journal}
  {Phys. Fluids}\ }\textbf {\bibinfo {volume} {28}},\ \bibinfo {pages} {082102}
  (\bibinfo {year} {2016})},\ \Eprint
  {http://arxiv.org/abs/http://arxiv.org/abs/1511.01167}
  {http://arxiv.org/abs/1511.01167} \BibitemShut {NoStop}%
\bibitem [{\citenamefont {Tewes}\ \emph {et~al.}(2019)\citenamefont {Tewes},
  \citenamefont {Wilczek}, \citenamefont {Gurevich},\ and\ \citenamefont
  {Thiele}}]{TWGT2019prf}%
  \BibitemOpen
  \bibfield  {author} {\bibinfo {author} {\bibfnamefont {W.}~\bibnamefont
  {Tewes}}, \bibinfo {author} {\bibfnamefont {M.}~\bibnamefont {Wilczek}},
  \bibinfo {author} {\bibfnamefont {S.~V.}\ \bibnamefont {Gurevich}}, \ and\
  \bibinfo {author} {\bibfnamefont {U.}~\bibnamefont {Thiele}},\ }\href
  {\doibase 10.1103/PhysRevFluids.4.123903} {\bibfield  {journal} {\bibinfo
  {journal} {Phys. Rev. Fluids}\ }\textbf {\bibinfo {volume} {4}},\ \bibinfo
  {pages} {123903} (\bibinfo {year} {2019})}\BibitemShut {NoStop}%
\bibitem [{\citenamefont {Praetorius}\ \emph {et~al.}(2018)\citenamefont
  {Praetorius}, \citenamefont {Voigt}, \citenamefont {Wittkowski},\ and\
  \citenamefont {L{\"o}wen}}]{praetorius2018active}%
  \BibitemOpen
  \bibfield  {author} {\bibinfo {author} {\bibfnamefont {S.}~\bibnamefont
  {Praetorius}}, \bibinfo {author} {\bibfnamefont {A.}~\bibnamefont {Voigt}},
  \bibinfo {author} {\bibfnamefont {R.}~\bibnamefont {Wittkowski}}, \ and\
  \bibinfo {author} {\bibfnamefont {H.}~\bibnamefont {L{\"o}wen}},\ }\href@noop
  {} {\bibfield  {journal} {\bibinfo  {journal} {Physical Review E}\ }\textbf
  {\bibinfo {volume} {97}},\ \bibinfo {pages} {052615} (\bibinfo {year}
  {2018})}\BibitemShut {NoStop}%
\bibitem [{\citenamefont {Doostmohammadi}\ \emph {et~al.}(2017)\citenamefont
  {Doostmohammadi}, \citenamefont {Shendruk}, \citenamefont {Thijssen},\ and\
  \citenamefont {Yeomans}}]{doostmohammadi2017onset}%
  \BibitemOpen
  \bibfield  {author} {\bibinfo {author} {\bibfnamefont {A.}~\bibnamefont
  {Doostmohammadi}}, \bibinfo {author} {\bibfnamefont {T.~N.}\ \bibnamefont
  {Shendruk}}, \bibinfo {author} {\bibfnamefont {K.}~\bibnamefont {Thijssen}},
  \ and\ \bibinfo {author} {\bibfnamefont {J.~M.}\ \bibnamefont {Yeomans}},\
  }\href@noop {} {\bibfield  {journal} {\bibinfo  {journal} {Nature
  communications}\ }\textbf {\bibinfo {volume} {8}},\ \bibinfo {pages} {1}
  (\bibinfo {year} {2017})}\BibitemShut {NoStop}%
\bibitem [{\citenamefont {Liehr}(2013)}]{liehr2013dissipative}%
  \BibitemOpen
  \bibfield  {author} {\bibinfo {author} {\bibfnamefont {A.~W.}\ \bibnamefont
  {Liehr}},\ }\href@noop {} {\emph {\bibinfo {title} {Dissipative solitons in
  reaction-diffusion systems}}}\ (\bibinfo  {publisher} {Springer},\ \bibinfo
  {year} {2013})\BibitemShut {NoStop}%
\bibitem [{\citenamefont {Nishiura}\ \emph {et~al.}(2005)\citenamefont
  {Nishiura}, \citenamefont {Teramoto},\ and\ \citenamefont
  {Ueda}}]{NTU_Chaos05}%
  \BibitemOpen
  \bibfield  {author} {\bibinfo {author} {\bibfnamefont {Y.}~\bibnamefont
  {Nishiura}}, \bibinfo {author} {\bibfnamefont {T.}~\bibnamefont {Teramoto}},
  \ and\ \bibinfo {author} {\bibfnamefont {K.-I.}\ \bibnamefont {Ueda}},\
  }\href {\doibase 10.1063/1.2087127} {\bibfield  {journal} {\bibinfo
  {journal} {Chaos: An Interdisciplinary Journal of Nonlinear Science}\
  }\textbf {\bibinfo {volume} {15}},\ \bibinfo {pages} {047509} (\bibinfo
  {year} {2005})},\ \Eprint
  {http://arxiv.org/abs/https://doi.org/10.1063/1.2087127}
  {https://doi.org/10.1063/1.2087127} \BibitemShut {NoStop}%
\bibitem [{\citenamefont {Moskalenko}\ \emph {et~al.}(2003)\citenamefont
  {Moskalenko}, \citenamefont {Liehr},\ and\ \citenamefont
  {Purwins}}]{MoskalenkoEPL2003}%
  \BibitemOpen
  \bibfield  {author} {\bibinfo {author} {\bibfnamefont {A.~S.}\ \bibnamefont
  {Moskalenko}}, \bibinfo {author} {\bibfnamefont {A.~W.}\ \bibnamefont
  {Liehr}}, \ and\ \bibinfo {author} {\bibfnamefont {H.-G.}\ \bibnamefont
  {Purwins}},\ }\href@noop {} {\bibfield  {journal} {\bibinfo  {journal}
  {Europhys. Lett.}\ }\textbf {\bibinfo {volume} {63}},\ \bibinfo {pages} {361}
  (\bibinfo {year} {2003})}\BibitemShut {NoStop}%
\bibitem [{\citenamefont {Buttinoni}\ \emph {et~al.}(2013)\citenamefont
  {Buttinoni}, \citenamefont {Bialk\'e}, \citenamefont {K\"ummel},
  \citenamefont {L\"owen}, \citenamefont {Bechinger},\ and\ \citenamefont
  {Speck}}]{speckPRL110}%
  \BibitemOpen
  \bibfield  {author} {\bibinfo {author} {\bibfnamefont {I.}~\bibnamefont
  {Buttinoni}}, \bibinfo {author} {\bibfnamefont {J.}~\bibnamefont {Bialk\'e}},
  \bibinfo {author} {\bibfnamefont {F.}~\bibnamefont {K\"ummel}}, \bibinfo
  {author} {\bibfnamefont {H.}~\bibnamefont {L\"owen}}, \bibinfo {author}
  {\bibfnamefont {C.}~\bibnamefont {Bechinger}}, \ and\ \bibinfo {author}
  {\bibfnamefont {T.}~\bibnamefont {Speck}},\ }\href {\doibase
  10.1103/PhysRevLett.110.238301} {\bibfield  {journal} {\bibinfo  {journal}
  {Phys. Rev. Lett.}\ }\textbf {\bibinfo {volume} {110}},\ \bibinfo {pages}
  {238301} (\bibinfo {year} {2013})}\BibitemShut {NoStop}%
\end{thebibliography}
\end{document}